\documentclass[]{interact}
\usepackage{xcolor} 
\usepackage{epstopdf}
\usepackage[caption=false]{subfig}
\usepackage[numbers,sort&compress]{natbib}

\usepackage[colorlinks=true,linkcolor=blue,citecolor=blue]{hyperref}
\usepackage{journals}  
\usepackage{hyperref}
\hypersetup{
    colorlinks=true,
    linkcolor=blue,
    filecolor=magenta,      
    urlcolor=cyan,
    pdfpagemode=FullScreen,
    }
\usepackage{booktabs}   
\usepackage{threeparttable} 

\bibpunct[, ]{[}{]}{,}{n}{,}{,}
\renewcommand\bibfont{\fontsize{10}{12}\selectfont}
\makeatletter
\def\NAT@def@citea{\def\@citea{\NAT@separator}}
\makeatother

\theoremstyle{plain}

\theoremstyle{definition}

\theoremstyle{remark}

\begin{document}
\articletype{ARTICLE TEMPLATE}

\title{Exploration of Exoplanet Atmospheres with the James Webb Space Telescope}


\author{
\name{Sasha Hinkley\textsuperscript{a}\thanks{Email: S.Hinkley@exeter.ac.uk} and David K. Sing\textsuperscript{b,c}}\thanks{Email: dsing@jhu.edu}
\affil{
\textsuperscript{a}University of Exeter, Physics Building, Stocker Road, Exeter EX4 4QL, UK; \textsuperscript{b}Department of Earth \& Planetary Sciences, Johns Hopkins University, Baltimore, MD, USA;
\textsuperscript{c}William H.\ Miller III Department of Physics \& Astronomy, Johns Hopkins University, 3400 N Charles St, Baltimore, MD, USA
}
}

\maketitle
\section{Abstract}


The James Webb Space Telescope (JWST) is providing transformative characterization capabilities for the study of both transiting and directly imaged extrasolar planets.  Observations deep into the infrared in the form of both spectroscopy and coronagraphic imaging means that JWST is uniquely sensitive to most of the physical processes that dictate the properties of exoplanetary atmospheres. The observatory’s thermal and pointing stability directly translates into the extreme photometric precision needed for exquisite transiting exoplanet photometry, as well as highly sensitive coronagraphic observations.  Over the mission’s 20-25 year lifetime, roughly 10$^4$ hours will be dedicated to observations of exoplanet atmospheres, providing a legacy data archive that will allow population level characterization of exoplanet atmospheres spanning terrestrial planets, sub-Neptunes and gas giants.  This effort will also assess the prevalence of atmospheres of rocky planets and thus the degree to which such planets may have conditions conducive to habitability.

\section{Introduction \& Scientific Background}
\subsection{History of the JWST Mission} 
The James Webb Space Telescope (JWST) was conceived with the goal of addressing the most fundamental questions in modern astrophysics. The broad science themes for the mission relate to determining the ionization history of the early universe and the assembly of the first galaxies, but also fundamental questions about the birth and evolution of stars and the chemical properties of planetary systems\cite{2006SSRv..123..485G}. The light from galaxies found in the early universe is redshifted to near- and mid-infrared wavelengths, and substellar objects (e.g, planet or brown dwarfs) emit at blackbody temperatures characteristic of this same wavelength range.  A large, stabilized, space-based infrared observatory was thus conceived to be able to spectrally and spatially resolve these objects that are extremely faint, especially compared to the infrared background levels of ground-based observatories.

\begin{figure*}[h]
  \includegraphics[width=1.0\linewidth]{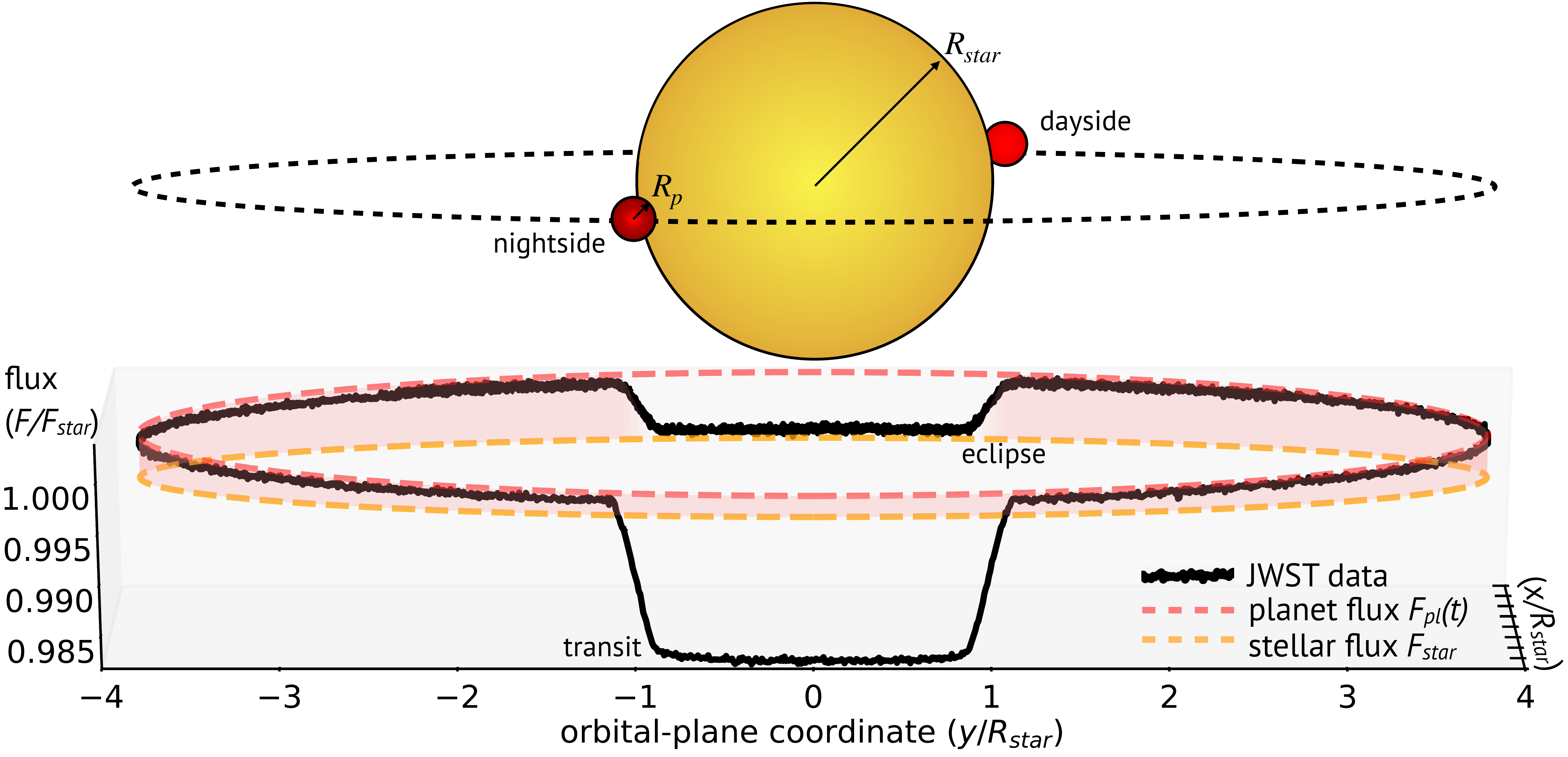}
  \vspace{-0.8cm}
  \caption{{\it Top:} Exoplanet orbital geometry for the transiting WASP-121A,b system. Plotted to scale. The dotted line traces the whole 1.27 day planetary orbit ($a/R_{star}$=3.8), with the planet depicted  during transit at ingress and secondary eclipse as it passes behind the star. {\it Bottom}: The total flux ($z$-axis) over the planetary orbit is plotted with the JWST NIRSpec/NRS2 phase curve data of WASP-121~b from \cite{2023ApJ...943L..17M,2025NatAs.tmp..123E} (black data), with the planet (light red dashed line) and stellar (gold dashed line) contributions indicated. The photometric data is pre-detrending/cleaning, so includes both photon and systematic noise, illustrating JWST's exquisite photometric precision. The planetary dayside flux is measured at eclipse, while the nightside flux is visible by-eye in the raw photometry as the planet flux during transit rises noticeably above the stellar flux level.}
  \label{fig:transit_geom}
\end{figure*}

As it happened, the stability and sensitivity requirements for characterizing faint cosmological objects at high redshift are essentially the same requirements as those for studying exoplanets, 
which was recognized by the forward-looking ``HST and beyond'' committee \cite{Dressler1996}, which 30 years ago made the case for JWST.
A thermally stable infrared observatory with a large aperture and very low infrared background leads naturally to exquisite photometry, perfect for detecting and characterizing exoplanets.  The broad wavelength coverage from $\sim$1\,$\mu$m extending deep into the mid-infrared to $\sim$20\,$\mu$m provides access to the numerous molecular spectral features that characterize exoplanetary atmospheres. Since the relative brightness between a planet and its host star is dramatically reduced in the mid- and thermal-infrared, this long wavelength coverage is also highly beneficial for \textit{directly imaging} gas giant exoplanets using a coronagraph, a set of optical masks used to suppress the overwhelming brightness of the host star \cite{2009ARA&A..47..253O, Galicher2024}.  
%
%
Exoplanet science influenced the design of JWST, but it was not the primary driver. The core design requirements---the large cold infrared-optimized telescope, the wavelength range, and the sensitivity---were predominantly driven by the need to detect and study the first luminous objects in the universe at high redshift. But it has only been in the last couple of decades that the exoplanet characterization potential of JWST has been fully realized.  Indeed, the study of \textit{transiting} exoplanets was only mentioned for two paragraphs in an entire 122-page introductory overview article 20 years ago \citep{2006SSRv..123..485G}. 
Indeed, JWST was not solely designed from the beginning to be optimized for characterizing exoplanets, but if it had been, the resulting observatory and instruments might not have been dissimilar from the present JWST. 

The study of exoplanets has expanded to be a major subfield in astrophysics, and with roughly one-third of JWST’s overall time allocation in each cycle so far dedicated to exoplanets, it has become a clear science driver for the mission. The highly successful launch by ESA at the end of 2021, combined with diligent mass management during fabrication, means that the remaining spacecraft propellant is more than expected, indicating JWST will have a lifetime hopefully longer than 20 years, much longer than initially planned \cite{2023PASP..135d8001R}. Indeed, it is likely that JWST’s lifespan will be limited by the lifetime of the instrument hardware, rather than the remaining spacecraft propellant. With such a long planned lifetime, and assuming the study of exoplanets remains a science driver, it can be easily estimated that $\sim$30,000 hours of observations will be dedicated to the discovery and characterization of exoplanets over JWST’s lifetime.

\subsection{Exoplanet Demographics, JWST's Characterization Capabilities, \& Major Science Themes} 

\begin{figure*}[h]
  \centering
  \includegraphics[width=0.75\linewidth]{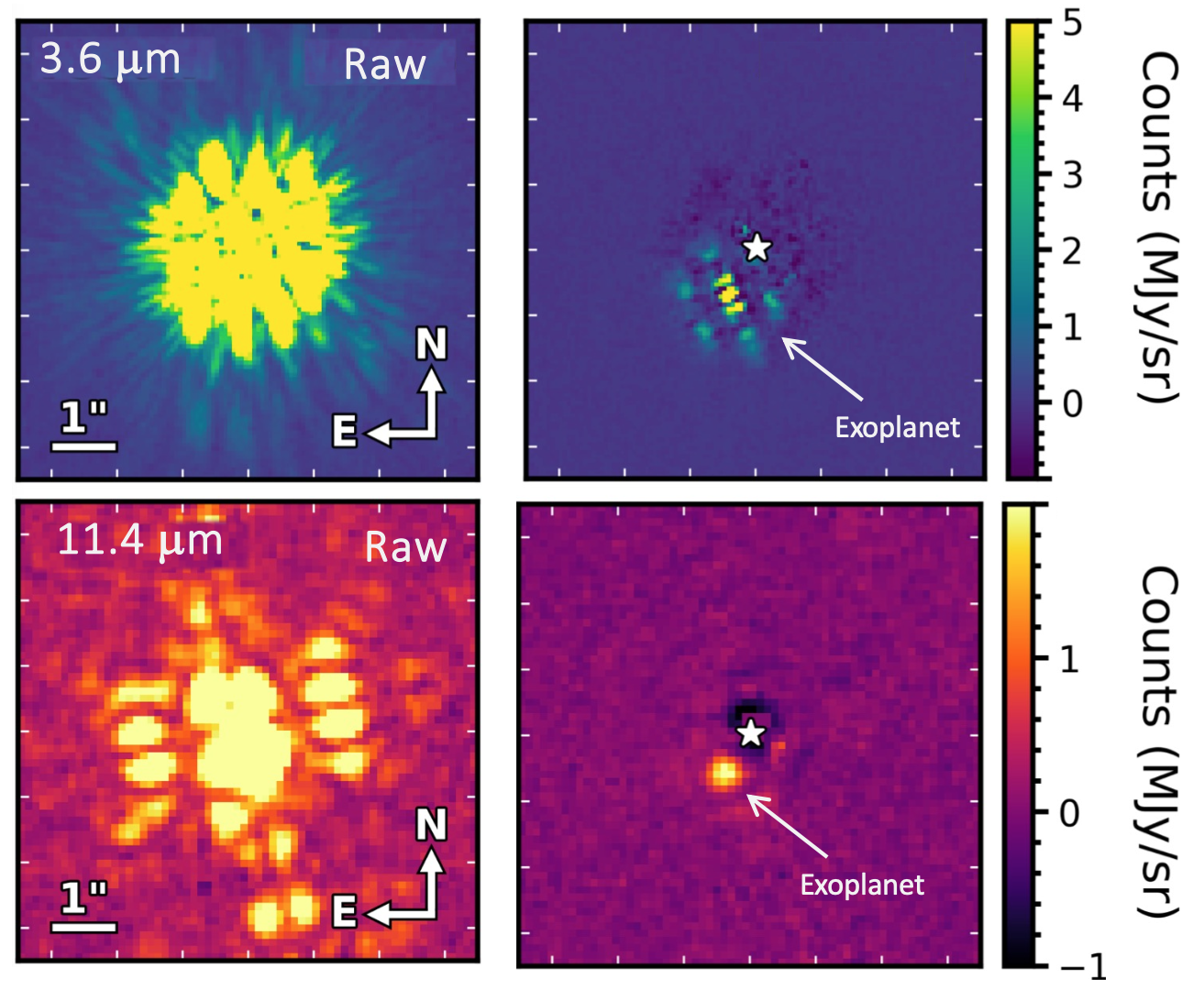}
  \caption{A figure adapted from \cite{Carter2023} showing JWST images of the planet-hosting star HIP\,65426 at 3.6\,$\mu$m using JWST/NIRCam (top row) and the MIRI instrument at 11.4\,$\mu$m (bottom row). In both of the left panels the coronagraphic mask is in place in the center of the image, and the residual uncorrected starlight can be clearly seen.  The right panels show the same data following the post-processing removal of this residual scattered starlight revealing detections of the exoplanet to the lower left of the coronagraphic mask.} 
  \label{fig:HIP65426b}
\end{figure*}

\begin{figure*}[h]
  \centering
  \includegraphics[width=1.0\linewidth]{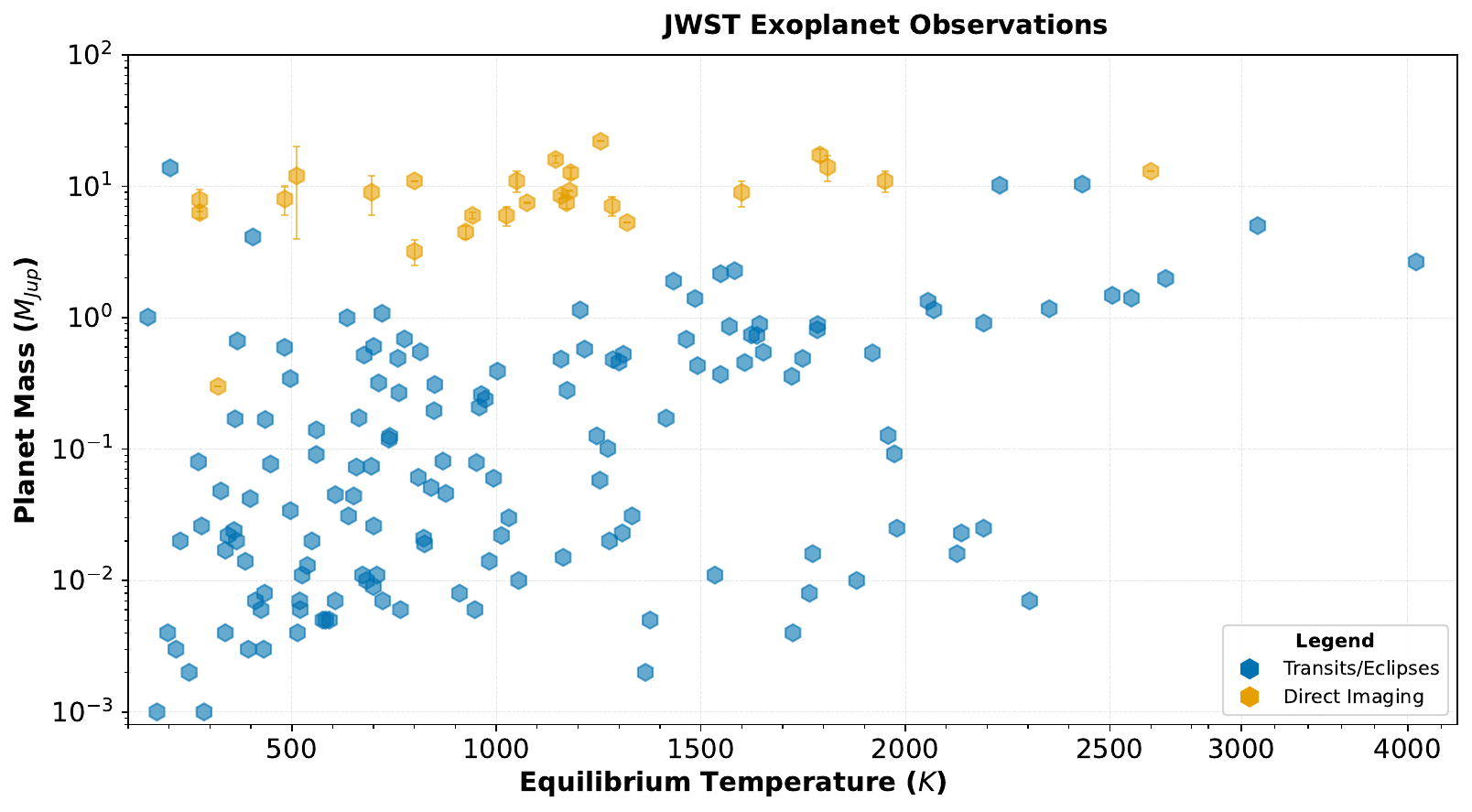}
  \vspace{-1cm}
  \caption{The distribution of exoplanets and planetary mass companions that have been observed so far with JWST via transits/eclipses (blue points) as well as direct imaging (orange points) expressed in terms of their masses and equilibrium temperatures. While these two populations span a similar range of temperatures, the masses sampled by each technique are starting to overlap, a trend that is expected to increase in future JWST cycles.} 
  \label{fig:mass_temp_jwst_targets}
  \vspace{-0.5cm}
\end{figure*}

The most prolific technique for the detection of exoplanets over the past $\sim$20 years has been the transit method \cite{2000ApJ...529L..45C},  in which an exoplanet crosses the face of its host star in an edge-on orbit providing a size and period of the orbiting planet. 
From the point of view of exoplanet atmospheric characterization, the overall challenge is a single one of contrast: separating the very faint signal of the exoplanet from the overwhelming brightness of the host star. 
In the case of transiting planets, in which the signal of the transit event is isolated from the stellar signal \textit{temporally}, the variation of the transit signal can be analyzed with wavelength, 
resulting in transit transmission spectroscopy \cite{2016Natur.529...59S, 2018arXiv180407357S, 2019AJ....158...91S}.
For the direct imaging case, Figure~\ref{fig:HIP65426b} highlights that the faint exoplanet light must be \textit{spatially} separated in the image from the very bright host star light using a coronagraph and image post processing techniques (e.g., \cite{2009ARA&A..47..253O, 2023PASP..135i3001F}).
One of the most powerful aspects of JWST is its ability to measure exoplanet properties across a wide variety of exoplanet ages, masses, and orbital separations, from the ubiquitous transiting ``sub-Neptune’’ and ``super-Earth’’ populations to warm young Jovians, as well as older, cold gas giant planets via direct imaging.  JWST is also unique in the sense that it is the only platform that is able to provide atmospheric characterization through spectroscopy of the transiting exoplanet population as well as those accessible only to direct imaging. Figure~\ref{fig:mass_temp_jwst_targets} shows the mass and temperature for transiting and direct imaging planetary mass companions targeted by JWST in the first four cycles. There is a near complete overlap in the temperatures probed between these populations, and some overlap in mass as well, which will be improved upon in future cycles.

JWST's remarkable diversity of characterization capabilities, as well as diversity of targets accessible for study shown in Figure~\ref{fig:mass_temp_jwst_targets}, means that JWST has the potential to address some of these most fundamental science questions in modern exoplanetary science.  Tables~\ref{tab:DI_table} and \ref{transit-table} list the targets that have been observed with JWST so far using direct imaging, or transit transmission spectroscopy, respectively.  The tables also list key parameters of the objects (mass, temperature, radius), their JWST program ID, and the publications so far that have showcased the observations of each target.  
Before beginning our discussion of the capabilities and accomplishments of JWST so far, it is worth pausing to assess where we stand with our overall understanding of extrasolar planets after only a few cycles of observations. 
Despite the outstanding promise of JWST's capabilities, major questions related to how to place our own solar system in context, as well as the physics of giant planet formation---the accretion of solids followed by the runaway buildup of a gaseous envelope \cite{Pollack1996} versus formation by an instability in a disk \cite{Kratter2016}---remain elusive.  
For example, while JWST is delivering precise spectroscopic elemental abundances of exoplanet atmospheres, which may be a clue to formation pathways, early works are finding a wide range of elemental abundances \cite[e.g.,][]{2025arXiv250916082G, 2025arXiv251026785L}. 
From the point of view of exoplanet ``demographics'', our own solar system appears to be unusual, with a lack of the commonly-found sub-Neptune/super-Earth exoplanets orbiting close to our Sun. Possible compositions of these planets range from rocky planets with H/He envelopes, to water worlds. Atmospheric characterization of these objects will thus be especially valuable, and while JWST is the only observatory capable of characterizing this type of planet, many sub-Neptunes are showing near-featureless spectra \cite{2024AJ....168...77W, 2025arXiv251215338F, 2024ApJ...974L..33S, 2024ApJ...974L..10P}. 
Perhaps the largest driving question is the degree to which rocky, terrestrial planets are capable of retaining an atmosphere, one of the prerequisite conditions for life as we know it to exist on any planet. Only JWST is capable of characterizing the atmospheres of transiting terrestrial exoplanets, and is providing our only opportunity to assess the commonality of such atmospheres. 
During its lifetime, it is very unlikely that a conclusive detection of a biosignature with JWST will be obtained, and we can expect many false positives along the way.  However, even before the exoplanet community selects the detection of biosignatures as a major observational goal of JWST, first establishing the broad landscape of exoplanet properties as measured through both transit transmission spectroscopy as well as direct imaging is critical, and will be a focus for roughly the next decade. 
%
	
In this review, in \S\,3 and 4 we give a broad overview of the advantages of carrying out exoplanet direct imaging with JWST, and highlight several recent results, respectively. In \S\,5 we discuss the relevant exoplanet atmospheric physics accessible using transmission spectroscopy of transiting systems, and present some recent results in \S\,6.  In \S\,7, we discuss expectations for the future of the mission, and the achieved science that we can expect at the end of JWST’s lifetime. Additional material on some of these related themes can also be found in \cite{Espinoza2026}.

\section{Directly Imaging Exoplanets with JWST}

\subsection{The Promise \& Challenges of Direct Imaging} 
As the indirect transit and radial velocity exoplanet detection methods are inherently much less sensitive to wide-separation planets with long orbital periods, the technique of exoplanet direct imaging will be the best method to fully define the outermost extents of planetary systems, and provide a more complete understanding of the true frequency of planetary mass companions to nearby stars \cite{Bowler2016, Nielsen2019, Squicciarini2025a}.  Perhaps most importantly, by spatially separating the light of the host star and the extremely faint planet, the direct imaging technique is naturally suited to direct spectroscopy of planets themselves \cite{Macintosh2015, Chauvin2017}, providing essential physical characteristics (e.g. temperature, surface gravity) and abundances of chemical species, making the directly imaged planets amongst the most valuable exoplanet population for atmospheric characterisation.  Further, upcoming exoplanet characterisation work using direct imaging in the next 2-5 years will be an essential step on the ``roadmap'' to detecting exo-earths and the search for biosignatures $\sim$20 years from now.  Consequently, direct imaging will be a leading technique going forward for the discipline of exoplanetary science, and is a key science driver for several current upcoming flagship space missions. For these reasons, the James Webb Space Telescope (JWST), which is so far only sensitive to thermal infrared emission from directly imaged planets, is proving to be transformative for directly characterizing exoplanet atmospheres in the mid and thermal infrared (3-28\,$\mu$m).  Recent observations with JWST are providing abundances of the dominant molecules in exoplanet atmospheres (e.g., CH$_4$, CO, CO$_2$, H$_2$O and NH$_3$), using multiple features and a much broader wavelength range \cite{Miles2023, Xuan2026, Ruffio2026, Whiteford2026} to distinguish between clouds and molecular absorption.  Furthermore, as discussed below, provided that the capability of JWST is maximised, the stability and sensitivity of space-based observing in the thermal-infrared ($\gtrsim$\,5\,$\mu$m) means that JWST will likely discover numerous more directly imaged exoplanets, including young Saturn and Neptune analogues \cite{Carter2021, Lagrange2025} at wide orbital separations ($>$10 AU).

\begin{figure*}[h]
  \includegraphics[width=1.05\linewidth]{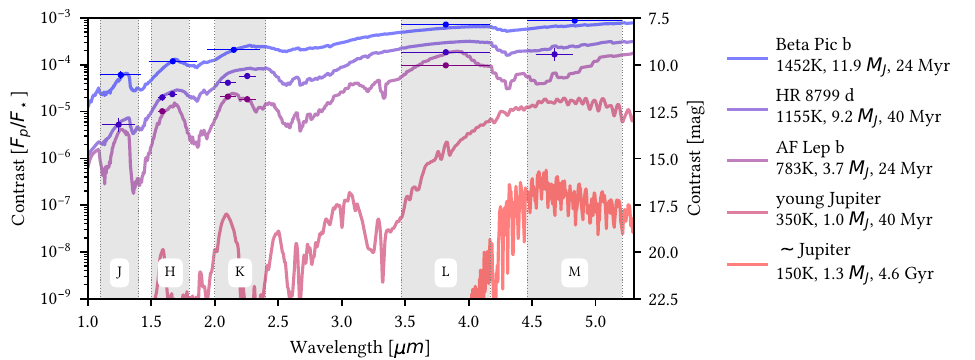}
  \vspace{-1cm}
  \caption{A figure showing the brightness contrast of various young ($\sim$24-40 million year old) planetary mass companions with a range of masses/temperatures (top three blue/purple curves), as well as Jupiter itself at young and old ages (lower two magenta and red curves).  In the mid-infrared regions (3-5\,$\mu$m) where JWST has superb sensitivity, a wide range of objects with varying mass and effective temperatures nonetheless have similar contrasts of $10^{-4}$ to $10^{-6}$, bringing them within reach to the JWST coronagraphs. The shaded regions indicate the ranges of wavelength accessible from the ground. Figure courtesy of Markus Bonse (ESO) from \cite{Bonse2025thesis}, and used with permission. 
  }  
  \label{fig:contrasts}
\end{figure*}

The singular task for achieving a direct image of an extrasolar planet is to overcome the very high planet-to-star brightness contrast ratio.  Depending on its age (and host star brightness), Figure~\ref{fig:contrasts} shows that the intrinsic thermal emission of a young Jovian mass extrasolar planet ($\sim$1 to $\sim$10 M$_{\rm Jup}$) in the near and mid-infrared will typically have a brightness relative to its host star of 10$^{-4}$ to 10$^{-6}$. This contrast ratio can be made more favourable primarily in two ways: the first is to observe younger planetary systems (typically tens to hundreds of millions of years) in which the intrinsic thermal emission from a planet may be elevated substantially \cite{Baraffe2015, Phillips2020} compared to a planet with an age comparable to our solar system (a few billion years). The second is to observe the exoplanet at a wavelength closer to the peak of its blackbody emission, typically at mid- and thermal-infrared wavelengths ($\gtrsim$3\,$\mu$m).  In addition to the increased brightness of the planet at these wavelengths, the brightness of the host star is rapidly decreasing with wavelength in the Rayleigh-Jeans tail, with both of these effects making for a more favourable contrast ratio. 
Figure~\ref{fig:contrasts} shows the brightness contrast of various substellar objects relative to a solar type star for several planetary mass objects at a variety of masses or evolutionary states, including: young ($\sim$20-40 million years) planetary mass companions ($\sim$4-12 Jupiter masses or $M_{Jup}$,  800-1450K), a young ($\sim$40 million year) analogue of our own Jupiter (350\,K), and a mature version of our own Jupiter ($\sim$4.6 billion years, 150\,K).  The expected contrasts of the younger objects are typically in the range of 10$^{-3}$ to $\sim$10$^{-6}$ at  3-5\,$\mu$m in the $L$ and $M$-bands.  In contrast, at optical or near infrared wavelengths (0.5 – 2\,$\mu$m), these objects span a much wider range of contrasts from 10$^{-4}$ to less than 10$^{-10}$.  The key message is that the wavelength range longward of 2\,$\mu$m not only provides access to a variety of different masses and temperatures, it is also sensitive to planets at a range of various \textit{evolutionary stages}. JWST is the only observing platform for the forseeable future that can achieve such contrasts at these wavelength ranges, and thus access such a wide variety of planetary mass companions.  

For the case of observing Jovian mass planets in \textit{reflected} light (typically at optical wavelengths $\sim$500\,nm, and thus out of reach to JWST), the contrast  of a such an object will typically be several orders of magnitude \textit{fainter} than the above contrasts, and thus reflected light observations have so far been out-of-reach to direct imaging efforts. 
However, for future imaging of terrestrial planets with platforms like the upcoming \textit{Nancy Grace Roman Space Telescope} and the \textit{Habitable Worlds Observatory (HWO)}, reflected light at visible wavelengths will be the  primary means for detecting such planets, demanding very ambitious contrasts of $\sim$10$^{-9}$ (or better) be reached, requiring active spacecraft wave front control via deformable mirrors to minimize small phase aberrations in the incoming stellar wave front.

To achieve contrasts of 10$^{-4}$ to 10$^{-6}$ for a space telescope such as JWST operating in the near- or mid-infrared, typically a two stage process is required. The first stage involves the use of a coronagraph (a sequence of optical masks within the imaging camera that suppress light in both focal planes and pupil planes) to suppress the overwhelming brightness of the host star \cite{Follette2023, Galicher2024}.  However, due primarily to small residual phase aberrations in the incoming stellar wave front typically at the level of tens or hundreds of nanometres which translates to an imperfect point spread function (PSF), the coronagraphic suppression is typically not perfect. This results in a substantial amount of residual scattered starlight that plagues the coronagraphic image.  This can be seen in the left panels of Figure~\ref{fig:HIP65426b}. 
The second stage of the direct imaging process involves calibrating this residual scattered starlight by creating an optimized model of it \cite{Marois2006, Lafreniere2007, Soummer2012, Bonse2025}, and subsequently removing it from the image.  It is only after these two steps that high contrast observations can be achieved.

 \subsection{Space-Based Direct Imaging: Advantages \& Limitations}
One of the most immediate advantages afforded by space-based observatories is the lack of phase aberrations in the incoming stellar wave front that are incurred due to turbulence in the Earth’s atmosphere.  Such turbulence results in variations in the refractive index of the atmosphere on a range of spatial and temporal scales.  In practice, such turbulence results in a wave front with an aberrated phase that is typically corrected by an Adaptive Optics (``AO'') system \cite{Oppenheimer2009, Davies2012, Follette2023}.  However, AO systems have varying levels of performance, depending on the strength and timescale of atmospheric turbulence, and the quality of the observing site.  Thus, given their relative stability and long timescale for variations in wave front phase aberrations, space-based observatories may be the best positioned to achieve the fundamental resolution ($\sim$$\lambda/D$) of a telescope which is governed solely by the wavelength of observation $\lambda$, and the diameter $D$ of the telescope aperture.  
On the other hand, next-generation ground-based observatories currently under construction have the distinct advantage that they can have very large primary mirror apertures of 25-40m, allowing never-before achieved angular resolution and light collecting capabilities. Building such an aperture for a space-based observatory would be prohibitively expensive, and thus space-based observatories are limited to a fraction of this, ranging from 0.85m in the case of the \textit{Spitzer Space Telescope}, to 2.4\,m for the \textit{Hubble Space Telescope (HST)}, and more recently 6.5\,m for JWST, inherently limiting their angular resolution. 

HST, which is optimized for observations in the ultraviolet and optical wavelengths, still provides a good level of angular resolution at optical wavelengths despite the rather small 2.4\,m aperture ($\lambda/D$\,$\simeq$\,40 miliarcseconds at a wavelength of $\sim$500\,nm).  This very good angular resolution coupled with outstanding sensitivity at optical wavelengths made HST highly sensitive to optical scattered light imaging of diffuse, extended circumstellar disks early in the HST mission \cite[e.g., ][]{Stapelfeldt1998, Clampin2003, Krist2005, Golimowski2006}. However, for directly imaging exoplanets that typically will have temperatures of a few hundred Kelvin up to $\sim$1500\,K, observations in the mid- and thermal infrared are necessary.  


\subsection{The Advantages of Direct Imaging with JWST}


In addition to observing planets closer to the peak of their thermal emission, obtaining exoplanet photometry and spectroscopy in the mid-infrared at 3–5\,$\mu$m provides access not only to markers of composition, but also many of the key atmospheric processes that shape the spectra of planetary-mass companions, e.g., vertical atmospheric mixing \cite{Hubeny2007, Barman2011}, cloud physics, and the presence of patchy cloud structures. The mid-infrared can thus dramatically improve differentiation between equilibrium and disequilibrium atmospheric chemistry \cite{Skemer2012, Phillips2020, Miles2020}, which cannot be entirely accessed through observations at 1–2\,$\mu$m alone. Thus, incorporating data at these wavelengths can dramatically help to break the persistent degeneracies (see e.g. the discussion in \cite{Chabrier2007}) in the interpretation of exoplanet spectra that might be in fact driven by enhanced elemental abundances, or various physical processes such as vertical atmospheric mixing \cite{Konopacky2013}.

Along with the sizeable collecting area delivered by its 6.5m diameter aperture compared to previous space telescopes such as \textit{HST} and \textit{Spitzer}, the most significant advantage delivered by JWST is its remarkable infrared sensitivity.  Any astronomical observations operating in the infrared will suffer from a level of noise in the form of infrared background radiation due to: 1) the intrinsic thermal blackbody emission of the Earth’s atmosphere ($\sim$220\,K) which is almost always the dominant source for ground-based observations; 2) faint emission zodiacal dust within our own solar system; 3) diffuse, unresolved faint stars in our galaxy; or 4) infrared cirrus.  For JWST, the lack of the ``telluric’’ atmospheric background is a paradigm-changing advantage compared to ground-based observatories.  Indeed, in the portion of JWST coronagraphic images that are ``background-limited'' (as opposed to the regions within $\simeq$1 arcsecond that are dominated by unsuppressed scattered starlight and thus ``contrast limited’’), JWST has $\sim50\times$ the sensitivity of dedicated exoplanet imaging platforms that operate at 2\,$\mu$m such as the Gemini Planet Imager \cite[``GPI,''][]{Macintosh2015}, or the SPHERE instrument at VLT \cite{Beuzit2019}, and $\sim500\times$ the sensitivity of an infrared imager at Mauna Kea operating at 4\,$\mu$m \cite{Perrin2018, 2023PASP..135d8001R}.  This much higher telluric infrared background that plagues ground-based observatories, especially in the mid- and thermal infrared ($\gtrsim$2\,$\mu$m) means that extrasolar planets had, prior to JWST, never been \textit{directly} observed at wavelengths longer than 5\,$\mu$m.  Further, these orders of magnitude increases in sensitivity mean that JWST has sensitivity to much colder planets such as those detected by precision radial velocity studies, as well as analogues of our own solar system ice-giant planets that have the peak of their intrinsic thermal emission shifted to much longer wavelengths into the mid- and thermal infrared ($\sim$5-15\,$\mu$m). JWST has even shown the necessary sensitivity to directly image wide, young analogues of our own Neptune \cite{Carter2023, Lagrange2025}. 
Indeed, this demonstrated sensitivity has motivated several coronagraphic surveys \cite{Carter2024, Biller2025} in the first three cycles to search for sub-Jupiter mass analogues of our own ice-giant planets that could be orbiting nearby young stars, as well as more distant, but younger stars in nearby star forming regions \cite{Kammerer2025}.  
These surveys will likely put the first constraints on the demographics of wide-separation ice giant planet analogues to our own Saturn and Neptune, complementing upcoming microlensing surveys with the \textit{Roman Space Telescope}. 

As described in \cite{Perrin2018}, to achieve the necessary high brightness contrast in the close vicinity to a host star, exquisite calibration of the residual, uncorrected starlight (left panels of Figure~\ref{fig:HIP65426b}) in the image plane is absolutely critical.  This residual scattered starlight that is not suppressed by the coronagraph can be traced to small phase aberrations incurred in the primary mirror or the rest of the optical train.   Indeed, one of the clear legacies of high contrast observations with HST is that, non-optimal coronagraphic masks that fail to suppress a significant amount of starlight can still be highly effective for achieving high contrast provided that the pattern and morphology of the starlight is highly \textit{stable} in time and can thus be calibrated \cite{Perrin2018}.  
This high-level of image-to-image stability, ultimately due to the stability of the optics and instruments residing in a space-based environment \cite{Telfer2024, Feinberg2024}, is what allows this calibration to be so effective.  

This level of wavefront stability can easily be demonstrated through maps of the optical aberrations (i.e. deviations from the intended ideal optical surface) in the JWST primary mirror. These are regularly measured and quantified, with the telescope primary mirror settling down to a wave front error of typically $\sim$70 nanometres \cite{Telfer2024, Feinberg2024}. This has in part been helped by avoidance of impacts of micrometeoroids on the primary mirror, which can obviously have a dramatic impact on the overall wavefront stability.  This image stability is also highly beneficial for minimizing the systematic errors inherent in spectroscopic studies of transiting planets in which the placement of the star on the detector is required to be stable to within 1/1000$^{th}$ of a pixel over the course of days.


\section{Direct Imaging Results}

\subsection{Early Release Science: The First Gifts from JWST}

Prior to the launch of JWST, with some conservative estimates placing the unknown lifetime of JWST at only 5-10 years, it was recognized that the optimal observing strategies needed to be identified as soon as possible soon after commissioning. This was especially true for both direct observations of extrasolar planets as well as observations of transiting exoplanets.  To address this, the JWST Early Release Science (ERS) Programs, made up of 500 hours of Director's Discretionary Time, were introduced with the goal of generating representative datasets in observation modes expected to be widely used by individual communities. From 2016-17, the exoplanet direct imaging and transiting communites self-organized to formulate programs designed to address key questions about JWST performance that will inform future proposal cycles. A total of 106 proposals were submitted, with 13 ultimately selected spanning a range of science disciplines, and two of these 13 were dedicated to extrasolar planet observations and characterization. 

The Early Release Science Program (GO Program 1386) focussed on direct imaging  coronagraphy and spectroscopy of planetary mass companions and circumstellar disks \cite{Hinkley2022} was ultimately awarded 77-hours to carry out coronagraphy, spectroscopy, as well as interferometric imaging of planetary systems. A majority of the time was dedicated to coronagraphic observations of two systems:  HIP65426, a very young ($\sim$16 million year old) 2 solar mass (M$_{\odot}$) star in the Scorpius Centaurus star forming region with a previously discovered \cite{Chauvin2017} wide-separation (92\,AU) massive (6-9\,M$_{Jup}$) super-Jovian planet \cite{Carter2023}, as well as HD\,141569A \cite{Millar-Blanchaer2025}, a $\sim$few million year old circumstellar disk with several concentric dust rings previously imaged with HST and VLT/SPHERE \cite{Mouillet2001, Clampin2003, Perrot2016}.  HIP65426b, a young, self-luminous giant planet with only modest contrast (3$\times$10$^{-4}$ at 4.44\,$\mu$m), and good amount of angular separation relative to its host star (0.8 arcseconds) served as an ideal early target to test the performance of the JWST coronagraphs.  This object was observed with a range of JWST filters ranging from 2.5 to 15.5\,$\mu$m.  Figure~\ref{fig:HIP65426b} shows examples of early coronagraphy of HIP 65426b from these observations at 4.44 and 11.4\,$\mu$m.  This was a ground breaking dataset for many reasons. First, prior to these observations, as mentioned earlier no direct observations of planetary mass companions had been obtained at wavelengths longward than 5\,$\mu$m. Second, these observations clearly indicated that contrast performance of JWST was reaching, and indeed exceeding, pre-launch predictions \cite{2021MNRAS.501.1999C, Hinkley2022}, verifying that JWST would be sensitive to Uranus and Neptune Mass objects at tens of AU around nearby stars. Lastly, the filters sampling the observations of HIP65426b spanned $\sim$97\% of the full-luminous range of this object providing precise measurements of its bolometric luminosity and effective temperature \cite{Carter2023}. 

\begin{figure*}[h]
  \vspace{-1cm}
  \includegraphics[width=1.0\linewidth]{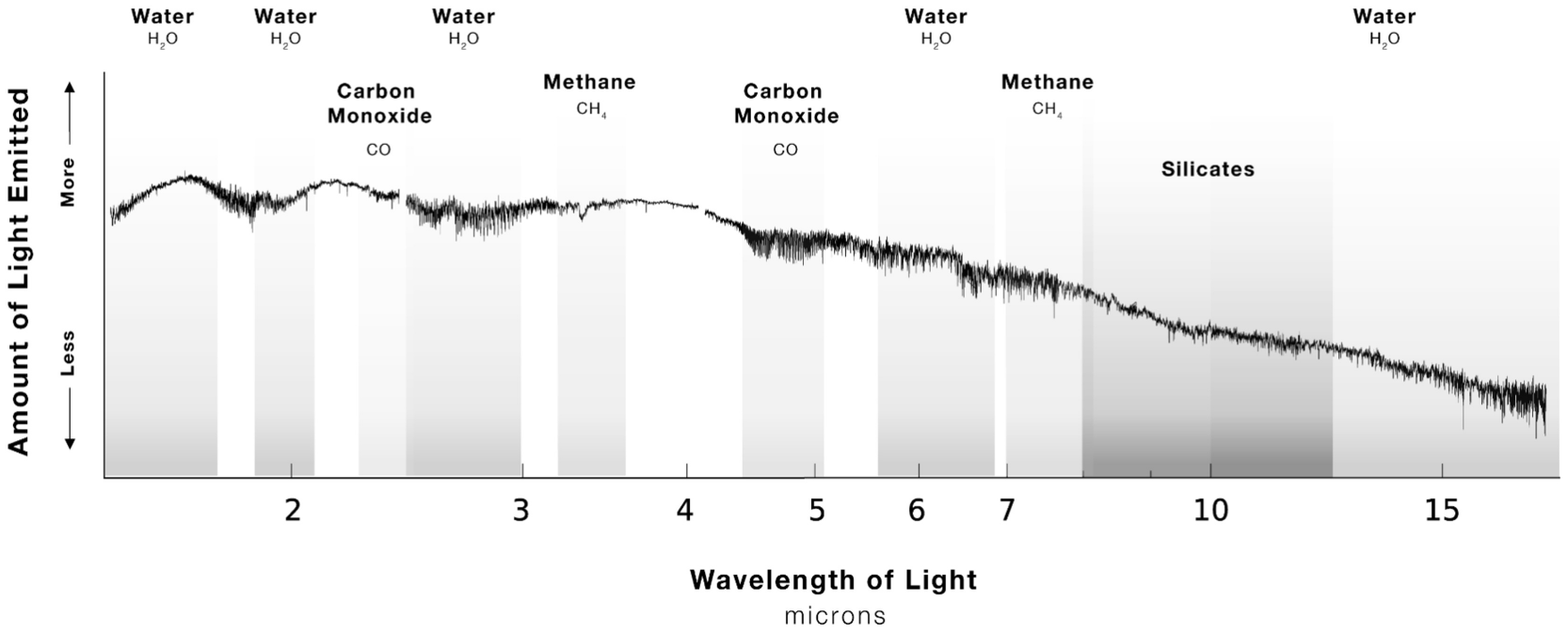}
  \vspace{-2.5cm}
  \caption{The full spectrum of the planetary mass companion VHS\,1256b \cite{Miles2023} obtained as part of the JWST Early Release Science Program for direct observations of exoplanetary systems \cite{Hinkley2022}. This is the first spectrum of a planetary mass companion extending in wavelength to $\sim$20\,$\mu$m, and covering the full luminous range of the object. 
  } 
  \label{fig:vhs1256b}
\end{figure*}

In addition to the ground-breaking coronagraphic observations of HIP65426b described above, the high contrast imaging ERS program also obtained the highest fidelity spectrum (Figure~\ref{fig:vhs1256b}) of a planetary mass object \cite{Miles2023, Whiteford2026}.  VHS\,J125601.92–125723.9\,b (hereafter ``VHS\,1256b'') is an ``adolescent'' ($\sim$300 million year) planetary mass companion orbiting a binary system \cite{Gauza2015}. Since none of the JWST spectrographs are equipped with coronagraphs, the wide angular separation ($\sim$8 arcseconds) of this companion from its host stars means that the spectrum would be vastly less susceptible contaminating light from the host stars. While the \textit{Spitzer} and \textit{Akari} missions obtained a handful of spectra of free-floating brown dwarfs \cite{Cushing2006, Sorahana2012}, never before had such a complete spectrum with such high spectral resolution been obtained for any substellar object.  This dataset, described in \cite{Espinoza2026} as ``nothing short of revolutionary'', was the first spectrum of a planetary mass companion extending in wavelength to $\sim$20\,$\mu$m, and covering the full luminous range of the object.  Prior to these measurements, no other telescope had ever simultaneously identified so many spectroscopic features from a single object.  From a practical point of view, with only a few hours of JWST observations, these observations of VHS\,1256b not only demonstrated the sensitivity to faint sources, but also the characteriztion potential of spectroscopy with reasonably high spectral resolution ($\lambda/\Delta\lambda\sim$2000-3500). VHS 1256b is a highly variable object \cite{Bowler2020, Zhou2020, Zhou2022}, and was the single most variable object known at the time, indicating a young, turbulent atmosphere characterized by disequilibrium chemistry and vigorous atmospheric mixing.  The spectrum shown in Figure~\ref{fig:vhs1256b} showed clear evidence for molecules such as H$_2$O, CH$_4$, CO$_2$, and even tentative detection of NH$_3$ obtained via the molecular mapping technique \cite{Malin2026}, that can thrive only in cooler atmospheres such as this.  In addition, a clear and unambiguous spectroscopic detection of silicate grains near 10-11\,$\mu$m was found as well.  
This detection with high significance of numerous spectroscopic features over such a long wavelength range early in the JWST mission was an extremely valuable precursor for several other spectroscopic programs targeting, cooler planetary mass companions that have numerous molecular and atomic spectroscopic features in the infrared due to their cooler atmospheres \cite[e.g.,][]{Matthews2025, Mang2026}.

\subsection{Other Early JWST Direct Imaging Results: Characterizing the Known Directly Imaged Planets at New Wavelengths}

\textbf{Unexpected Detections, Enhanced Atmospheric Abundances:}

Obvious targets for early JWST observations were the HR\,8799 system, a close ($\sim$40\,parsecs), $\sim$2\,$M_\odot$ star known to host four directly imaged planetary mass companions orbiting between $\sim$16 to $\sim$70 astronomical units (AU) \cite{Marois2008, Marois2010}, as well as the $\beta$ Pictoris planetary system \cite{Kammerer2024} which recently revealed a third directly imaged planet \cite{Sutlieff2026, Gibbs2026}. While numerous ground-based observations of the four HR\,8799 planets (HR\,8799bcde) have been obtained at 1-2\,$\mu$m, and even using narrow-band filters around $\sim$3\,$\mu$m \cite{Skemer2014}, characterizing these planets has been extremely challenging at wavelengths longer than 4\,$\mu$m. 
Early JWST  observations of HR\,8799 from 3-5\,$\mu$m \cite{Balmer2025} with the NIRCam bar coronagraphic mask \cite[e.g.,][]{Girard2022}, and at 10-11\,$\mu$m with the MIRI Four-Quadrant Phase Masks \cite[``4QPMs,''][]{Boccaletti2022, Boccaletti2024} coronagraphs revealed clear detections of the three outer planets with orbits ranging from $\sim$0.6 to $\sim$1.7 arcseconds ($\sim$24–68 AU orbital separation projected onto the sky). The clear detection of the innermost planet HR\,8799e at a separation of only $\sim$0.35 arcseconds with both instruments, as well as a detection of the close-in planet 51 Eri b at only $\sim$0.3\,arcseconds, are very notable since these angular separations are only $\sim$2-3$\lambda/D$ for the NIRCam bandpasses, and $\sim$1$\lambda/D$ for the MIRI bandpasses, well inside the nominal inner working angles for the JWST MIRI 4QPM coronagraphs.  Excellent infrared sensitivity was always expected in the portion of JWST's field-of-view that is background limited ($\gtrsim$1-2 arcseconds), but detections of warm, young gas giant extrasolar planets at only a few diffraction widths from the host star is testament to the remarkable stability of the space-based environment, allowing calibration of the residual scattered starlight discussed earlier. The 3-5\,$\mu$m  wavelength coverage of the NIRCam observations presented in \cite{Balmer2025} for both 51 Eridani b and HR\,8799bcde revealed clear evidence that the atmospheres of both sets of planets have a high ``metallicity'' (enrichment in elements heavier than Helium) compared to the solar value, and about 5$\times$ compared to solar in the case of 51 Eridani b. 
Observations of the even younger HD\,95086b planet \cite{Malin2024} not only clearly detected the planet itself at 10.65 and 11.4$\mu$m for the first time, but also the system's warm inner circumstellar disk at these same wavelengths. 

\textbf{
Circumplanetary Disks:~A Unique Niche for JWST:
}~
Other early observations with JWST showcased the incredible characterization power that can be delivered with the observatory's long wavelength coverage. 
The observing bandpasses offered by MIRI also allowed a deep characterization of the GJ\,504b planetary mass companion returning the first significant detection of NH$_3$ \cite{Malin2025}. Long wavelength spectroscopy of the YSES\,1b and YSES\,1c system of planets \cite{Hoch2025} revealed clear evidence for silicate clouds, but also potentially the presence of a circumplanetary disk (CPD). Indeed, the remarkable spectroscopic sensitivity that can be achieved by JWST's NIRSpec and MIRI spectrographs have allowed groundbreaking detections of CPDs in the case of GQ\,Lup\,b \cite{Cugno2024} as well as a Carbon-rich CPD surrounding the Delorme ABb system \cite{Malin2025b}. 

\textbf{
High Contrast Spectroscopy as a Probe of Formation:
}
Perhaps the most transformative example of the exoplanet atmospheric characterization that can be returned from JWST spectroscopy is not using the coronagraphic instruments, but rather with the NIRSpec Integral Field Spectrograph (an imaging spectrograph), which does not come equipped with any kind of coronagraphic suppression capabilities. Nonetheless, the technique of ``high-contrast spectroscopy'' \cite[e.g.,][]{Ruffio2024} can be achieved with NIRSpec by carefully modeling the distribution of scattered starlight directly on the NIRSpec detector, and differentiating that from the signal of any orbiting planets.  A spectacular demonstration of this has been carried out for the four HR\,8799 planets \cite{Ruffio2026, Xuan2026} as well as for the 51 Eri\,b planet \cite{Madurowicz2025}, returning exquisite spectroscopy of the atmospheres from 3-5\,$\mu$m at spectral resolution of a few thousand.  In the case of the HR\,8799 planets, such high-fidelity spectra have revealed that; 1) similar to Jupiter and Saturn, the atmospheres of these planets are highly enriched in heavy elements compared to the host star, suggesting efficient accretion of solids during their formation; and 2) there is a tentative trend with orbital distance of increasing Sulfur abundance \cite{Xuan2026}, a tracer of refractory solids.

\textbf{
Imaging Planets Near the Diffraction Limit \& Under the Occulting Mask:
}
By leveraging advanced image post-processing as well as the exquisite observatory stability of JWST to achieve remarkable sensitivity at very small inner working angles, other recent JWST results have continued the trend of detecting, and indeed characterizing, directly imaged planets.
Among these results are the detection of at least two of the protoplanets in the PDS\,70 protoplanetary system, a very young ($\sim$few million year old) system with a cleared inner circumstellar disk and at least two accreting protoplanets \cite{Keppler2018, Haffert2019}.  
Detections of the two outermost planets were made using the Aperture Masking Interferometry (AMI) mode with the NIRISS instrument \cite{Blakely2025}, confirming that JWST is able to make precise and robust measurements of the position and flux of young, massive Jovian protoplanets in its interferometric mode as well. 
Although early examples of the AMI mode \cite{Sallum2024, Ray2025} as part of the JWST ERS program \cite{Hinkley2022} showcased the first examples of carrying out interferometry in space, the achieved contrast performance of this mode was somewhat poorer than initially predicted, most likely due to systematics in the NIRISS detector.  Subsequent forward models of the optics, detector physics, and readout electronics \cite{Charles2025, Desdoigts2025} allowed the remarkable detection of the HD \,206893c planetary mass companion \cite{Hinkley2023} at $\sim$10 magnitudes and only 100 mas.  The detection of companions with JWST at such small separations would have been predicted to be impossible prior to launch. 
Detections using classical non-coronagraphic imaging of two of the PDS\,70 planets were reported in \cite{Christiaens2024} using NIRCam, and co-located at the expected locations of $\sim$150 – 200 milliarcseconds for protoplanets b and c based on previous ground-based works, as well as the tentative detection of a possible third planet at the remarkably small angular separation of $\sim$115\,milliarcseconds at 1.87\,$\mu$m. These detections were made without a coronagraphic mask, providing access to the system's innermost regions, and again points to the fact that JWST, even when operating without a coronagraph, can make high contrast detections very close to the observatory's diffraction limit.

Another dramatic example of a high contrast detection of an exoplanet, the detection of which might have been met with scepticism prior to launch, is the detection of AF Lep\,b, a young giant planet with a dynamically measured mass of 3\,M$_{\rm Jup}$ \cite{Franson2023} and an angular separation of only 320 miliarcseconds.  This object was detected using Director’s Discretionary Time with NIRCam coronagraphic imaging at 4.4\,$\mu$m \cite{Franson2024}. This angular separation of $2.3\lambda/D$ at 4.4\,$\mu$m, is well within the nominally quoted inner working angle for NIRCam, and the angular separation corresponds to a coronagraphic transmission of only $\sim$7\%. 
All of these detections of previously known planets indicated that JWST has been performing beyond its expected capabilities in terms of achieving remarkable sensitivity  to young, self-luminous extrasolar giant planets at very small angular separations from their host stars.  

\begin{figure*}[h]
  \centering
  \includegraphics[width=0.75\linewidth]{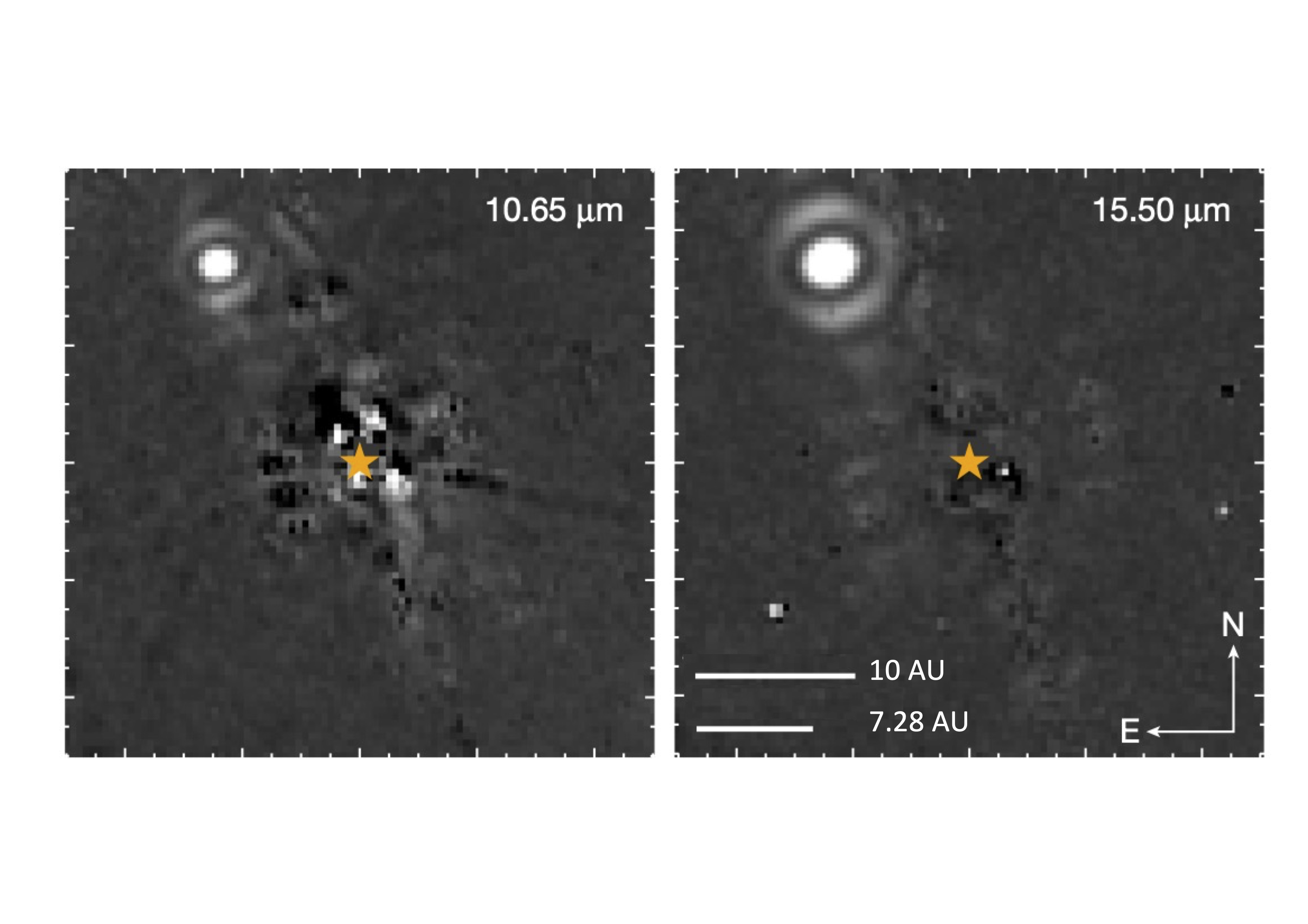}
  \vspace{-1cm}
  \caption{A figure adapted from \cite{Matthews2024} showing images of the $\epsilon$ Indi Ab system at 10.65 and 15.50\,$\mu$m obtained with the MIRI Four Quadrant Phase Mask coronagraph. The location of the host star is indicated by the orange star symbol, and the exoplanet is clearly resolved to the Northeast, corresponding to a projected orbital separation of $\sim$15\,AU.  The top and bottom scale bars in the right panel indicate 2.75 and 2.0 arcseconds, respectively ($\sim$10 and $\sim$7.3\,AU).} 
  \label{fig:epsIndi}
\end{figure*}

\subsection{Mature Planets: Detections of True Solar System Analogues}

In addition to the dramatic capabilities of JWST to characterize previously known extrasolar giant planets at very small angular separations from their host stars, JWST has also demonstrated its ability to make pure \textit{discoveries} of extrasolar giant planets, especially when previously obtained dynamical information hint at the presence of an orbiting companion. A notable recent example of this is the discovery and confirmation of Eps Indi Ab \cite{Matthews2024, Matthews2026}, a giant planet orbiting a roughly stellar age (3-5 billion year) host star.  This star exhibited an astrometric acceleration indicating the presence of a massive substellar companion bound to the system. The system's close proximity to the Sun (3.6 parsecs), combined with the unprecedented sensitivity of the MIRI imager allowed a clear initial detection at 10.65, 11.3 and 15.5\,$\mu$m of this object, shown in Figure~\ref{fig:epsIndi}.   
A somewhat surprising aspect of observations of this system is the suppressed flux in the 4-5\,$\mu$m NIRCam images \cite{Sanghi2026b}, likely due to CO$_2$ absorption resulting from an enhanced metallicity. An emerging hypothesis that would describe the relative faintness of such planets at NIRCam wavelengths relative to the longer MIRI wavelengths is likely due to enhanced metallicity or possibly clouds. \cite{Crotts2025, Sanghi2026a}
The system 14 Herculis is another close ($\sim$18\,parsecs), old ($\sim$4-5 billion year), K-type star with previous evidence from precision radial velocity studies for a companion orbiting at 15 – 35\,AU. NIRCam observations detected the known planet \cite{Bardalez-Gagliuffi2025} within 1$\sigma$ of its predicted location from fits to absolute astrometry and radial velocity measurements.  
Until recently, theoretical atmospheric models of cooler, mature objects could only be compared to direct spectroscopy of objects in our own solar system.  However, JWST is now providing photometry and spectroscopy of true exoplanet analogues to our own solar system objects.  
Especially when detailed spectroscopy of objects such as Eps Indi Ab and 14 Her c (and other similar objects to be discovered) can be obtained, these objects will serve as extremely valuable laboratories for calibration of atmospheric models.  
Lastly, JWST is already demonstrating that it can probe the habitable zones of the closest stars to return intriguing candidate companions, a clear goal for the field going forward \cite{Beichman2025, Sanghi2025}.

\subsection{Young Analogues to Solar System Ice Giants}

Early results from direct imaging surveys \cite[e.g.,][]{Nielsen2019, Wagner2019, Vigan2021} suggest that an abundant population of low mass planets (0.1 – 1.0\,M$_{\rm Jup}$) exists at wide orbital separations (tens to hundreds of AU) from their host stars.  These results are consistent with estimates from microlensing studies, as well \cite{Poleski2021}. JWST’s mid-infrared background that is 2-3 orders of magnitude lower than previous ground-based instruments, and even more so in the thermal infrared, means JWST has coronagraphic sensitivity to young analogues of our own ice giant planets at wide separations (tens to hundreds of AU) from their host stars. This sensitivity was first quantified in \cite{Carter2021}, and the first coronagraphic datasets \cite{Carter2023} from the ERS program discussed above affirmed this sensitivity.  
While not part of a survey, 
the recent detection of a Saturn-mass candidate companion \cite{Lagrange2025, Crotts2025} whose presence was predicted based on the morphology of the nearly face-on circumstellar disk surrounding the 7-13 million year old TWA 7 system is potentially a harbinger for many more sub-Jupiter discoveries to come.   

\subsection{Early JWST Results on Circumstellar Debris Disks }

No discussion of JWST’s capacity to directly characterize nearby exo-planetary systems would be complete without some discussion of JWST’s ability to characterize the diffuse scattered light emission from nearby circumstellar debris disks \cite{Wyatt2008, Hughes2018}. Such disks are comprised of belts of kilometre-sized planetesimals, remnants of the planet formation process, that may be colllisionally excited onto eccentric orbits and forced to collide, resulting in dust that is usually detected through scattered light emission.  With its outstanding infrared sensitivity, and the large continuous fields of view with both NIRCam and MIRI, JWST has provided breathtaking images of such disks \cite{Gaspar2023, Lawson2023, Lawson2024, Lazzoni2025}, giving information about not just the location and extent of the exo-Kuiper belt, constraints on planets sculpting this dust \cite{Bendahan-West2026}, as well as the detection of ice populations in the dust \cite{Millar-Blanchaer2025}. Most notably, NIRCam and MIRI coronagraphic imaging from 1.8 to 23\,$\mu$m of the canonical $\beta$\,Pictoris edge-on disk \cite{Rebollido2024} have shown surprising never-before-seen asymmetries in the dust, as well as an extended, curved secondary disk misaligned with the plane of the disk.  At the same time, observations of the close debris disk systems Fomalhaut and Vega using MIRI without a coronagraph \cite{Gaspar2023, Su2024} showed a much broader-than-expected dust belt, indicating a more dynamically complex system.  This large cavity of small dust grains is likely being replenished via Poynting-Robertson drag from the outer belt, indicating that such a cavity is unlikely to also be hosting planets. One class of stars hosting debris disks to which JWST may be uniquely sensitive are the M-dwarf stars. The intrinsic faintness of such stars in the optical makes them challenging for ground-based AO systems, but their redder colors translate directly over to higher surface brightnesses in the near-infrared. Observations of the AU Mic system \cite{Lawson2023} showcased an early example of NIRCam's sensitivity to planets with masses down to 0.1\,M$_{\rm Jup}$, and observations of the Fomalhaut C system \cite{Lawson2024} were the first detections of the disk that was previously not detected with HST and VLT/SPHERE \cite{Cronin-Coltsmann2021}. 

\subsection{JWST Exoplanet Direct Imaging: Lessons Learned So Far }
It would not be an overstatement to say that, after only the first few years of JWST observations, we are truly witnessing a revolution in the field of exoplanet direct imaging. The substantial list of targets presented in Table~\ref{tab:DI_table} that have benefitted from the direct imaging technique highlights the rapid progress made in the last few years. 
Detailed direct imaging spectroscopy over JWST's wide wavelength bandpasses is providing very precise atmospheric abundances \cite[e.g.,][]{Miles2023, Ruffio2026, Xuan2026, Whiteford2026}, which is not only our best chance to potentially constrain the complex physics of planet formation, but such measurements may even be hinting at compositional gradients for multi-planet systems in the primordial disk \cite{Xuan2026}, indicating possible formation locations for individual planets. JWST's unparalleled sensitivity deep into the infrared means that it is delivering the first detections of analogues of our own ice-giant planets \cite[][]{Lagrange2025, Crotts2025}, and circumplanetary disks \cite[e.g.,][]{Hoch2025, Cugno2024, Malin2025b}.  The remarkable stability of the observatory means that it is sensitive to young planets and protoplanets near to, or even inside, the diffraction limit of the telescope \cite[][]{Ray2025, Blakely2025, Desdoigts2025}, as well as ``under'' the observatory's coronagraphic masks \cite{Franson2024}. Lastly, the JWST coronagraphs are delivering detections of true analogues of our own cold, mature solar system giants \cite[e.g.,][]{Matthews2024, Matthews2026, Bardalez-Gagliuffi2025}, which will be ideal for further spectroscopic follow-up observations. None of these breakthroughs would have been possible with ground-based instrumentation or previous space observatories, and heralds an exciting next 20 years for exoplanet direct imaging. 

%
%

\section{Transiting Planets}
With directly imaged planets, the faint planetary light is separated and resolved from the bright host star {\it spatially} on the sky (see Figs. \ref{fig:HIP65426b} and \ref{fig:epsIndi}). For transiting exoplanets, the separation is accomplished in the {\it time} dimension instead. Transiting exoplanets have favourably aligned orbits and are observed to pass directly in front of their host star. During this time, a small percentage of light is blocked by the planet and can be detected in time series (spectro)-photometry. As Figure~\ref{fig:transit_geom} shows, for a typical transiting gas giant planet, about 1\% of the host star's light is blocked during a transit, which lasts a few hours.

While only a small fraction of exoplanets have a favourable geometry to transit, the information content in the light curves is immense. Transits allow many fundamental astrophysical parameters of the planet to be measured in a near model-independent manner. Chief among these are the planet's mass and radius. These are typically uncertain for non-transiting planets but can be measured to accuracies $\sim$1\% using the latest techniques \cite{2025ApJ...993..235R}. A transit directly measures the planet-to-star radius ratio, as the blocked flux is proportional to the projected area between the planet and star,
\begin{equation}
\frac{\Delta f_{tr}}{f} =  \left(\frac{R_{pl}}{R_{star}}\right)^2,
\end{equation}
where ${\Delta f_{tr}}/{f}$ is the fractional flux deficit measured from the transit light curve (i.e. the transit depth), $R_{pl}$ is the planetary radius and $R_{star}$ the stellar radius. Since Sun-like stellar radii can generally be determined to within a few percent, the exoplanet radius can be measured to comparable precision. The transit light curve also contains precise orbital information, including the inclination of the planet's orbit, $i$. For a planet to transit, the orbit must be near edge-on ($i\sim$ 90 degrees), and modelling the transit shape and duration allows a precise inclination measurement. As radial-velocity measurements can be used to measure the minimum mass of the planet, $M_{min}$, a precise inclination allows the sin($i$) degeneracy to be lifted, giving access to the true planetary mass, $M_{true} = M_{min}/\sin(i)$.

Beyond bulk mass and radius, transiting planets allow transmission and emission spectra to be measured from time series observations, enabling detailed atmospheric characterisation. Historically, this has used three geometries: primary transit (planet passes in front of the star), secondary eclipse (planet passes behind the star), and the phase curve (flux over the entire orbit is measured). Each provides unique information on the planet's atmosphere, chemistry, and temperature structure, with all three measurements possible for some favourable exoplanets.

\subsection{Primary Transit Transmission Spectra:}
During primary transit, an exoplanet's transmission spectrum can be measured. Light from the star passes through the atmosphere and is absorbed or scattered out of the line-of-sight, leading to an increase in the planet's apparent size at spectrally active wavelengths. By spectroscopically measuring a transit, the transit depth at each wavelength ($\lambda$) in the time-series can be measured precisely, the planetary radius at each wavelength determined, and the transmission spectrum $R_p(\lambda)$ extracted. The transmission spectrum can be thought of as an absorption spectrum, as wavelengths where atoms and molecules have strong spectral features absorb or scatter incoming starlight, producing large absorption signals in the transmission spectrum. However, as a measure of wavelength-dependent radius (or altitude), a transmission spectrum is quite different from traditional astrophysical spectra of stars and directly imaged planets (with units of flux), with the closest analogous measurement technique being solar system occultation measurements. Typical transmission-spectrum measurements are sensitive to $\sim$millibar pressure levels in the optical and near-IR, though the strongest spectral lines and species have been detected out to the exosphere and beyond \cite{2003Natur.422..143V,2010A&A...514A..72L,2019AJ....158...91S}.

An estimate of the transmission-signal size can be calculated by assuming the extra flux blocked by the atmosphere, $\Delta f_{atmo}$, above an opaque planet of radius $R_{pl}$, is proportional to the area ratio of the atmospheric annulus to the stellar disc. Assuming the atmospheric annulus has thickness $H$ and $H \ll R_{pl}$, the transmission-spectrum signal is then,
\begin{equation}
\frac{\Delta f_{atmo}}{f} =  \frac{2 R_{pl} H}{R_{star}^2}.
\end{equation}
$H$ can be estimated as the atmospheric pressure scale height, $H = k_B T/(\mu g)$, where $k_B$ is the Boltzmann constant, $T$ is the atmospheric temperature, $\mu$ is the mean mass of atmospheric particles, and $g$ is the surface gravity. Typical transmission signals for various atoms and molecules span a few scale heights, and a typical hot gas giant has signal sizes of order 100 ppm. This level of precision makes transmission spectra challenging to measure, as spectrophotometry with comparable or better precision must be obtained during the short, $\sim$hours-long transit event. An analytic transmission-spectrum formula can also be derived from first principles, with details in Refs. \cite{2005ApJ...627L..69F}, \cite{2008A&A...481L..83L}, \cite{2013Sci...342.1473D}, \cite{2017MNRAS.467.2834B}, \cite{2017MNRAS.470.2972H}, and \cite{2018arXiv180407357S}. For a given absorbing species with mixing ratio $\xi_{abs}$, the measured transit-depth altitude, $z(\lambda)$, is
\begin{equation}
z(\lambda)= H \ln \left(  \frac{\xi_{abs}\sigma_{abs}(\lambda)P_0}{\tau_{eq}}\sqrt{\frac{2 \pi R_{pl}}{k_B T \mu g}} \right),
\label{eq:Lec08}
\end{equation}
where $\sigma_{abs}(\lambda)$ is the cross-section of the absorbing species, $P_0$ is the reference pressure, and $\tau_{eq}$ is the equivalent optical depth at which the transmission spectrum is optically thick. $\tau_{eq}=0.56$ is an accurate approximation for most circumstances \cite{2008A&A...481L..83L}.
Hot H-rich exoplanets with low surface gravities have the largest signals, as the scale height is maximised; the host star's brightness typically limits noise, favouring bright stars. However, since stars are bright across a wide wavelength range, measurements have been made from X-rays and FUV through the optical to the mid-infrared around A- to M-type stars.

\subsection{Secondary Eclipse:}
On the opposite side of a planet's transit, with a favourable inclination the planet can pass behind its star and be eclipsed. As show in in Figure~\ref{fig:transit_geom}, when fully behind the star, the planet contributes no flux; only starlight is detected. The eclipse effectively separates the light from planet and star, allowing the planetary flux to be isolated and measured. The fractional flux deficit at secondary eclipse, $\Delta f_{se}/f$, can be calculated as
\begin{equation}
\frac{\Delta f_{se}}{f} = \frac{F_{pl}}{F_{star}}\times \left(\frac{R_{pl}}{R_{star}}\right)^2.
\end{equation}
where $F_{pl}/F_{star}$ is the planet-to-star flux ratio. Similar to transmission spectra, a planet's emission spectrum, $F_{pl}(\lambda)$, can be extracted by measuring the secondary eclipse depth as a function of wavelength. The stellar contribution, $F_{star}(\lambda)$, can be measured using absolute flux-calibrated data or, as often is the case, assumed from stellar models; the transit depth fixes $({R_{pl}}/{R_{star}})^2$, isolating $F_{pl}(\lambda)$. A planet's emission spectrum has two components: reflected light, detectable primarily at optical wavelengths, and thermal emission, dominant at infrared wavelengths. The secondary-eclipse geometry measures the planet's day-side. Because the star is hotter than the planet, eclipse depths tend to increase with wavelength, with a typical hot gas giant having depths of order $\sim$0.1 \% at H-band wavelengths. Compared to the grazing geometry of transit, the emission spectrum probes deeper pressures, as thermal planetary flux emanates from the photosphere, with pressures between about 1 millibar and 1 bar typically probed. The emission spectrum is very sensitive to the temperature–pressure ($T$–$P$) profile of the troposphere and stratosphere. In the planet's emission spectrum, spectrally active atoms and molecules absorb or emit (depending on the $T$–$P$ profile) relative to the continuum, producing absorption and emission features that can be used to constrain the atmospheric composition. The main C-, N-, and O-bearing molecules have their strongest rotational–vibrational features in the near-infrared, making that region particularly constraining for the lower atmosphere of warm to hot planets.
 
\subsection{Phase Curve:} A planet's phase curve is an extension of the secondary eclipse, except the time series observation spans an entire planetary orbit such that the planet's brightness contribution is continually measured as varying portions of the dayside and nightside rotate in and out of view.  Continuous measurements are practical for short-period ($\lesssim$1 day) planets. Short-period planets are tidally locked, so the entire planet is observed over a complete phase curve: the night-side around transit, the half-illuminated planet and terminator at quadrature(s), and the dayside around secondary eclipse. As with secondary eclipse, the fraction of planetary light at a given phase is separated from the combined light using the stellar-only contribution isolated during the eclipse event(s). Phase curve measurements can be challenging, as instrumental or stellar flux changes over the days-long curve complicate extracting the phase-dependent planetary flux. In principle, an observational set-up capable of detecting the secondary eclipse should also measure the phase curve, provided stability and precise photometry can be maintained over the long time series observation.
 
\subsection{Eclipse Mapping \& Transit Morning/Evening Limb-Asymmetries:} 
During ingress and egress of a secondary eclipse, the planet is partly blocked by the star. The stellar disc `scans across' the planet, obscuring slices. This scanning makes it possible to detect changes in brightness across its dayside hemisphere and build a spatially resolved picture of the dayside hemisphere, in a technique called eclipse mapping. Pioneered with \textit{Spitzer} and \textit{HST}, the technique requires extremely high spectro-photometric precision during the brief ingress and egress phases (often $\sim$10 minutes or shorter), and thus large telescopes like JWST.

Similar to eclipse mapping, during the ingress and egress of a transit only part of the terminator --- the limb of the planet dividing day from night --- blocks the star. This makes it possible to time resolve the contribution of the leading and trailing terminators, detect differences between them, and spatially resolve the transmission spectra for the morning and evening terminators. For tidally locked exoplanets, the evening terminator is expected to be hotter than the morning terminator, as atmospheric gas cools during the night while being advected by winds around the planet. The hotter evening limb produces a larger spectral signal, while the cooler morning limb may exhibit muted features due to increased cloud or haze formation. As with eclipse mapping, large telescopes are typically required, as the technique is limited by how precisely the transit can be measured during the brief ingress and egress phases.

\subsection{Science return from transiting planets:} 
\begin{figure*}[h]
  \includegraphics[width=1.0\linewidth]{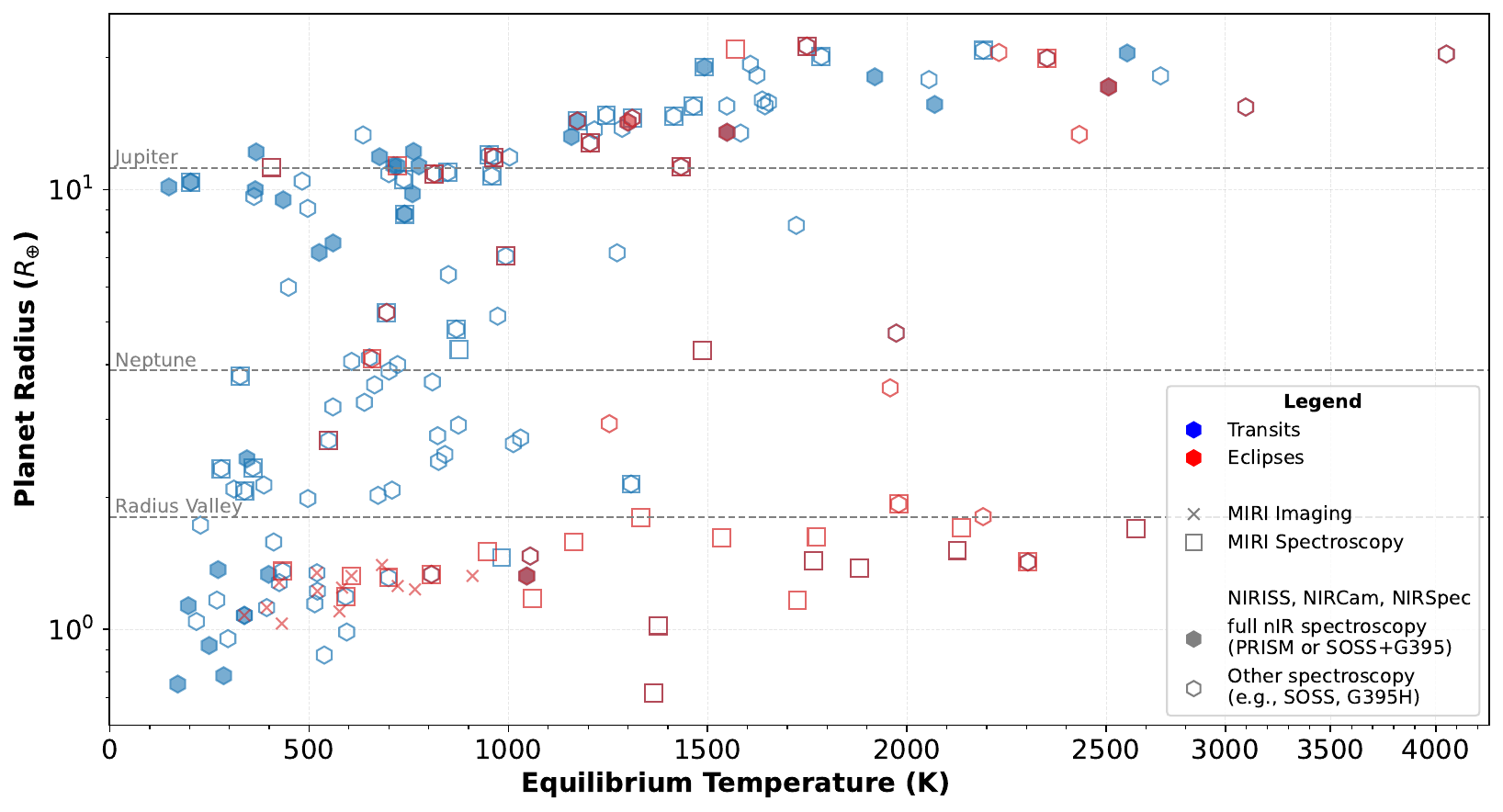}
  \vspace{-1cm}
  \caption{Transiting planets observed by JWST, data from TrExoLiSTS \cite{2022RNAAS...6..272N}. Equilibrium temperature is plotted against the planetary radii, with transit and eclipse data indicated. Planets with full JWST nIR spectroscopy are shown as filled markers.}
  \label{fig:tr_radius}
\end{figure*}

There are distinct advantages for studying transiting exoplanets for atmospheric characterisation. While many diverse science cases exist, we focus on four key advantages. First, the mass and radius are often known, fixing fundamental parameters and alleviating degeneracies. In particular, the $T$-$P$ structure, cloud opacity, metallicity, radius, and surface gravity exhibit complex degeneracies when interpreting exoplanet spectra. Second, the technique uniquely enables study of short-period exoplanets, providing access to hot and warm conditions. The physical effects of XUV stellar irradiation on processes such as photochemistry and atmospheric escape can be studied, and transmission spectroscopy remains sensitive down to temperate and cold ($<$500 K) temperatures. Third, with thousands of transiting exoplanets, it is the first characterisation technique enabling statistical studies of entire populations (see Fig. \ref{fig:tr_radius}). This includes sub-Neptunes, the most common planet type known, yet absent from our Solar System. Studying planetary populations as a whole can uncover universal processes, including those relevant to planet formation, atmospheric dynamics, and evolution, as the technique is sensitive to both young and old planets. Finally, until the Habitable Worlds Observatory (or similar telescope) is operational, it is the only method capable of characterising atmospheres of planets smaller than Neptune, with JWST sensitive down to terrestrial-sized planets.


\subsection{Advantages with JWST} 
For transiting exoplanets, there are several distinct advantages compared to other facilities. Compared to HST, JWST has superior infrared wavelength coverage, spectro-photometric precision, spectroscopic resolution, time resolution, and pointing stability. The superior pointing stability has proven to be a game-changer. Originally, it was expected that telescope pointing variations and jitter could interact with intra-pixel sensitivity variations, contributing a sizable, if not limiting, systematic noise source in time-series spectro-photometry. Pre-flight estimates of intrapixel sensitivity with NIRSpec were as large as 400 ppm, assuming 7 milliarcsecond pointing variations \citep{2022A&A...661A..83B}. However, on-sky jitter has been measured at $\sim$1 milliarcsecond \cite{2023PASP..135d8001R}, better than design specifications even over very long durations, with 1/4 milliarcsecond measured during the 37.8 hour NIRSpec phase-curve observation of WASP-121 b \citep{2025NatAs.tmp..123E}. This low jitter compares favourably with HST ($\sim$2 mas) and with the upcoming ARIEL mission's design upper limit of $<$70 mas. For JWST, this rock-solid stability directly translates into increased photometric precision, as intra-pixel sensitivity variations are dramatically reduced, if not undetectable, in most observations. In addition, variable slit losses are essentially non-existent. For NIRISS, NIRCam, and MIRI, slitless spectrograph modes bypass slit losses, while NIRSpec has a wide $1.6$ arcsecond square aperture that also mitigates slit losses (provided the target star is well placed during acquisition). Sub-10 ppm photometric precision is now routinely achieved with JWST, with no clear instrument noise floor yet found for the near-IR instruments.

The ability to obtain precise near- to mid-IR transmission and emission spectra is a major advantage for JWST. The infrared is where molecular rotational-vibrational transitions are found, so sensitivity to these features is required to study the rich set of exoplanet species (e.g. H$_2$O, CO, CO$_2$, CH$_4$, etc.) and chemistry. HST does have limited IR spectroscopic capability, with WFC3/IR sensitive to wavelengths out to 1.6 $\mu$m. This opened study of the 1.4 $\mu$m water feature, used to detect water in the atmospheres of many dozens of planets. However, HST is sensitive to just one molecule, H$_2$O, making carbon chemistry inaccessible. Ground-based high-resolution spectroscopy can measure species in the near-IR (1-2\,$\mu$m), but has historically been limited to the brightest and/or hottest exoplanets, hindering broad surveys. For transiting exoplanets, JWST's mid-IR capabilities are unique, with no other facility able to measure spectra longward of about 2.3 $\mu$m. This opens exoplanet studies to important features such as CH$_4$, CO$_2$, SO$_2$, and cloud-condensate signatures, in addition to the thermal emission of temperate planets.

\section{Transiting Planet Results}
\subsection{Early Release Science: Out-of-the-box Gifts With New Molecular Species}
The Transiting Exoplanet Community Early Release Science Program (Program 1366, 86.9 hours){\footnote{\url{https://ers-transit.github.io}}} aimed to provide the community with publicly available JWST transit data early in the mission. It also included pre-launch activities, coordinated collaborations, trained the community to work with JWST data, established best practices early, and developed an inclusive open-science collaboration to foster transiting exoplanet science \citep{2016PASP..128i4401S, 2018PASP..130k4402B}. The programme gathered over 300 participants worldwide, with three main science observations: a full NIR panchromatic transmission spectrum of WASP-39~b to compare instrument modes; a bright-star near-saturation NIRISS test with WASP-18b's eclipse; and a test of long-stare capabilities with a MIRI phase curve of WASP-43~b. The Transit ERS programme data have catalysed a plethora of results from both the ERS team and the broader community
\citep{2025A&A...694A.249R,2025A&A...694A.233B,2024A&A...692A.113W,2024MNRAS.535...27F,2024AJ....168..227C,2024AJ....168..111A,2024NatAs...8.1008C,2024Natur.632.1017E,2024MNRAS.532.4350H,2024NatAs...8..929K,2024MNRAS.532..460Y,2024ApJ...970...71X,2024A&A...687A.110L,2024AJ....168....4H,2024NatAs...8..879B,2024MNRAS.531.2731S,2024A&A...685A..64K,2024ApJ...966..189Y,2024MNRAS.530.3252C,2024ApJ...963L..37M,2024A&A...683A.231T,2023AJ....166..251C,2023ApJ...956L..19L,2023ApJ...955L..19E,2023MNRAS.524..361Z,2023MNRAS.524..377H,2023Natur.620..292C,2023ApJ...950L..17N,2023ApJ...949L..15G,2023Natur.617..483T,2023A&A...672A.110L,2023ApJ...943L..10C,2023Natur.614..670F,2023Natur.614..664A,2023Natur.614..659R,2023Natur.614..653A,2023Natur.614..649J}
and a special collection of results is available at Nature{\footnote{\url{https://www.nature.com/collections/adhibjjgid}}}.

\begin{figure*}[h]
  \includegraphics[width=1.0\linewidth]{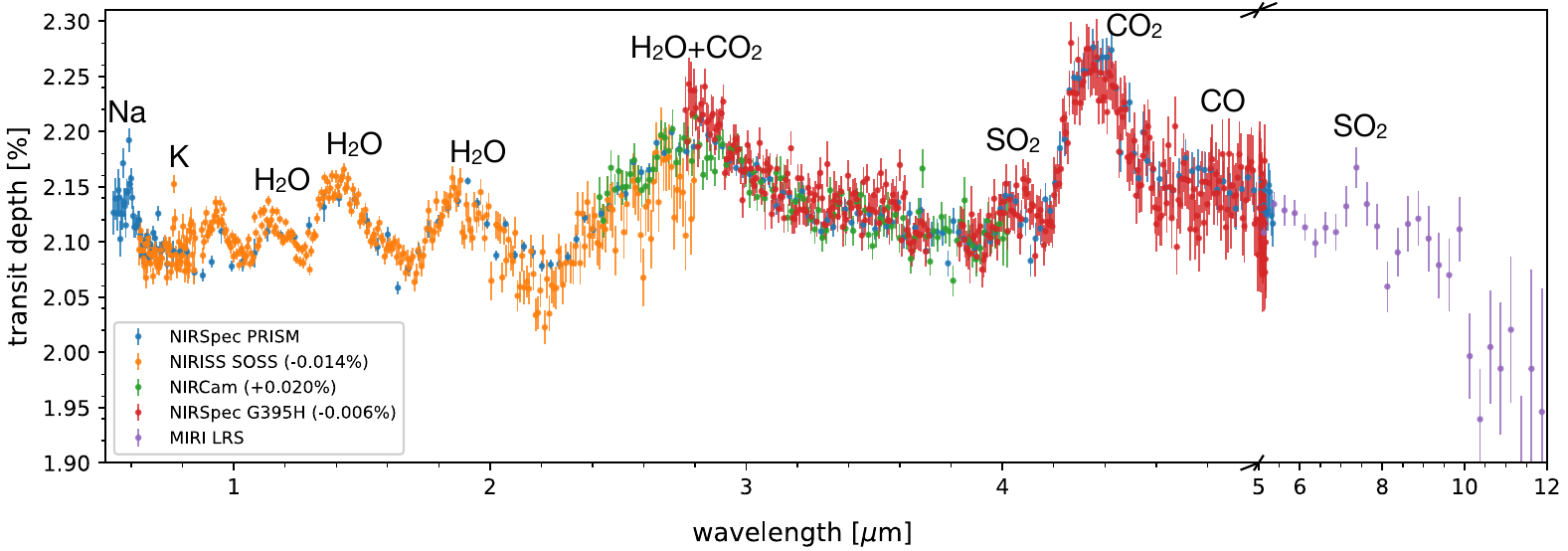} 
  \vspace{-0.8cm}
  \caption{The full spectrum of WASP-39~b obtained from the Transit ERS program \cite{2023Natur.614..649J,2023Natur.614..653A,2023Natur.614..659R,2023Natur.614..664A,2023Natur.614..670F,2024Natur.626..979P}. Offsets have been applied to account for the non-uniform reductions \cite{2024NatAs...8.1008C}. The feature-rich spectrum includes strong absorption features from H$_2$O, CO, CO$_2$ and SO$_2$ with the later two species first identified from these spectra.
  }
  \label{fig:w39}
\end{figure*}

One of the first datasets released after the telescope commissioning was transit data of WASP-39 b\cite{2011A&A...531A..40F}. The target was selected as a known feature-rich exoplanet from ground-based and HST spectroscopy \cite{2016ApJ...827...19F,2016ApJ...832..191N,2018AJ....155...29W,2019AJ....158..144K} and is sufficiently faint to be observable by NIRSpec/PRISM. Complementary spectroscopic modes were chosen with NIRSS, NIRSpec and NIRCam such that a full 0.6 to 5.5 micron transmission spectra could be built up, with multiple overlapping spectral regions where the instrument performance between modes could be compared. The extreme high-quality of the data and power of precision IR spectroscopy for exoplanets was immediately realized, as the very first dataset spectroscopically resolved and detected the important CO$_2$ molecule for the first time in an exoplanetary atmosphere \cite{2023Natur.614..649J} (see Fig. \ref{fig:w39}). Previously, the best evidence for CO$_2$ was from \textit{Spitzer} photometry \cite{2021MNRAS.500.4042S}.  In addition, an unanticipated detection of SO$_2$ was also made\citep{2023Natur.614..659R,2023Natur.614..664A} indicating ongoing photochemistry in an exoplanet atmosphere \citep{2023Natur.617..483T}. The detection and spectroscopic idenification was later confirmed by MIRI observations \cite{2024Natur.626..979P}. The infrared capabilities of JWST thus has ushered in the era of characterizing disequilibrium processes such as photochemistry in exoplanet atmospheres, with the MIRI phase curve of WASP-43b also exhibiting signatures of kinetic chemistry under day–night horizontal transport disequilibrium.

\subsection{Characterization Highlights by Planet Type}
With transiting exoplanets, a wide variety of planets have been atmospherically characterised, with masses and radii spanning planet types from terrestrial through sub-Neptunes, Neptunes, Saturns, and Jupiters. The temperature range has been wide as well, with measurements from temperate habitable-zone exoplanets to ultra-hot ($>$2000 K) conditions. In Table \ref{transit-table} we provide an extensive list of JWST atmospheric results for transiting exoplanets from the first few cycles. Below, we briefly discuss a few self-selected highlights to illustrate the broad exoplanet science being done.

\subsubsection{Giant Planets} 
\textbf{Methane:} 
A long-standing mystery from pre-JWST transiting exoplanet spectroscopy was the lack of CH$_4$ detections. HST/WFC3 is sensitive out to 1.6 $\mu$m, sufficient to detect a strong CH$_4$ feature in cooler ($<$1000 K) exoplanets; however, despite strong H$_2$O features across dozens of gas giants, CH$_4$ was never found. With JWST, CH$_4$ was detected early in Cycle 1, namely in K2-18~b\cite{2023ApJ...956L..13M} and WASP-80~b\cite{2023Natur.623..709B}. The wider wavelength coverage across multiple strong CH$_4$ bands, including the fundamental Q-branch band head, together with greatly increased photometric precision, has enabled robust detections. Curiously, both planets had prior HST data where a 1.4 $\mu$m feature was interpreted as H$_2$O but later shown to be substantially CH$_4$ \cite{2019ApJ...887L..14B,2022AJ....164...30W}. The HST ambiguity stemmed from detector-edge effects combined with limited near-IR coverage, where only the single 1.4 $\mu$m absorption feature could be measured. Moreover, near-complete overlap with H$_2$O at 1.4 $\mu$m makes unique identification of CH$_4$ from a single band head challenging. Subsequently, CH$_4$ has also been detected in other exoplanets \cite{2024Natur.625...51D,2024Natur.630..831S,2024Natur.630..836W}. However, for hot Jupiters where methane should be detectable under equilibrium ($\sim$1000–1300 K), features are typically weaker than predicted, with firm detections so far requiring cooler conditions.\\

\textbf{Photochemistry:} 
After the initial surprise of SO$_2$ detected in WASP-39~b, the molecule has appeared in a number of other planets as well \cite{2025arXiv250102081F}. As a relatively common new feature, SO$_2$ will be a key photochemical by-product to test models across a range of XUV irradiation and atmospheric conditions \cite{2023Natur.617..483T}. Photochemical models indicate SO$_2$ is sensitive to both temperature, $T$, and metallicity, $Z$.  Maximum Volume Mixing Ratios (VMR) are predicted near 1100 K, with a steep drop-off below 900 K. There is also a strong increase in VMR with increasing $Z$, with VMR rising by a factor of 10$^6$ when $Z$ increases from 1$\times$ to 100$\times$ solar. These sensitivities provide independent probes of $Z$ and $T$, helping to lift degeneracies in spectral-retrieval modelling. Early trends comparing the first 8 available 3–5 $\mu$m JWST spectra find SO$_2$ strength correlated with both $Z$ and $T$, supporting these predictions \cite{2025arXiv250102081F}.\\
  
\textbf{Disequilibrium Kinetic Chemistry:} 
Disequilibrium chemistry arises when atmospheric flow transports molecular species to pressures and temperatures where reaction rates are slower than transport timescales, leaving the gas out of equilibrium. This process has long been known in the Solar System, as Jupiter's observed CO abundance requires vertical transport from hotter conditions \cite{1977Sci...198.1031P}. A comprehensive recent theoretical analysis of disequilibrium chemistry for irradiated atmospheres can be found in \cite{2025ApJ...985..209M}. With JWST, clear signatures have now been found for several exoplanets. WASP-107~b is among the coolest, most spectroscopically accessible gas giants, and was targeted by many JWST Cycle 1 programmes. These observations found strong methane depletion \cite{2024Natur.625...51D}, with only the strongest Q-branch band head near 3.3 $\mu$m detectable in transmission spectroscopy \cite{2024Natur.630..831S,2024Natur.630..836W}. Methane is depleted from equilibrium expectations by a factor of $1000\times$, with strong vertical mixing from a much hotter interior than predicted being the driving mechanism \cite{2024Natur.630..831S,2024Natur.630..836W}. 
In what has been called the 'methane thermometer', the disequilibrium chemistry of CH$_4$ has proven to be sensitive to the pressures and temperatures of the deeper interior, as the chemical abundances  are frozen in place when transported to the higher altitudes and lower pressures probed with spectroscopy. For WASP-107~b, the intrinsic temperature was measured at a hot $T_{int}=460\pm40$ K \cite{2024Natur.630..831S}, compared to Jupiter at $\sim$100 K. The extra interior heat is thought to be caused by tidal heating \cite{2024Natur.630..836W,2024Natur.630..831S}, as interactions with another planet in the system may be increasing the eccentricity. With firm measurements of composition, mass, radius, and $T_{int}$, interior models have been used to measure an exoplanet's core mass for the first time, with WASP-107~b found to have $11.5^{+3.0}_{-3.6}$ Earth Masses ($M_{\oplus}$), in line with core-accretion expectations \cite{2024Natur.630..831S, 2024Natur.630..836W}.

Strong vertical mixing was also found on a much hotter exoplanet, WASP-121~b \cite{2025NatAs.tmp..123E}. A full phase curve revealed substantial CH$_4$ on the planet's night-side, a very surprising result given the very hot dayside ($\sim$2700 K) with a strong thermal inversion. Horizontal transport from the hot near-equilibrium dayside to the cooler night-side brings CH$_4$-depleted material. Thus, to  
explain the high CH$_4$ night-side levels, strong vertical mixing from the interior is required \cite{2025NatAs.tmp..123E}. 
For the 1500 K planet WASP-43b, the dominant mechanisms are reversed as a MIRI phase curve found a lack of CH$_4$ on the night-side \cite{2024NatAs...8..879B}. Here, horizontal mixing of hot CH$_4$-depleted material dominates over vertical mixing from the interior, yielding a planet-wide CH$_4$-depleted state.  These first few JWST measurements show horizontal and vertical advection combine with chemistry in complex, differing ways depending on planetary conditions. Future measurements can tease out the physical and chemical mechanisms driving these differences and chart the diversity across the planetary population.\\

\textbf{Limb-Limb Transit Asymmetries:}
Highly irradiated gas giant planets have long been thought to be dominated by strong day-night atmospheric advection \cite{2002A&A...385..166S}, as short orbital periods cause tidal locking with permanent day and night sides. The resulting strong day/night temperature gradients produce equatorial jets that circulate around the planet. This atmospheric flow pattern is a fundamental distinction between hot Jupiters and their direct-imaging giant-planet cousins, which have similar temperatures but lack large day/night temperature contrasts. With clouds long found to be dominant on hot Jupiters (e.g. \cite{2002ApJ...568..377C,2008MNRAS.385..109P,2016Natur.529...59S}), theoretical studies predicted strong cloud differences between morning and evening limbs \cite{2013A&A...558A..91P,2016ApJ...820...78L}, which could lead to important biases in interpretation if not accounted for \cite{2016ApJ...820...78L}. These differences arise when the day-night temperature difference straddles a cloud's condensation temperature. In that case, clouds can form on the cooler night side of the planet, cover the morning terminator, and evaporate on the hotter dayside, leaving the evening terminator largely cloud-free.

\begin{figure*}[h]
  \includegraphics[width=1\textwidth]{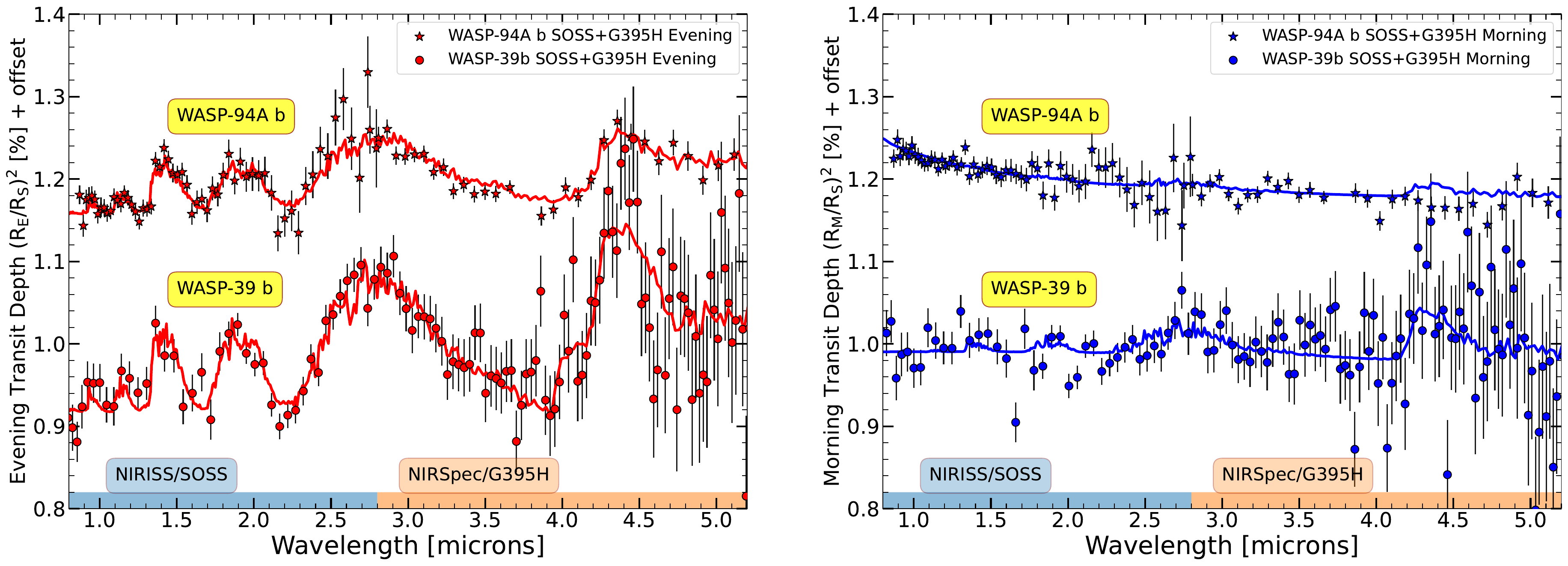}
  \vspace{-0.9cm}
  \caption{Morning/Evening limb-asymmetry results for WASP-39~b \cite{2024Natur.632.1017E,2025ApJ...989L..17F} and WASP-94A~b \cite{2025MNRAS.tmp..824A,2025arXiv250510910M} from JWST NIRISS and NIRSpec data. The transmission spectra for these planets have been spatially resolved between the evening terminator ({\it left}), and the morning terminator ({\it right}). Both planets show a stark dichotomy in their cloud coverage between the morning and evening planetary limbs. The evening terminators are rich in molecular absorption features, showing multiple strong H$_2$O and CO$_2$ peaks, being largely cloud-free. The evening terminators both show strong high-altitude aerosols which form on the planetary night-side which largely covers the molecular features. Best-fit models (red \& blue) using PICASO\cite{2023ApJ...942...71M} are shown [S. Mukherjee, priv.~comm.].}
  \label{fig:limbasym}
\end{figure*}

Before JWST, limb-asymmetries were difficult to study, as HST has large gaps in its observability window due to its Earth orbit, so the tell-tale limb-asymmetric signatures detectable during ingress and egress \cite{2021AJ....162..165E} are largely unobservable in a single transit. With JWST, even the first transit observation showed limb-asymmetries \cite{2023Natur.614..659R}, with the evening terminator of WASP-39~b measurably hotter than the morning terminator (See Figure~\ref{fig:limbasym} and \cite{2024Natur.632.1017E}). The predicted cloud inhomogeneity was detected as well. For WASP-94A\,b, the morning terminator was found to have extremely strong optically thick clouds at very high altitudes and low pressures ($<$0.01 millibar), covering all molecular H$_2$O features \cite{2025arXiv250510910M}. However, on the evening terminator the spectrum is nearly cloud-free, with very strong H$_2$O features. The limb-asymmetry effectively separates the evening molecular-rich spectrum from the cloudy morning, revealing a strong bias in abundance measurements if not resolved. A spherically symmetric analysis gives abundances $\sim$2 dex higher than a limb–limb asymmetric analysis. Unfortunately, \cite{2025arXiv250510910M} found one cannot simply set up a retrieval on the spherically symmetric transmission spectrum accounting for patchy clouds, as the observed cloud cover could not be recovered and metallicities remained heavily biased. Fundamentally, limb–limb asymmetry dilutes molecular features in a transmission spectrum (when interpreted spherically, as clear and cloudy spectra are summed), which cannot be easily distinguished from symmetric clouds that mute covered features\cite{2016ApJ...820...78L}. The first population study found the effect is widespread across the transiting-planet population \cite{2025ApJ...989L..17F}. This indicates strong limb–limb bias has been present in past results (e.g. HST), so most abundance measurements need to be revisited to either rule out limb-asymmetries or account for them. 

A major challenge for non-JWST observatories is that most of the limb-asymmetric signal occurs during ingress and egress, which typically last only 10--15 minutes. In this short span, very precise $\sim$100 ppm spectrophotometry must be achieved. While this is relatively easy with JWST due to its large aperture, providing many photons in the needed short time frame, for smaller observatories such as HST or ARIEL the measurement is much more difficult.

\textbf{Elemental ratios:}
Spectroscopy of transiting gas giant planets has long promised insights into complex processes such as planet formation by measuring key elemental ratios in their atmospheres. The Carbon-to-Oxygen (C/O) ratio, in particular, has been considered a tracer of formation conditions \cite{2011ApJ...743L..16O,2012ApJ...758...36M}. Barring substantial atmospheric escape in some lower-mass gas giants, these planets are largely expected to retain their initial primordial accreted abundances, measurable in the atmosphere. Hot and very hot conditions uniquely access refractory species, largely unavailable for Solar System gas giants. Pre-JWST, routine C/O measurements across diverse planets were not possible, with notable exceptions from high-resolution ground-based spectroscopy \cite{2021Natur.598..580L}. JWST gives access to the major C- and O-carrying species—CO, H$_2$O, CH$_4$, CO$_2$—and others such as SO$_2$ and OCS. While no clear trends have yet emerged, a wide range of planetary C/O ratios is observed \cite{2025arXiv251026785L}. WASP-121b was found to have a high C/O of 0.92$^{+0.02}_{-0.03}$, more than 1.63 times the stellar C/O, indicating a very C-rich planet. Conversely, other planets show sub-stellar C/O ratios, e.g. HAT-P-26b \citep{2025arXiv250916082G}. Such a wide range should not be surprising, as stellar C/O—and thus the initial inventory at formation—can vary star-to-star by a factor of about 2 in the stellar neighbourhood \cite{2022AJ....163..159R}.\\

\textbf{Clouds:} 
With sensitivity to broadband spectrophotometry from the optical to IR, JWST was poised to make significant contributions to planetary cloud studies. Aerosol scattering slopes have been observed in transiting planets since the first spectral results \cite{2008MNRAS.385..109P} and across the hot Jupiter parameter space \cite{2016Natur.529...59S}. However, pre-JWST the aerosol composition and the mechanisms producing them remained largely unconstrained, with photochemical hazes and condensate clouds both possible. With JWST, optical scattering features can be detected at short wavelengths (e.g. SOSS), while IR rotational–vibrational features with MIRI can pinpoint aerosol composition \cite{2015A&A...573A.122W}. Thus far, results for transiting planets show evidence for SiO$_2$ (quartz) clouds in both WASP-17~b \cite{2023ApJ...956L..32G} and HD~189733b \cite{2024ApJ...973L..41I}. These findings are somewhat unexpected, as the most prevalent condensates were widely expected to be Mg-Si clouds such as enstatite (MgSiO$_3$), found in brown dwarfs and directly imaged planets at similar temperatures. As the Mg-Si cloud precursor molecule SiO has been detected in the ultra-hot Jupiter WASP-121b \cite{2025NatAs.tmp..123E}, an anticipated enstatite detection will require MIRI observations of a somewhat cooler ($\sim$1500 K) hot Jupiter.
    
\subsubsection{Sub-Neptunes:}
Sub-Neptune planets have masses of approximately 2–20 $M_{E}$ and radii of 1.5–4 $R_{E}$ \cite{2017AJ....154..109F}. These planets have significant H/He envelopes and are highly amenable to transmission spectroscopy, as their low mean molecular weight atmospheres provide large pressure scale heights. They are particularly interesting for characterisation, as those on long periods represent the most common known outcome of planet formation \cite{2012ApJS..201...15H}, yet our Solar System lacks such a planet. Atmospheric characterisation is therefore especially valuable, with these planets in largely unexplored territory in both chemistry and overall composition. Notably, predicted compositions range from rocky planets with H/He envelopes to water worlds. 

Thus far, results exist for a handful of sub-Neptunes. Several show near-featureless transmission spectra, e.g. TOI-836 c \cite{2024AJ....168...77W} and TOI-1685~b \cite{2025arXiv251215338F}, or small-amplitude features like GJ 1214~b and GJ 9827~b, possibly due to heavy aerosol coverage or metal-rich atmospheres \cite{2024ApJ...974L..33S, 2024ApJ...974L..10P}. These small-amplitude spectra typically occur at warm ($\sim$400–1000 K) temperatures, with photochemical hazes a prime suspect for muting features. At both hotter and cooler temperatures, several planets exhibit large-amplitude features, including TOI-270~d and K2-18~b at cool temperatures, and TOI-421~b at hot temperatures. TOI-270~d and K2-18~b both show strong CH$_4$ features \cite{2024arXiv240303325B,2023ApJ...956L..13M}, broadly consistent with chemistry expected at 300–400 K. However, temperature is not the only driver: LP 791-18c, intermediate between TOI-270~d and K2-18~b, shows much stronger haze features and no CO$_2$ \cite{2025arXiv251210876R}.

{\it Biomarker False-Positives}: As a mini-Neptune in the habitable zone, K2-18~b was claimed to show biomarker signatures from dimethyl sulfide (DMS) in JWST NIRSpec and MIRI data \cite{2023ApJ...956L..13M,2025ApJ...983L..40M}. Subsequent studies found the claims suffered from several issues: statistical errors converting Bayes factors to frequentist $\sigma$ confidence levels \cite{2025arXiv251000169T} that inflated probabilities; spectral features depending on data handling and binning \cite{2025arXiv250118477S}; instrument systematics \cite{2025AJ....170..257S}; and incomplete modelling, with alternative models providing equivalent or better fits \cite{2025arXiv250421788W}. Additional NIRSpec transits did not find DMS \cite{2025arXiv250712622H}; the marginal evidence reported \cite{2025arXiv250712622H} still used inaccurate statistics, inflating the $\sigma$ levels \cite{2025arXiv251000169T}. Although in the habitable zone, liquid-water oceans were also found to be implausible for K2-18~b without significant fine-tuning \cite{2025arXiv250118477S}, and would in any case be unstable due to a runaway greenhouse effect
\cite{2023ApJ...953..168I}, further undermining biomarker interpretations. Thus, while biomarkers remain an intriguing possibility for sub-Neptunes accessible to JWST now, these specific claims have been thoroughly overturned.


\subsubsection{Rocky Planets:}

JWST is the first facility capable of detecting secondary atmospheres of terrestrial exoplanets. Secondary atmospheres form after loss of the primordial H/He envelope; Venus and Mars, with their CO$_2$-rich atmospheres, are prime examples. JWST is capable of detecting a secondary atmosphere via either transmission or emission spectroscopy. Even with JWST, atmospheric investigations for terrestrial planets have been confined to those orbiting dwarf stars with M Spectral Types ($\sim$0.1-0.6\,$R_\odot$, $\sim$0.08-0.58\,$M_\odot$, $\sim$2300-3900\,K), as their small stellar sizes considerably increase the SNR of transit and eclipse observations -- making such studies feasible. The terrestrial planets studied thus far span a wide range, and for simplicity we categorise them into two broad groups: hot lava worlds with temperatures above about 2000 K, and terrestrial planets below 2000 K.\\

{\bf Lava worlds:} Lava worlds are extreme planets. Planetary daysides can be hot enough for molten rock, with 55 Cnc~e ($T_{eq}$=1949 K) a well-known example. For this planet, there was initially evidence for CO or CO$_2$ absorption \cite{2024Natur.630..609H}. However, subsequent observations indicate variability \cite{Patel2024}, suggesting stochastic outgassing tied to volcanism. Instrumental systematics may also be at play, as the star is extremely bright, complicating the analysis. Having a lower density than expected for a terrestrial planet, there are indications that 55 Cnc~e is not rocky, but rather the stripped core of a mini-Neptune with significant hydrogen, helium, water, and/or other volatiles in its interior \cite{2025ApJ...993..235R}. Another similar lower-density lava world is TOI-561~b ($T_{eq}$=2292 K), which also shows evidence for volatiles from JWST, as eclipse depths are lower than expected \cite{2025arXiv250917231T}, consistent with atmospheric volatile absorption.\\

\begin{figure*}[h]
  \includegraphics[width=1\textwidth]{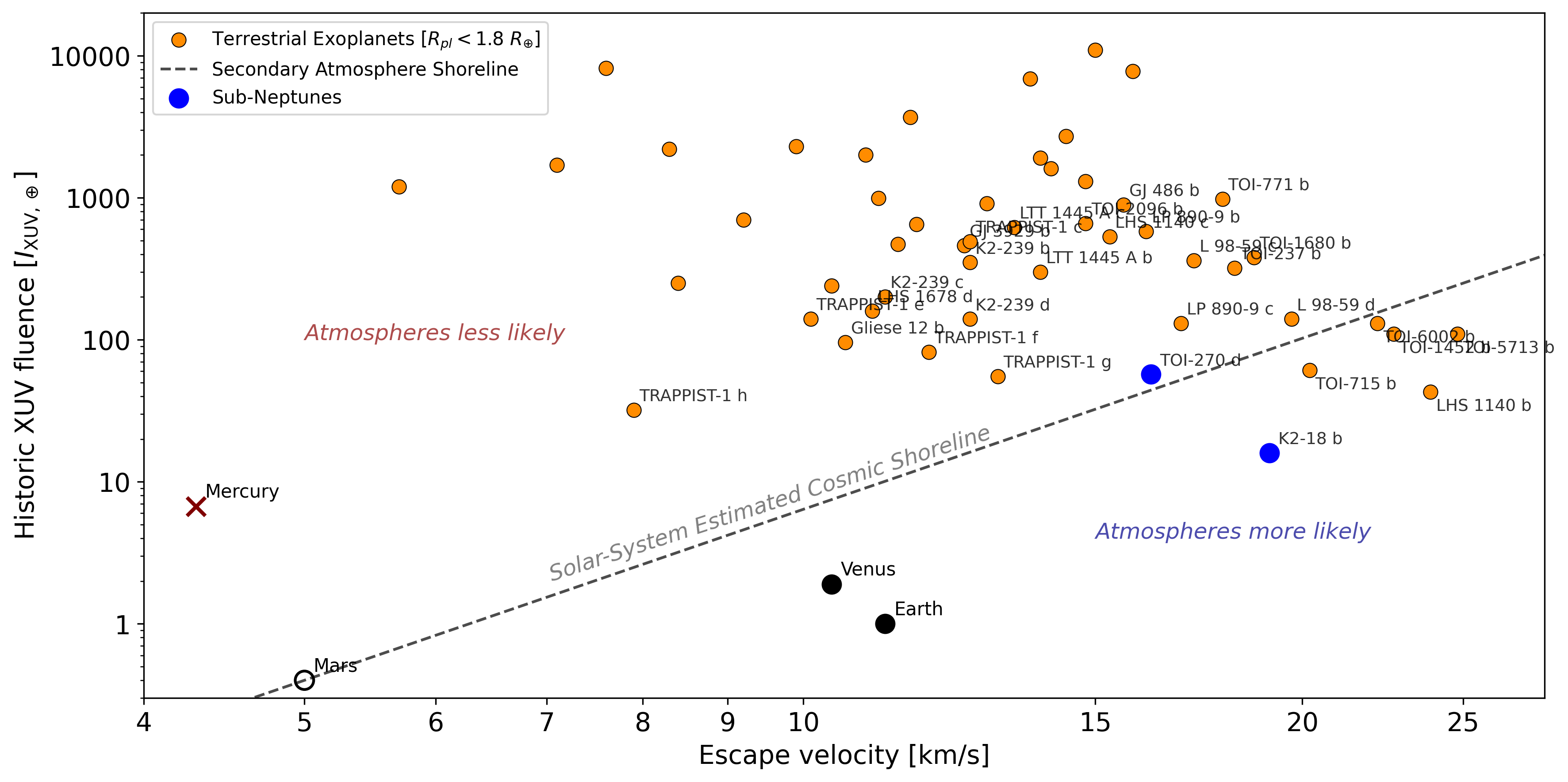}
  \vspace{-0.8cm}
  \caption{From \cite{2025ApJ...986L...3P}. 
  Escape velocity versus historic XUV influence for terrestrial exoplanets ($R_{\rm pl} < 1.8$ R$_\oplus$; orange circles), compared to Solar System bodies. The dashed line represents the ``cosmic shoreline'' calibrated to pass through Mars, following the relation $I_{\rm XUV} \propto v_{\rm esc}^4$. For context, two sub-Neptunes (blue circles) known to have primary atmospheres (H/He-rich) are also shown.
  }
  \label{fig:shoreline}
\end{figure*}

{\bf Terrestrial planets \& the cosmic shoreline:}
For cooler terrestrial exoplanets, where molten rock is not expected, great efforts are being made to detect secondary atmospheres. To date, no definitive detection has been found. These searches are primarily restricted to planets orbiting small M-dwarf stars, which poses several challenges. First, M-dwarfs are very magnetically active, greatly complicating interpretation of transmission spectra, as evolving spots and faculae can mimic planetary spectral features. Second, the planets are tidally locked. Like Mercury, they can have very cold night-side temperatures that may freeze atmospheric gases. Finally, magnetically active M-dwarfs have high XUV flux levels, which may drive vigorous atmospheric escape. 

Current searches are guided by the concept of the ``cosmic shoreline,'' which originated in the Solar System as an empirical boundary separating bodies with atmospheres from those without \cite{2017ApJ...843..122Z} (see Fig. \ref{fig:shoreline}). The shoreline depends on planetary escape velocity ($v_{\rm esc}$) and insolation ($I$): planets with higher $v_{\rm esc}$ more easily retain atmospheres, while those with higher stellar radiation lose them more readily. In the Solar System, the dividing line scales as $I \propto v_{\rm esc}^4$ \cite{2017ApJ...843..122Z}. Since insolation and escape velocity can be estimated for exoplanets, the shoreline is used to identify which may host atmospheres and to test its universality. For M-dwarf hosts—those to which JWST is most sensitive for terrestrial atmospheres—recent findings suggest only the most massive terrestrial planets are likely to have retained atmospheres \cite{2025ApJ...986L...3P}.

There are two main JWST methods for detecting secondary atmospheres on terrestrial exoplanets: broadband transmission spectroscopy and secondary eclipse primarily at 15 $\mu$m, where CO$_2$ has a strong feature \cite{2023Natur.620..746Z}. These methods are highly synergistic. Transmission spectroscopy is sensitive down to thin atmospheres with Mars-like surface pressures. However, the signals become very small for high mean molecular weight compositions like CO$_2$-rich atmospheres. Conversely, IR eclipse data are sensitive to thicker, high mean molecular weight CO$_2$-dominated atmospheres, though not to thin ones. Together, they cover each other's ``blind spots'' and can confirm the presence or lack of an atmosphere. An example is the rocky planet GJ 1132~b: eclipse photometry placed stringent limits on thick atmospheres, ruling out surface pressures of $\sim$1\,bar or more for most scenarios \cite{2024ApJ...973L...8X}, and CO$_2$-dominated atmospheres at millibar-or-more pressures. Thinner atmospheres remained possible. From transits, it was initially unclear whether signals were due to stellar activity or the planet’s atmosphere (with H$_2$O a possibility) \cite{2023ApJ...959L...9M}; however, combining four JWST transits provided strong limits, and the eclipse plus transit together essentially rule out most plausible atmospheres \cite{2025AJ....170..205B}. Thus far, no definitive signatures of a secondary atmosphere have been found for terrestrial planets cooler than lava-world temperatures (e.g. \cite{2023NatAs...7.1317L, gordon2025jwstcompassinsightssystematic}).\\

{\bf Stellar activity and the active M-dwarf problem:}
M-dwarf stars have long been known to exhibit persistent stellar activity \cite{1996AJ....112.2799H}. Such activity is a source of false-positive signals for transiting planets, as occulted and unocculted active regions can mimic transmission spectral features \cite{2008MNRAS.385..109P, 2011MNRAS.416.1443S,2018ApJ...853..122R}. The effects are stronger for later-type stars \cite{2024AJ....168...82R}; for M-dwarfs with H$_2$O in the stellar photosphere, planetary H$_2$O signatures can be affected as well. JWST observations have largely confirmed expectations \cite{2024AJ....168...82R}, with active M-dwarfs showing tell-tale activity-induced features in transit light curves. The late M-dwarf TRAPPIST-1A, which hosts at least seven terrestrial planets, is particularly problematic, with transit-to-transit differences in transmission spectra often exceeding hundreds of parts-per-million \cite{2025ApJ...990L..52E}, while expected atmospheric features are about an order of magnitude smaller. Attempts are underway to use near-simultaneous transits of planet b—likely airless \cite{2023Natur.618...39G}—with transits of other planets to calibrate and correct activity empirically, in situ \cite{2025arXiv251207695A}. Stars with FGK spectral types ($\sim$0.6-1.7\,$R_\odot$, $\sim$0.6-1.6\,$M_\odot$, $\sim$3900-7200\,K), even with moderate activity, do not show obvious unocculted spot effects on transmission spectra (e.g. K-dwarf WASP-107~A \cite{2024Natur.630..831S,2024Natur.630..836W}), making active M-dwarf hosts by far the most problematic class. With most high-value M-dwarf-hosting targets showing high levels of activity and affected transmission spectra, including TRAPPIST-1A and LHS\,1140A \cite{2024ApJ...970L...2C}, M-dwarf stellar activity is the chief obstruction to detecting terrestrial-planet atmospheres with JWST—assuming they host them.

\begin{figure*}[h]
  \includegraphics[width=1.0\linewidth]{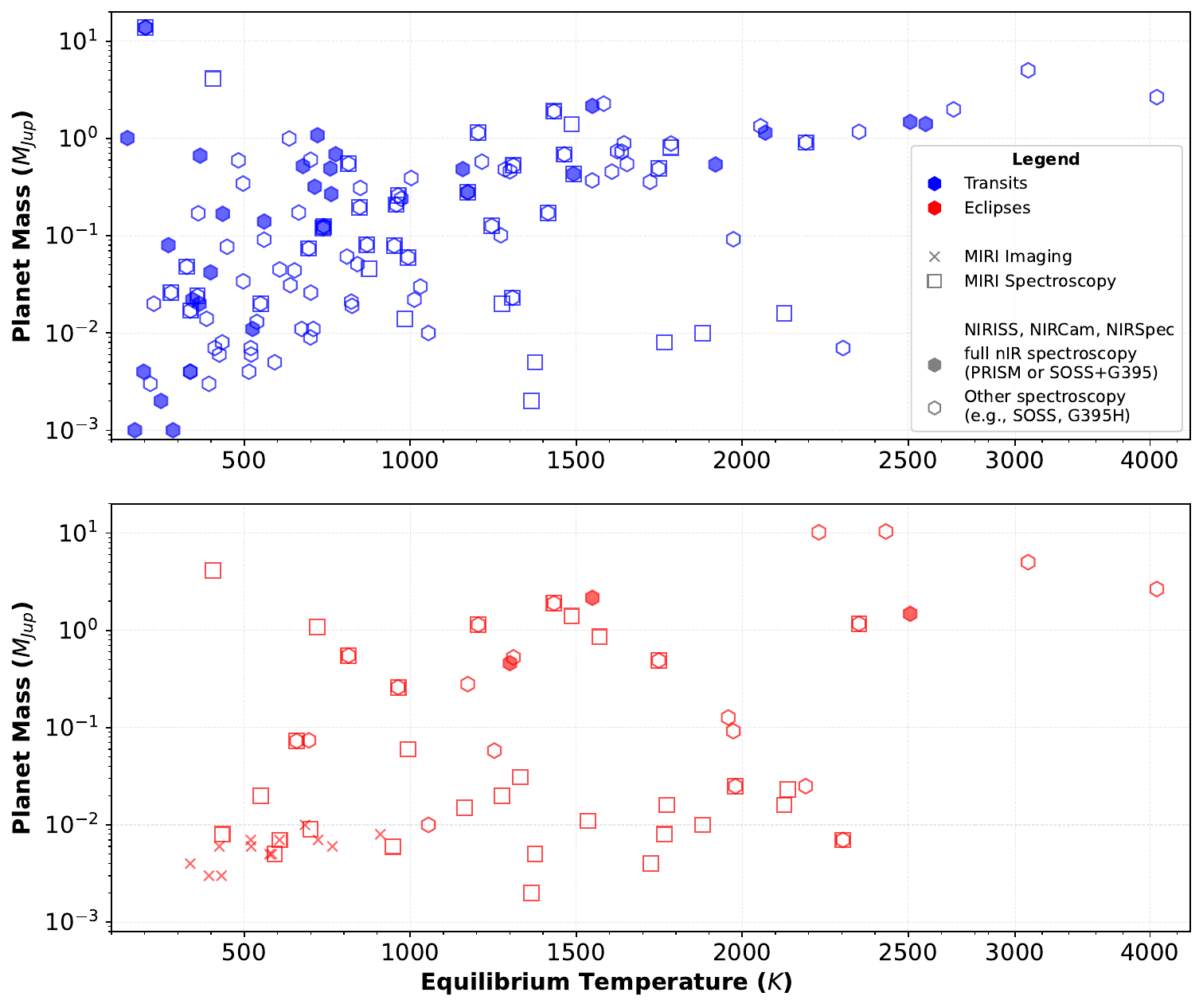}
  \vspace{-1cm}
  \caption{Transiting planets observed by JWST. Data from TrExoLiSTS \cite{2022RNAAS...6..272N}. Note the change in the X-axis linear scaling at 2700 K.}
  \label{fig:trexo}
\end{figure*}

\section{Expectations for the Future}

In April 2023, The Space Telescope Science Institute (STScI) convened the Working Group on Strategic Exoplanet Initiatives with HST and JWST, with the charge to solicit input from the community on key science areas that should be prioritized for future JWST and HST observing programs, but also to develop a concept for a $\sim$500 hour Director’s Discretionary program to begin in Cycle 3.  As described in \cite{2024arXiv240402932R}, 
a clear outcome of the Working Group was the identification of three key science themes that should be addressed with JWST. The first key science theme, described in more detail below, is to assess the commonality and diversity of atmospheres hosted by terrestrial planets orbiting low mass stars (M Spectral types). This first key science theme formed the basis for the recommendation of a 500-hour Director’s Discretionary Time (DDT) allocation to carry out a survey for thermal emission from a collection of rocky exoplanets orbiting M-dwarfs, with the goal of identifying which targets show evidence that their atmospheres have been retained, and which are consistent with being essentially bare rocks. The second theme takes a population-level approach, focussing on understanding the  trends of exoplanet atmospheres spanning sub-Neptunes to gas giants. This recommendation formed the basis for a strategic approach for how to support and structure the science that will result from the inevitable $10^4$ hour survey of exoplanet atmospheres that will be returned by the end of JWST’s lifetime over roughly twenty cycles. Recognizing that such a large dataset will inevitably be obtained represents a remarkable strategic planning opportunity now to maximize the science related to the characterization of exoplanet atmospheres in the next 15-20 years. The last theme is related to understanding exoplanets in the context of their stellar environments, and will focus on synergies with the short-wavelength capabilities of HST.  

The first major recommendation of the WG was for STScI to dedicate a planned 500-hour DDT allocation towards a comprehensive survey of thermal emission from rocky exoplanets spanning a range of temperatures and orbiting M-dwarf stars. M-dwarfs account for roughly 75\% of all stars in the galaxy, and results from the Kepler and TESS missions have shown that rocky exoplanets are relatively common \cite{Dattilo2023}, making rocky exoplanets orbiting M-dwarfs the most common class of planet. Such a program would have the goal of not only searching for atmospheres on these rocky exoplanets, but also identifying the ``cosmic shoreline’’, the boundary that delineates between planets with and without atmospheres, defined primarily by the escape velocity of the planet and the level of insolation from their host stars (See Figure~\ref{fig:shoreline}).   In addition to shedding light on the physical processes involved with atmospheric retention, such an effort would serve as the first crucial step to determine if, and which, rocky planets orbiting M dwarfs have conditions conducive to habitability or not. As outlined in \cite{2024arXiv240402932R}, such an effort needs to be carried out early in the mission to ensure adequate follow-up observations of high-priority targets in subsequent cycles. Surveying a wide sample will also be imperative so that the cosmic shoreline can indeed be identified. 

Roughly one-third of the observing time in the first few JWST cycles have been dedicated to exoplanet science of some kind.  Further, approximately one-third of this fraction has been dedicated to atmospheric characterization of exoplanets, typically via either transit spectroscopy or direct imaging observations.  Assuming this fraction dedicated to exoplanet atmospheric characterization (roughly 10\%) of JWST’s overall time allocation is maintained going forward, and assuming a mission lifetime of 20 cycles with roughly 5000 hours of total time available each cycle, this suggests that $\sim10^4$ hours could, and likely \textit{will}, be dedicated to exoplanet atmospheric characterization over JWST’s lifetime.  With some planning and guidance, such an allocation of time could form a substantially large survey over all classes of exoplanets, spanning terrestrial planets to directly imaged gas giant planets. 
\cite{2024arXiv240402932R} gave recommendations for how best to ensure a comprehensive survey is achieved, as it is unlikely to occur organically through the time allocation committees. From the first four JWST Cycles, for transiting exoplanets there remains an overall lack of statistics in temperature-mass parameter space, especially when considering full wavelength coverage (see Fig. \ref{fig:trexo}).
This ``$10^4$-hour survey’’ is an unparalleled opportunity for strategic planning for the broader exoplanet community.  With the appropriate mechanisms in place that encourage a well-structured and statistically robust survey to be built up ensuring multiple areas of parameter space are covered, such an effort will ensure that the scientific legacy of JWST is maximized.  

While it is often assumed that the interiors and atmospheres of terrestrial planets are inherently diverse, it is becoming increasingly clear that the atmospheres of extrasolar \textit{giant} planets showcase an exceedingly diverse range of properties as well.   Indeed, even before the early JWST cycles it was increasingly clear that exoplanet atmospheric abundances spanned a wide range, which likely reflects the large variety of ways such planets form and migrate through their circumstellar disks. Given this diversity, a large number of planets, spanning a wide range of effective temperatures, compositions, surface gravities, and insolation will need to be analyzed to begin to identify the physical and chemical processes that sculpt planetary atmospheres. 
Recognizing now that $\sim10^4$ hours will inevitably be dedicated to atmospheric characterization of a wide range of exoplanets is a monumental strategic planning opportunity, with the end goal of forming a legacy database in which the eventual impact is greater than the sum of its parts \cite{2024arXiv240402932R}. The NIRSPEC Integral Field Unit (IFU) utilizing the G395H spectral disperser, with spectral resolution of a few thousand from 3-5\,$\mu$m, would provide the requisite high resolution spectroscopy and wavelength coverage to probe the numerous spectral features due to the most important Carbon and Oxygen bearing molecules (e.g., H$_2$O, CO, CO$_2$, CH$_4$).  Studying the trends of the precisely-measured C/O ratios across a range of planet demographics from terrestrials to gas giants may give the first most important constraints on the processes dictating planet formation.

\section{Conclusions \& Final Remarks}
Exoplanetary science is now a major theme within astrophysics, and roughly one-third of the observing time in the initial JWST cycles has been dedicated to exoplanet detection and characterization.  The exoplanet science themes for the JWST programs in the first few cycles have ranged from topics related to planet formation, atmospheric physics, as well as the search for conditions on terrestrial planets conducive to habitability. 
Indeed, the same technical requirements for the detection of high redshift objects have proven to be ideal also for exoplanet characterization and detection. For transiting planets, JWST is providing unprecedented characterization of exoplanets using transit time series spectroscopic observations, eclipse mapping, and phase curves. In the case of the direct imaging modes, JWST is allowing us to directly characterize planetary mass companions at totally new wavelengths \cite{Carter2023, Miles2023} and obtain detections of analogues of our own mature Jupiter as well as solar system ice-giant analogues at wide orbital separations.  
With the remarkable stability of JWST and exquisite low background levels, JWST has really been our first opportunity to carry out precision coronagraphy from space, and the ``lessons learned’’ from this effort will invariably inform the community’s efforts for how to optimally carry out the science goals of the Nancy Grace Roman and Habitable Worlds Observatory missions.  Assuming the observatory continues to perform well over the next 20+ years and that roughly one-third of the observing time continues to be dedicated to exoplanetary science, the exoplanet community can expect roughly 10,000 hours of observations dedicated to atmospheric characterization. With some coordination, this could lead to a substantially large survey of exoplanet atmospheres over all classes of exoplanets from young/old gas giants to terrestrial planets.  JWST will no doubt give us the opportunity to assess the degree to which the transiting planet population as well as those detectable only via direct imaging perhaps form a continuum in which their intrinsic chemical abundances and histories of formation and migration may be imprinted on their atmospheric compositions.

\begin{table}[]
    \centering
    \tiny
    \begin{tabular}{llllll}
\toprule
Object                  & {\rm Mass (M$_{\rm Jup}$)} & $T_{\rm eff} (K)$ & Object Reference & JWST Program ID & JWST Obs.~Ref \\
\midrule
HIP\,65426b		        & 7.1 $\pm$	1.2		& 1283 $\pm$ 25  & \cite{Chauvin2017}              & ERS 1386                       & \cite{Carter2023}\\ 
VHS\,1256b		        & 16  $\pm$	1		& 1145 $\pm$ 25  & \cite{Gauza2015}                & ERS 1386                       & \cite{Miles2023} \\ 
HR\,8799b		        & 6   $\pm$	0.3		& 942  $\pm$ 12  & \cite{Marois2008,Marois2010}    & 4829, 6139, 7268, GTO1188, &  \cite{Balmer2025, Xuan2026, Ruffio2026} \\ 
        		        &             		&                &                                 & GTO1194, GTO1200, GTO4535  &  \\ 
HR\,8799c		        & 8.5 $\pm$	0.4		& 1158 $\pm$ 11  & \cite{Marois2008,Marois2010}    & `` ''                      &  \cite{Balmer2025, Xuan2026, Ruffio2026} \\
HR\,8799d		        & 9.19$\pm$	0.08	& 1179 $\pm$ 38  & \cite{Marois2008,Marois2010}    & `` ''                      &  \cite{Balmer2025, Xuan2026, Ruffio2026} \\ 
HR\,8799e		        & 7.5 $\pm$	0.7		& 1172 $\pm$ 29  & \cite{Marois2008,Marois2010}    & `` ''                      &  \cite{Balmer2025, Xuan2026, Ruffio2026} \\
AF\,Lep\,b		        & 3.2 $\pm$	0.7		& 800  $\pm$ 100 & \cite{Franson2023}              &  DD4558, 5342              & \cite{Franson2024}\\ 
$\beta$\,Pic\,b	        & 9   $\pm$	2	    & 1600 $\pm$ 150 & \cite{Lagrange2010}             &  GTO1241, 4758             & \cite{Kammerer2024} \\ 
PDS\,70b		        & 5.3     			& 1320 $\pm$ 80  & \cite{Keppler2018, Haffert2019} &  1282                      &  \cite{Christiaens2024} \\ 
PDS\,70c		        & 7.5     			& 1075 $\pm$ 75  & \cite{Keppler2018, Haffert2019} &  1282                      &  \cite{Christiaens2024} \\
HR\,2562b 		        & 22  		    	& 1255 $\pm$ 14  & \cite{Konopacky2016}            &  GTO1241                   & \cite{Godoy2024, Ruffio2024}\\ 
$\epsilon$\,Indi\,Ab	& 6.31 $\pm$ 0.60	& 275            & \cite{Matthews2024}             &  2243, 5037, 8438, 8714    & \cite{Matthews2024, Matthews2026, Sanghi2026b} \\ 
14\,Her\,c		        & 7.9 $\pm$  1.5 	& 275            & \cite{Bardalez-Gagliuffi2021}   &  3337                      & \cite{Bardalez-Gagliuffi2025}  \\ 
YSES-1b			        & 14  $\pm$  3  	& 1810 $\pm$ 210 & \cite{Bohn2020}                 &  2044                      & \cite{Hoch2025} \\ 
YSES-1c			        & 6   $\pm$	1		& 1025 $\pm$ 75  & \cite{Bohn2020}                 &  2044                      & \cite{Hoch2025}   \\ 
51\,Eri\,b		        & 11  $\pm$	2		& 800  $\pm$ 39  & \cite{Macintosh2015}            &  GTO1241, GTO1412, 3522    & \cite{Balmer2025} \\ 
HD\,95086b		        & 4.5 $\pm$	0.5		& 925  $\pm$ 125 & \cite{rameau95086}              &  GTO1195, GTO1277, GTO1200 & \cite{Malin2024}\\ 
GJ\,504b		        & 12  $\pm$	8		& 512  $\pm$ 10  & \cite{Kuzuhara2013}             &  GTO1277, 2778, 3647, 6005 & \cite{Malin2024} \\ 
TWA\,7b			        & 0.3     			& 320  $\pm$ 15  & \cite{Lagrange2025}             &  3662                      & \cite{Lagrange2025, Crotts2025}\\ 
Gl\,229\,Bab            & 38.1$\pm$1.0, 34.4$\pm$1.5 & 900$^{+78}_{-29}$, 775$^{+20}_{-33}$    & \cite{nok95, Xuan2024} &  3762 & \cite{Xuan2024b} \\
HD\,106906b		        & 11  $\pm$	2		& 1950 $\pm$ 200 & \cite{Bailey2014}               &  GTO1277                   &  \\ 
Delorme\,1 AB\,b		& 13  $\pm$	5		& 1725 $\pm$ 134 & \cite{Delorme2013}              &  2778                      &  \cite{Malin2025b}\\ 
Coconuts\,2b		    & 8   $\pm$	2		& 483  $\pm$ 49  & \cite{Zhang2021Coconuts}        &  2124                      & \cite{Ravet2026} \\  
2M\,1207\,b		        & 5.5 $\pm$	0.5		& 1300 $\pm$ 100 & \cite{chauvin2004}              &  3181                      & \cite{Luhman2023, Patapis2025}  \\ 
$\kappa$\,And\,b		& 17.3 $\pm$ 1.8	& 1791 $\pm$  68 &  \cite{Carson2013}              &  GTO1241                   & \cite{Godoy2025} \\ 
HD\,206893\,B		    & 28.0 $\pm$ 2.2	& 1429 $\pm$ 6   & \cite{Milli2017}                &  1843, 5485                & \cite{Desdoigts2025} \\ 
HD\,206893\,c 		    & 12.7 $\pm$ 1.2	& 1182 $\pm$ 37  & \cite{Hinkley2023}              &  1843, 5485                & \cite{Desdoigts2025} \\ 
HIP\,99770              & 15 $\pm$ 5        & 1300$\pm$30    & \cite{Currie2023, Winterhalder2025} & 6905                   & \cite{Balmer2026} \\
2M0359+2009\,B		    & 15    			& 2400           & \cite{Martinez2019}             &  2311                      & \\ 
FU\,Tau\,B		        & 15    			& 2400           & \cite{Luhman2009}               &  2311                      & \\ 
DH\,Tau\,B		        & 15  $\pm$	7		& 2400 $\pm$ 50  & \cite{Bonnefoy2014}             &  2311                      & \\ 
1RXS\,1609\,B		    & 8      			& 1800 $\pm$ 200 & \cite{Lafreniere2008}           &  2311                      & \\ 
CT\,Cha\,B		        & 17    			& 2600 $\pm$ 250 & \cite{Schmidt2008}              &  1958, 2311                & \\ 
GSC\,6214-210\,B 	    & 15  $\pm$	2		& 2400 $\pm$     & \cite{Kraus2008, Ireland2011}   &  2311                      & \\ 
SR\,12\,c		        & 13     			& 2600           & \cite{Kuzuhara2011}             &  2311, 6086                & \\ 
AB\,Pic b		        & 13.5 $\pm$ 0.5	& 2000 $\pm$ 200 & \cite{Chauvin2005}              &  6005                      & \\ 
Ross\,458c 		        & 9   $\pm$	3		& 695  $\pm$ 60  & \cite{Burningham2011}           &  GTO1277                   & \\ 
GU\,Psc\,b		        & 11  $\pm$	2		& 1050 $\pm$ 50  & \cite{Naud2014}                 &  GTO1188                   & \\ 
GJ\,758			        & 42  $\pm$	13		& 675  $\pm$ 75  & \cite{Thalmann2009, Bowler2018} &  GTO1413                   & \\ 
HD\,19467B		        & 56.7 $\pm$ 5.9	& 978  $\pm$ 32  & \cite{Crepp2014, Crepp2015}     &  GTO1414                   & \\ 
2MASS\,J2236b		    & 12.5 $\pm$ 1.5	& 1050 $\pm$ 50  & \cite{Bowler2017}               &  GTO1188                   & \\ 
GQ\,Lup\,B		        & 25  $\pm$	15		& 2600 $\pm$ 300 & \cite{Neuhauser2005, McElwain2007} & 7792                    & \\ 
\bottomrule
    \end{tabular}
    \vspace{0.3cm}
    \caption{A representative table of exoplanets and substellar companions that have been observed with JWST in the first few cycles. The masses and effective temperatures listed are those plotted in Figure~\ref{fig:mass_temp_jwst_targets}. The last column lists the publications so far that have showcased the observations of each object.}  
    \label{tab:DI_table}
\end{table}

\begin{table*}
\caption{Summary of JWST transiting exoplanet target results with physical parameters.}
\tiny
\begin{tabular}{l c c r@{\,$\pm$\,}l r@{\,$\pm$\,}l c l}
\toprule
Planet Name & Type & JWST Observation &
\multicolumn{2}{c}{$M_{pl}$ [$M_\oplus$]} &
\multicolumn{2}{c}{$R_{pl}$ [$R_\oplus$]} &
$T_{\rm eq}$ [K] & JWST Obs.\ Ref.\\
\midrule
55 Cancri e & Super-Earth & Eclipse & \multicolumn{2}{c}{7.99 (+0.32/-0.33)} & 1.88 & 0.03 & 1949\textsuperscript{*} & \cite{2024Natur.630..609H,Patel2024} \\
GJ 357 b & Super-Earth & Transit & \multicolumn{2}{c}{--} & 1.20 & 0.06 & 518 & \cite{2025MNRAS.540.3677T,2025arXiv250707165A} \\
GJ 341 b & Terrestrial & Transit & \multicolumn{2}{c}{$<$4.00~~~~~~~~~~~~~~~~~~~} & 0.88 & 0.05 & 539\textsuperscript{*} & \cite{2024AJ....167...90K} \\
GJ 367 b & Terrestrial & Phase Curve & 0.63 & 0.05 & 0.70 & 0.02 & 1365 & \cite{2024ApJ...961L..44Z} \\
GJ 436 b & Warm Neptune & Eclipse & 22.10 & 2.30 & 4.17 & 0.17 & 686 & \cite{2025ApJ...982L..39M} \\
GJ 486 b & Super-Earth & Eclipse & \multicolumn{2}{c}{2.77 (+0.08/-0.07)} & \multicolumn{2}{c}{1.29 (+0.02/-0.01)} & 696 & \cite{2024ApJ...975L..22W} \\
GJ 486 b & Super-Earth & Transit & \multicolumn{2}{c}{2.77 (+0.08/-0.07)} & \multicolumn{2}{c}{1.29 (+0.02/-0.01)} & 696 & \cite{2023ApJ...948L..11M} \\
GJ 1132 b & Terrestrial & Transit & 1.84 & 0.19 & 1.19 & 0.04 & 584 & \cite{2023ApJ...959L...9M, 2025AJ....170..205B} \\
GJ 1132 b & Terrestrial & Eclipse & 1.84 & 0.19 & 1.19 & 0.04 & 584 & \cite{2024ApJ...973L...8X} \\
GJ 1214 b & Sub-Neptune & Phase Curve & \multicolumn{2}{c}{8.41 (+0.36/-0.35)} & 2.73 & 0.03 & 567 & \cite{2023Natur.620...67K} \\
GJ 1214 b & Sub-Neptune & Transit & \multicolumn{2}{c}{8.41 (+0.36/-0.35)} & 2.73 & 0.03 & 567 & \cite{2024ApJ...974L..33S,2025ApJ...979L...7O} \\
GJ 3470 b & Sub-Neptune & Transit & 13.90 & 1.50 & 4.57 & 0.18 & 594 & \cite{2024ApJ...970L..10B} \\
GJ 3090 b & Sub-Neptune & Transit & 3.34 & 0.72 & 2.13 & 0.11 & 693 & \cite{2025ApJ...985L..10A} \\
GJ 3929 b & Terrestrial & Eclipse & \multicolumn{2}{c}{1.75 (+0.44/-0.45)} & 1.09 & 0.04 & 568 & \cite{2025arXiv250812516X} \\
GJ 9827 d & Sub-Neptune & Transit & \multicolumn{2}{c}{3.02 (+0.58/-0.57)} & \multicolumn{2}{c}{1.98 (+0.11/-0.10)} & 675 & \cite{2024ApJ...974L..10P} \\
HAT-P-14 b & Hot Jupiter & Transit & 1093.34 & 200.23 & 15.92 & 1.35 & 1564\textsuperscript{*} & \cite{2023PASP..135a8002E,2025AJ....169..335L} \\
HAT-P-18 b & Warm Jupiter & Transit & 62.61 & 4.13 & 11.15 & 0.58 & 852 & \cite{2022ApJ...940L..35F,2024MNRAS.528.3354F} \\
HAT-P-26 b & Warm Neptune & Transit & 22.25 & 6.36 & 7.06 & 0.45 & 993\textsuperscript{*} & \cite{2025arXiv250916082G} \\
HD 149026 b & Hot Jupiter & Transit & 120.78 & 19.07 & 8.29 & 0.22 & 1676\textsuperscript{*} & \cite{2023Natur.618...43B} \\
HD 189733 b & Hot Jupiter & Transit & 359.15 & 25.43 & 12.67 & 0.11 & 1201\textsuperscript{*} & \cite{2024Natur.632..752F,2025AJ....169...38Z} \\
HD 189733 b & Hot Jupiter & Eclipse & 359.15 & 25.43 & 12.67 & 0.11 & 1201\textsuperscript{*} & \cite{2024ApJ...973L..41I,2025ApJ...983L..13L} \\
HD 209458 b & Hot Jupiter & Transit & 232.02 & 12.71 & 15.58 & 0.22 & 1451\textsuperscript{*} & \cite{2024ApJ...963L...5X} \\
HD 80606 b & Hot Jupiter & Eclipse & 1323.47 & 1.49 & 11.57 & 0.17 & 405\textsuperscript{*} & \cite{2024arXiv240712456S} \\
HIP 67522 b & Hot Jupiter & Transit & \multicolumn{2}{c}{--} & \multicolumn{2}{c}{9.99 (+0.24/-0.22)} & 1175 & \cite{2024AJ....168..297T} \\
KELT-7 b & Ultra-Hot Jupiter & Transit & 441.78 & 69.92 & 17.93 & 0.67 & 2041\textsuperscript{*} & \cite{2025MNRAS.543.2442A} \\
Kepler-51 d & Sub-Neptune & Transit & 7.60 & 1.10 & 9.70 & 0.50 & 381 & \cite{2025arXiv250521358L} \\
K2-18 b & Sub-Neptune & Transit & \multicolumn{2}{c}{8.92 (+1.70/-1.60)} & 2.37 & 0.22 & 284 & \cite{2023ApJ...956L..13M,2025arXiv250118477S} \\
K2-22 b & Terrestrial & Transit & \multicolumn{2}{c}{--} & 2.30 & 0.10 & 1584\textsuperscript{*} & \cite{2025arXiv250108301T} \\
L 98-59 b & Super-Earth & Transit & 0.46 & 0.11 & 0.84 & 0.02 & 620 & \cite{2025ApJ...980L..26B} \\
L 98-59 c & Super-Earth & Transit & 2.00 & 0.13 & 1.33 & 0.03 & 526 & \cite{2024AJ....168..276S} \\
L 98-59 d & Terrestrial & Transit & 1.64 & 0.07 & 1.63 & 0.04 & 416 & \cite{2024ApJ...975L..11B} \\
L 168-9 b & Super-Earth & Transit & 4.07 & 0.45 & 1.63 & 0.14 & 998 & \cite{2025AJ....169...15A} \\
LHS 475 b & Terrestrial & Transit & \multicolumn{2}{c}{--} & 0.99 & 0.05 & 586 & \cite{2023NatAs...7.1317L} \\
LHS 1140 b & Sub-Neptune & Transit & 5.60 & 0.19 & 1.73 & 0.03 & 226 & \cite{2024ApJ...970L...2C} \\
LHS 1140 c & Super-Earth & Eclipse & 1.91 & 0.06 & 1.27 & 0.03 & 422 & \cite{Fortune2025} \\
LHS 1478 b & Super-Earth & Eclipse & 2.33 & 0.20 & 1.24 & 0.05 & 595 & \cite{2025AA...695A.171A} \\
LP 791-18c & Sub-Neptune & Transit & 7.10 & 0.70 & 2.44 & 0.10 & 324 & \cite{2025arXiv251210876R} \\
LTT 3780 b & Super-Earth & Eclipse & 2.46 & 0.19 & 1.32 & 0.06 & 903 & \cite{2025arXiv250814210A} \\
LTT 1445 A b & Super-Earth & Eclipse & \multicolumn{2}{c}{2.73 (+0.25/-0.23)} & \multicolumn{2}{c}{1.34 (+0.11/-0.06)} & 431 & \cite{2025AJ....169..311W} \\
LTT 9779 b & Hot-Neptune & Transit & \multicolumn{2}{c}{29.32 (+0.78/-0.81)} & 4.72 & 0.23 & 1978 & \cite{2024ApJ...962L..20R} \\
LTT 9779 b & Hot-Neptune & Phase Curve & \multicolumn{2}{c}{29.32 (+0.78/-0.81)} & 4.72 & 0.23 & 1978 & \cite{2025arXiv251004863A} \\
TOI-270 b & Super-Earth & Transit & 1.48 & 0.18 & \multicolumn{2}{c}{1.28 (+0.05/-0.04)} & 600 & \cite{2025AJ....170..226C} \\
TOI-270 d & Sub-Neptune & Transit & 4.20 & 0.16 & 2.00 & 0.05 & 383 & \cite{2024arXiv240303325B} \\
TOI-421 b & Sub-Neptune & Transit & 6.70 & 0.60 & 2.64 & 0.08 & 922 & \cite{2025ApJ...984L..44D} \\
TOI-561 b & Super-Earth & Eclipse & 2.02 & 0.23 & 1.40 & 0.03 & 2319 & \cite{2025arXiv250917231T} \\
TOI-776 b & Super-Earth & Transit & 5.00 & 1.60 & 1.80 & 0.08 & 520 & \cite{2025AJ....169..142A} \\
TOI-776 c & Sub-Neptune & Transit & \multicolumn{2}{c}{6.90 (+2.60/-2.50)} & 2.05 & 0.08 & 420 & \cite{2025AJ....169..249T} \\
TOI-836 b & Super-Earth & Transit & \multicolumn{2}{c}{4.53 (+0.92/-0.86)} & 1.70 & 0.07 & 871 & \cite{2024AJ....167..216A} \\
TOI-836 c & Sub-Neptune & Transit & \multicolumn{2}{c}{9.60 (+2.70/-2.50)} & 2.59 & 0.09 & 665 & \cite{2024AJ....168...77W} \\
TOI-1468 b & Super-Earth & Eclipse & 3.04 & 0.46 & 1.40 & 0.06 & 684 & \cite{2025AA...698A..68M} \\
TOI-1685 b & Super-Earth & Phase Curve, Transits & \multicolumn{2}{c}{3.07 (+0.34/-0.33)} & 1.42 & 0.06 & 1000 & \cite{2024arXiv241203411L, 2025arXiv251215338F} \\
TOI-5205 b & Warm Jupiter & Transit & \multicolumn{2}{c}{343.00 (+18.00/-17.00)} & 11.60 & 0.30 & 737 & \cite{2025arXiv250206966C} \\
TRAPPIST-1 b & Terrestrial & Eclipse, Phase Curve & 1.37 & 0.07 & 1.12 & 0.01 & 397\textsuperscript{*} & \cite{2023Natur.618...39G,2025NatAs...9..358D,2025arXiv250902128G} \\
TRAPPIST-1 c & Terrestrial & Transit , Phase Curve & 1.31 & 0.06 & 1.10 & 0.01 & 340\textsuperscript{*} & \cite{2023ApJ...955L...7L,2025ApJ...979L...5R,2025arXiv250902128G} \\
TRAPPIST-1 c & Terrestrial & Eclipse & 1.31 & 0.06 & 1.10 & 0.01 & 340\textsuperscript{*} & \cite{2023Natur.620..746Z} \\
TRAPPIST-1 d & Terrestrial & Transit & 0.39 & 0.01 & 0.79 & 0.01 & 286\textsuperscript{*} & \cite{2025arXiv250808416P} \\
TRAPPIST-1 e & Terrestrial & Transit & 0.69 & 0.02 & 0.92 & 0.01 & 250\textsuperscript{*} & \cite{2025ApJ...990L..52E,2025arXiv251207695A} \\
TrES-4 b & Hot Jupiter & Transit & 247.91 & 60.39 & 18.05 & 2.02 & 1784\textsuperscript{*} & \cite{2025MNRAS.539.1381M} \\
V1298 Tau b & Sub-Neptune & Transit & \multicolumn{2}{c}{--} & \multicolumn{2}{c}{10.27 (+0.58/-0.53)} & 677 & \cite{2025arXiv250708837B} \\
WASP-15 b & Hot Jupiter & Transit & 171.63 & 92.17 & 15.80 & 4.04 & 1650\textsuperscript{*} & \cite{2025MNRAS.537.3027K} \\
WASP-17 b & Hot Jupiter & Transit & 247.91 & 73.10 & 20.96 & 2.69 & 1546\textsuperscript{*} & \cite{2023ApJ...956L..32G,2025AJ....169...86L} \\
WASP-17 b & Hot Jupiter & Eclipse & 247.91 & 73.10 & 20.96 & 2.69 & 1546\textsuperscript{*} & \cite{2024AJ....168..123V,2025AJ....169...57G} \\
WASP-18 b & Ultra-Hot Jupiter & Eclipse & 3241.85 & 111.24 & 13.90 & 0.89 & 2429 & \cite{2023Natur.620..292C} \\
WASP-39 b & Hot Jupiter & Transit & 89.31 & 10.17 & 14.34 & 0.45 & 1166 & \cite{2023Natur.614..649J,2023Natur.614..653A,2023Natur.614..659R,2023Natur.614..664A,2023Natur.614..670F,2024Natur.626..979P,2024NatAs...8.1008C} \\
WASP-43 b & Hot Jupiter & Eclipse & 565.71 & 31.78 & \multicolumn{2}{c}{10.42 (+0.79/-1.01)} & 1427 & \cite{2024NatAs...8..879B,2024AJ....168....4H} \\
WASP-43 b & Hot Jupiter & Phase Curve & 565.71 & 31.78 & \multicolumn{2}{c}{10.42 (+0.79/-1.01)} & 1427 & \cite{2024ApJ...969L..32C} \\
WASP-52 b & Hot Jupiter & Transit & 146.20 & 6.36 & 14.24 & 0.34 & 1315 & \cite{2025MNRAS.539..422F} \\
WASP-69 b & Warm Jupiter & Eclipse & 82.58 & 5.37 & 11.21 & 0.22 & 971 & \cite{2024AJ....168..104S} \\
WASP-77 A b & Hot Jupiter & Eclipse & 529.31 & 0.14 & 13.56 & 0.22 & 1695\textsuperscript{*} & \cite{2023ApJ...953L..24A} \\
WASP-80 b & Warm Jupiter & Eclipse & \multicolumn{2}{c}{170.99 (+11.12/-11.44)} & \multicolumn{2}{c}{11.20 (+0.34/-0.35)} & 825 & \cite{2025arXiv250601800W} \\
WASP-80 b & Warm Jupiter & Transit & \multicolumn{2}{c}{170.99 (+11.12/-11.44)} & \multicolumn{2}{c}{11.20 (+0.34/-0.35)} & 825 & \cite{2023Natur.623..709B} \\
WASP-94 A b & Hot Jupiter & Transit & \multicolumn{2}{c}{196.41 (+8.90/-9.22)} & 17.7 & 1.5 & 1604 & \cite{2025MNRAS.tmp..824A,2025arXiv250510910M} \\
WASP-96 b & Hot Jupiter & Transit & 152.55 & 9.53 & 13.45 & 0.67 & 1285 & \cite{2023MNRAS.524..835R,2023MNRAS.524..817T,2025arXiv251116771W} \\
WASP-107 b & Warm Neptune & Transit & 30.50 & 1.70 & 10.54 & 0.22 & 770 & \cite{2024Natur.625...51D,2024Natur.630..831S,2024Natur.630..836W,2024Natur.632.1017E,2024NatAs...8.1562M,2025NatAs.tmp..236K} \\
WASP-121 b & Ultra-Hot Jupiter & Phase Curve & 371.86 & 13.67 & 19.53 & 0.07 & 2409 & \cite{2023ApJ...943L..17M,2024AJ....168..231S,2025NatAs.tmp..123E,2025arXiv251009809A} \\
WASP-121 b & Ultra-Hot Jupiter & Transit & 371.86 & 13.67 & 19.53 & 0.07 & 2409 & \cite{2025AJ....169..341G} \\
WASP-166 b & Hot Neptune & Transit & 32.10 & 1.59 & 7.06 & 0.34 & 1270 & \cite{2025arXiv250100609M} \\
WASP-178 b & Ultra-Hot Jupiter & Transit & 527.60 & 38.14 & 20.29 & 1.01 & 2470 & \cite{2025AJ....169..274L} \\
\bottomrule
\end{tabular}
\begin{tablenotes}
\tiny
\item[$^*$]  Planetary and stellar parameters from the NASA Exoplanet Archive \cite{2025PSJ.....6..186C}. $^*$$T_{\rm eq}$ calculated from $T_{\rm eff}$ and $a/R_*$ values assuming zero albedo. 
\end{tablenotes}
\label{transit-table}
\end{table*}

\bibliographystyle{tfnlm}
\bibliography{bibtexabs}{}

@inproceedings{Girard2022,
	adsurl = {https://ui.adsabs.harvard.edu/abs/2022SPIE12180E..3QG},
	archiveprefix = {arXiv},
	author = {{Girard}, Julien H. and {Leisenring}, Jarron and {Kammerer}, Jens and {Gennaro}, Mario and {Rieke}, Marcia and {Stansberry}, John and {Rest}, Armin and {Egami}, Eiichi and {Sunnquist}, Ben and {Boyer}, Martha and {Canipe}, Alicia and {Correnti}, Matteo and {Hilbert}, Bryan and {Perrin}, Marshall D. and {Pueyo}, Laurent and {Soummer}, Remi and {Allen}, Marsha and {Bushouse}, Howard and {Aguilar}, Jonathan and {Brooks}, Brian and {Coe}, Dan and {DiFelice}, Audrey and {Golimowski}, David and {Hartig}, George and {Hines}, Dean C. and {Koekemoer}, Anton and {Nickson}, Bryony and {Nikolov}, Nikolay and {Kozhurina-Platais}, Vera and {Pirzkal}, Nor and {Robberto}, Massimo and {Sivaramakrishnan}, Anand and {Sohn}, Sangmo Tony and {Telfer}, Randal and {Wu}, Chi Rai and {Beatty}, Thomas and {Florian}, Michael and {Hainline}, Kevin and {Kelly}, Doug and {Misselt}, Karl and {Schlawin}, Everett and {Sun}, Fengwu and {Williams}, Christina and {Willmer}, Christopher and {Stark}, Christopher and {Ygouf}, Marie and {Carter}, Aarynn and {Beichman}, Charles and {Greene}, Thomas P. and {Roellig}, Thomas and {Krist}, John and {Adams Redai}, J{\'e}a. and {Wang}, Jason and {Clark}, Charles R. and {Lewis}, Dan and {Ferry}, Malcolm},
	booktitle = {Space Telescopes and Instrumentation 2022: Optical, Infrared, and Millimeter Wave},
	doi = {10.1117/12.2629636},
	editor = {{Coyle}, Laura E. and {Matsuura}, Shuji and {Perrin}, Marshall D.},
	eid = {121803Q},
	eprint = {2208.00998},
	month = aug,
	pages = {121803Q},
	primaryclass = {astro-ph.IM},
	series = {Society of Photo-Optical Instrumentation Engineers (SPIE) Conference Series},
	title = {{JWST/NIRCam coronagraphy: commissioning and first on-sky results}},
	volume = {12180},
	year = 2022}

@article{Boccaletti2022,
	adsurl = {https://ui.adsabs.harvard.edu/abs/2022A&A...667A.165B},
	archiveprefix = {arXiv},
	author = {{Boccaletti}, A. and {Cossou}, C. and {Baudoz}, P. and {Lagage}, P.~O. and {Dicken}, D. and {Glasse}, A. and {Hines}, D.~C. and {Aguilar}, J. and {Detre}, O. and {Nickson}, B. and {Noriega-Crespo}, A. and {G{\'a}sp{\'a}r}, A. and {Labiano}, A. and {Stark}, C. and {Rouan}, D. and {Reess}, J.~M. and {Wright}, G.~S. and {Rieke}, G. and {Garcia Marin}, M. and {Lajoie}, C. and {Girard}, J. and {Perrin}, M. and {Soummer}, R. and {Pueyo}, L.},
	doi = {10.1051/0004-6361/202244578},
	eid = {A165},
	eprint = {2207.11080},
	journal = {\aap},
	month = nov,
	pages = {A165},
	primaryclass = {astro-ph.IM},
	title = {{JWST/MIRI coronagraphic performances as measured on-sky}},
	volume = {667},
	year = 2022}

@article{Winterhalder2025,
	adsurl = {https://ui.adsabs.harvard.edu/abs/2025A&A...700A...4W},
	archiveprefix = {arXiv},
	author = {{Winterhalder}, T.~O. and {Kammerer}, J. and {Lacour}, S. and {M{\'e}rand}, A. and {Nowak}, M. and {Stolker}, T. and {Balmer}, W.~O. and {Marleau}, G.-D. and {Abuter}, R. and {Amorim}, A. and {Asensio-Torres}, R. and {Berger}, J.-P. and {Beust}, H. and {Blunt}, S. and {Bonnefoy}, M. and {Bonnet}, H. and {Bordoni}, M.~S. and {Bourdarot}, G. and {Brandner}, W. and {Cantalloube}, F. and {Caselli}, P. and {Charnay}, B. and {Chauvin}, G. and {Chavez}, A. and {Choquet}, E. and {Christiaens}, V. and {Cl{\'e}net}, Y. and {Coud{\'e} du Foresto}, V. and {Cridland}, A. and {Davies}, R. and {Dembet}, R. and {Dexter}, J. and {Drescher}, A. and {Duvert}, G. and {Eckart}, A. and {Eisenhauer}, F. and {F{\"o}rster Schreiber}, N.~M. and {Garcia}, P. and {Garcia Lopez}, R. and {Gardner}, T. and {Gendron}, E. and {Genzel}, R. and {Gillessen}, S. and {Girard}, J.~H. and {Grant}, S. and {Haubois}, X. and {Hei{\ss}el}, G. and {Henning}, Th. and {Hinkley}, S. and {Hippler}, S. and {Houll{\'e}}, M. and {Hubert}, Z. and {Jocou}, L. and {Keppler}, M. and {Kervella}, P. and {Kreidberg}, L. and {Kurtovic}, N.~T. and {Lagrange}, A.-M. and {Lapeyr{\`e}re}, V. and {Le Bouquin}, J.-B. and {Lutz}, D. and {Maire}, A.-L. and {Mang}, F. and {Molli{\`e}re}, P. and {Mordasini}, C. and {Mouillet}, D. and {Nasedkin}, E. and {Ott}, T. and {Otten}, G.~P.~P.~L. and {Paladini}, C. and {Paumard}, T. and {Perraut}, K. and {Perrin}, G. and {Pourr{\'e}}, N. and {Pueyo}, L. and {C Ribeiro}, D. and {Rickman}, E. and {Rustamkulov}, Z. and {Shangguan}, J. and {Shimizu}, T. and {Sing}, D. and {Stadler}, J. and {Straub}, O. and {Straubmeier}, C. and {Sturm}, E. and {Tacconi}, L.~J. and {van Dishoeck}, E.~F. and {Vigan}, A. and {Vincent}, F. and {von Fellenberg}, S.~D. and {Wang}, J.~J. and {Widmann}, F. and {Woillez}, J. and {Yazici}, S.},
	doi = {10.1051/0004-6361/202554766},
	eid = {A4},
	eprint = {2507.00117},
	journal = {\aap},
	month = aug,
	pages = {A4},
	primaryclass = {astro-ph.EP},
	title = {{Orbit and atmosphere of HIP 99770 b through the eyes of VLTI/GRAVITY}},
	volume = {700},
	year = 2025
    }

@article{Balmer2026,
	adsurl = {https://ui.adsabs.harvard.edu/abs/2026ApJ..1001L..26B},
	archiveprefix = {arXiv},
	author = {{Balmer}, William O. and {Pueyo}, Laurent and {Messier}, Ashley and {Bruinsma}, Evelyn and {Jones}, Jeremy and {Matuszewska}, Klara and {Perrin}, Marshall D. and {Girard}, Julien H. and {Leisenring}, Jarron M. and {Lawson}, Kellen and {van der Marel}, Roeland P. and {Kammerer}, Jens and {Carter}, Aarynn and {M{\^a}lin}, Mathilde and {Ward-Duong}, Kimberly and {Hoch}, Kielan K.~W. and {Rickman}, Emily and {Seager}, Sara},
	doi = {10.3847/2041-8213/ae374a},
	eid = {L26},
	eprint = {2604.09785},
	journal = {\apjl},
	month = apr,
	number = {2},
	pages = {L26},
	primaryclass = {astro-ph.EP},
	title = {{Direct Images of CO$_{2}$ Absorption in the Atmosphere of a Super-Jupiter: Enhanced Metallicity Suggestive of Formation in a Disk}},
	volume = {1001},
	year = 2026}

@article{Currie2023,
	adsurl = {https://ui.adsabs.harvard.edu/abs/2023Sci...380..198C},
	archiveprefix = {arXiv},
	author = {{Currie}, Thayne and {Brandt}, G. Mirek and {Brandt}, Timothy D. and {Lacy}, Brianna and {Burrows}, Adam and {Guyon}, Olivier and {Tamura}, Motohide and {Liu}, Ranger Y. and {Sagynbayeva}, Sabina and {Tobin}, Taylor and {Chilcote}, Jeffrey and {Groff}, Tyler and {Marois}, Christian and {Thompson}, William and {Murphy}, Simon J. and {Kuzuhara}, Masayuki and {Lawson}, Kellen and {Lozi}, Julien and {Deo}, Vincent and {Vievard}, Sebastien and {Skaf}, Nour and {Uyama}, Taichi and {Jovanovic}, Nemanja and {Martinache}, Frantz and {Kasdin}, N. Jeremy and {Kudo}, Tomoyuki and {McElwain}, Michael and {Janson}, Markus and {Wisniewski}, John and {Hodapp}, Klaus and {Nishikawa}, Jun and {He{\l}miniak}, Krzysztof and {Kwon}, Jungmi and {Hayashi}, Masahiko},
	doi = {10.1126/science.abo6192},
	eprint = {2212.00034},
	journal = {Science},
	month = apr,
	number = {6641},
	pages = {198-203},
	primaryclass = {astro-ph.EP},
	title = {{Direct imaging and astrometric detection of a gas giant planet orbiting an accelerating star}},
	volume = {380},
	year = 2023
    }

@article{Lazzoni2025,
	adsurl = {https://ui.adsabs.harvard.edu/abs/2025A&A...704A.176L},
	archiveprefix = {arXiv},
	author = {{Lazzoni}, C. and {Bendahan-West}, R. and {Marino}, S. and {Lawson}, K.~D. and {Carter}, A. and {Squicciarini}, V. and {Strampelli}, G. and {Hinkley}, S. and {Kennedy}, G. and {James}, A.~D. and {Milli}, J. and {Ray}, S.},
	doi = {10.1051/0004-6361/202555602},
	eid = {A176},
	eprint = {2511.07561},
	journal = {\aap},
	month = dec,
	pages = {A176},
	primaryclass = {astro-ph.EP},
	title = {{JWST/NIRCam observations of HD 92945 debris disk: An asymmetric disk with a gap}},
	volume = {704},
	year = 2025
    }

@article{Ravet2026,
	adsurl = {https://ui.adsabs.harvard.edu/abs/2026A&A...710A..46R},
	archiveprefix = {arXiv},
	author = {{Ravet}, Matthieu and {Bonnefoy}, Micka{\"e}l and {Chauvin}, Ga{\"e}l and {Zhang}, Zhoujian and {Faherty}, Jacqueline K. and {Voyer}, Ma{\"e}l and {Phillips}, Mark W. and {Tremblin}, Pascal and {Kiman}, Rocio and {Copeland}, Jessica and {Mang}, James J. and {Morley}, Caroline V. and {K{\"u}hnle}, Helena and {Charnay}, Benjamin and {de Regt}, Sam and {Molli{\`e}re}, Paul and {Petrus}, Simon and {Denis}, Allan and {Radcliffe}, Alice and {Palma-Bifani}, Paulina and {Vigan}, Arthur and {M{\^a}lin}, Mathilde and {Marleau}, Gabriel-Dominique and {Manjavacas}, Elena and {Hoy}, Kevin and {Matthews}, Elisabeth C. and {Henning}, Thomas K.},
	doi = {10.1051/0004-6361/202658927},
	eid = {A46},
	eprint = {2604.07176},
	journal = {\aap},
	month = may,
	pages = {A46},
	primaryclass = {astro-ph.EP},
	title = {{Panchromatic view of the frigid Jovian exoplanet COCONUTS-2 b}},
	volume = {710},
	year = 2026
    }

@article{Xuan2024,
	adsurl = {https://ui.adsabs.harvard.edu/abs/2024Natur.634.1070X},
	archiveprefix = {arXiv},
	author = {{Xuan}, Jerry W. and {M{\'e}rand}, A. and {Thompson}, W. and {Zhang}, Y. and {Lacour}, S. and {Blakely}, D. and {Mawet}, D. and {Oppenheimer}, R. and {Kammerer}, J. and {Batygin}, K. and {Sanghi}, A. and {Wang}, J. and {Ruffio}, J.-B. and {Liu}, M.~C. and {Knutson}, H. and {Brandner}, W. and {Burgasser}, A. and {Rickman}, E. and {Bowens-Rubin}, R. and {Salama}, M. and {Balmer}, W. and {Blunt}, S. and {Bourdarot}, G. and {Caselli}, P. and {Chauvin}, G. and {Davies}, R. and {Drescher}, A. and {Eckart}, A. and {Eisenhauer}, F. and {Fabricius}, M. and {Feuchtgruber}, H. and {Finger}, G. and {F{\"o}rster Schreiber}, N.~M. and {Garcia}, P. and {Genzel}, R. and {Gillessen}, S. and {Grant}, S. and {Hartl}, M. and {Hau{\ss}mann}, F. and {Henning}, T. and {Hinkley}, S. and {H{\"o}nig}, S.~F. and {Horrobin}, M. and {Houll{\'e}}, M. and {Janson}, M. and {Kervella}, P. and {Kral}, Q. and {Kreidberg}, L. and {Le Bouquin}, J.-B. and {Lutz}, D. and {Mang}, F. and {Marleau}, G.-D. and {Millour}, F. and {More}, N. and {Nowak}, M. and {Ott}, T. and {Otten}, G. and {Paumard}, T. and {Rabien}, S. and {Rau}, C. and {Ribeiro}, D.~C. and {Sadun Bordoni}, M. and {Sauter}, J. and {Shangguan}, J. and {Shimizu}, T.~T. and {Sykes}, C. and {Soulain}, A. and {Spezzano}, S. and {Straubmeier}, C. and {Stolker}, T. and {Sturm}, E. and {Subroweit}, M. and {Tacconi}, L.~J. and {van Dishoeck}, E.~F. and {Vigan}, A. and {Widmann}, F. and {Wieprecht}, E. and {Winterhalder}, T.~O. and {Woillez}, J.},
	doi = {10.1038/s41586-024-08064-x},
	eprint = {2410.11953},
	journal = {\nat},
	month = oct,
	number = {8036},
	pages = {1070-1074},
	primaryclass = {astro-ph.SR},
	title = {{The cool brown dwarf Gliese 229 B is a close binary}},
	volume = {634},
	year = 2024
    }

@article{nok95,
	adsurl = {https://ui.adsabs.harvard.edu/abs/1995Natur.378..463N},
	author = {{Nakajima}, T. and {Oppenheimer}, B.~R. and {Kulkarni}, S.~R. and {Golimowski}, D.~A. and {Matthews}, K. and {Durrance}, S.~T.},
	doi = {10.1038/378463a0},
	journal = {\nat},
	month = nov,
	number = {6556},
	pages = {463-465},
	title = {{Discovery of a cool brown dwarf}},
	volume = {378},
	year = 1995}

@article{Xuan2024b,
	adsurl = {https://ui.adsabs.harvard.edu/abs/2024ApJ...977L..32X},
	archiveprefix = {arXiv},
	author = {{Xuan}, Jerry W. and {Perrin}, Marshall D. and {Mawet}, Dimitri and {Knutson}, Heather A. and {Mukherjee}, Sagnick and {Zhang}, Yapeng and {Hoch}, Kielan K.~W. and {Wang}, Jason J. and {Inglis}, Julie and {Wallack}, Nicole L. and {Ruffio}, Jean-Baptiste},
	doi = {10.3847/2041-8213/ad92f9},
	eid = {L32},
	eprint = {2411.10571},
	journal = {\apjl},
	month = dec,
	number = {2},
	pages = {L32},
	primaryclass = {astro-ph.SR},
	title = {{Atmospheric Abundances and Bulk Properties of the Binary Brown Dwarf Gliese 229Bab from JWST/MIRI Spectroscopy}},
	volume = {977},
	year = 2024
    }

@article{Whiteford2026,
	adsurl = {https://ui.adsabs.harvard.edu/abs/2026arXiv260806583W},
	archiveprefix = {arXiv},
	author = {{Whiteford}, Niall and {Faherty}, Jacqueline K. and {Burningham}, Ben and {Vos}, Johanna M. and {Petrus}, Simon and {Patapis}, Polychronis and {Biller}, Beth A. and {Skemer}, Andrew and {Hinkley}, Sasha and {Calamari}, Emily and {Su{\'a}rez}, Genaro and {Cruz}, Kelle L. and {Miles}, Brittany E. and {Carter}, Aarynn L. and {Martinez}, Francisco A. and {Rowland}, Melanie J. and {Absil}, Olivier and {Adams}, Arthur D. and {Balmer}, William O. and {Boccaletti}, Anthony and {Bonavita}, Mariangela and {Bonnefoy}, Micka{\"e}l and {Booth}, Mark and {Bowler}, Brendan P. and {Briesemeister}, Zackery W. and {Bryan}, Marta L. and {Calissendorff}, Per and {Cantalloube}, Faustine and {Charnay}, Benjamin and {Chauvin}, Ga{\"e}l and {Chen}, Christine H. and {Choquet}, Elodie and {Christiaens}, Valentin and {Cugno}, Gabriele and {Currie}, Thayne and {Danielski}, Camilla and {De Furio}, Matthew and {Dupuy}, Trent J. and {Factor}, Samuel M. and {Fitzgerald}, Michael P. and {Fortney}, Jonathan J. and {Franson}, Kyle and {Girard}, Julien H. and {Gonzales}, Eileen C. and {Grady}, Carol A. and {Henning}, Thomas and {Hines}, Dean C. and {Hood}, Callie E. and {Hoch}, Kielan K.~W. and {Howe}, Alex R. and {Janson}, Markus and {Kalas}, Paul and {Kammerer}, Jens and {Kennedy}, Grant M. and {Kervella}, Pierre and {Kim}, Minjae and {Kitzmann}, Daniel and {Kraus}, Adam L. and {Kuzuhara}, Masayuki and {Lagage}, Pierre-Olivier and {Lagrange}, Anne-Marie and {Lawson}, Kellen and {Lazzoni}, Cecilia and {Leisenring}, Jarron M. and {Lew}, Ben W.~P. and {Liu}, Michael C. and {Liu}, Pengyu and {Llop-Sayson}, Jorge and {Lloyd}, James P. and {Lueber}, Anna and {Macintosh}, Bruce and {M{\^a}lin}, Mathilde and {Manjavacas}, Elena and {Marino}, Sebasti{\'a}n and {Marley}, Mark S. and {Marois}, Christian and {Martinez}, Raquel A. and {Matthews}, Elisabeth C. and {Matthews}, Brenda C. and {Mawet}, Dimitri and {Mazoyer}, Johan and {McElwain}, Michael W. and {Metchev}, Stanimir and {Meyer}, Michael R. and {Millar-Blanchaer}, Maxwell A. and {Molli{\`e}re}, Paul and {Moran}, Sarah E. and {Morley}, Caroline V. and {Mukherjee}, Sagnick and {Palma-Bifani}, Paulina and {Pantin}, Eric and {Perrin}, Marshall D. and {Pueyo}, Laurent and {Quanz}, Sascha P. and {Quirrenbach}, Andreas and {Ray}, Shrishmoy and {Rebollido}, Isabel and {Adams Redai}, Jea and {Ren}, Bin B. and {Rickman}, Emily and {Sallum}, Steph and {Samland}, Matthias and {Sargent}, Benjamin and {Schlieder}, Joshua E. and {Stapelfeldt}, Karl R. and {Stone}, Jordan M. and {Sutlieff}, Ben J. and {Tamura}, Motohide and {Tan}, Xianyu and {Theissen}, Christopher A. and {Tremblin}, Pascal and {Uyama}, Taichi and {Vasist}, Malavika and {Vigan}, Arthur and {Wagner}, Kevin and {Wang}, Jason J. and {Ward-Duong}, Kimberly and {Wolff}, Schuyler G. and {Worthen}, Kadin and {Wyatt}, Mark C. and {Ygouf}, Marie and {Zurlo}, Alice and {Zhang}, Xi and {Zhang}, Keming and {Zhang}, Zhoujian and {Zhou}, Yifan},
	doi = {10.48550/arXiv.2608.06583},
	eid = {arXiv:2608.06583},
	eprint = {2608.06583},
	journal = {arXiv e-prints},
	month = aug,
	pages = {arXiv:2608.06583},
	primaryclass = {astro-ph.EP},
	title = {{The JWST Early Release Science Program for Direct Observations of Exoplanetary Systems VIII: patchy forsterite and enstatite clouds in the atmosphere of VHS 1256 b, retrieval lessons learned and outlook to the future}},
	year = 2026
    }

@article{Beichman2025,
	adsurl = {https://ui.adsabs.harvard.edu/abs/2025ApJ...989L..22B},
	archiveprefix = {arXiv},
	author = {{Beichman}, Charles and {Sanghi}, Aniket and {Mawet}, Dimitri and {Kervella}, Pierre and {Wagner}, Kevin and {Quarles}, Billy and {Lissauer}, Jack J. and {Sommer}, Max and {Wyatt}, Mark and {Godoy}, Nicolas and {Balmer}, William O. and {Pueyo}, Laurent and {Llop-Sayson}, Jorge and {Aguilar}, Jonathan and {Akeson}, Rachel and {Belikov}, Ruslan and {Boccaletti}, Anthony and {Choquet}, Elodie and {Fomalont}, Edward and {Henning}, Thomas and {Hines}, Dean and {Hu}, Renyu and {Lagage}, Pierre-Olivier and {Leisenring}, Jarron and {Mang}, James and {Ressler}, Michael and {Serabyn}, Eugene and {Tremblin}, Pascal and {Marie Ygouf} and {Zilinskas}, Mantas},
	doi = {10.3847/2041-8213/adf53f},
	eid = {L22},
	eprint = {2508.03814},
	journal = {\apjl},
	month = aug,
	number = {2},
	pages = {L22},
	primaryclass = {astro-ph.EP},
	title = {{Worlds Next Door: A Candidate Giant Planet Imaged in the Habitable Zone of {\ensuremath{\alpha}} Centauri A. I. Observations, Orbital and Physical Properties, and Exozodi Upper Limits}},
	volume = {989},
	year = 2025
    }

@article{Sanghi2025,
	adsurl = {https://ui.adsabs.harvard.edu/abs/2025ApJ...989L..23S},
	archiveprefix = {arXiv},
	author = {{Sanghi}, Aniket and {Beichman}, Charles and {Mawet}, Dimitri and {Balmer}, William O. and {Godoy}, Nicolas and {Pueyo}, Laurent and {Boccaletti}, Anthony and {Sommer}, Max and {Bidot}, Alexis and {Choquet}, Elodie and {Kervella}, Pierre and {Lagage}, Pierre-Olivier and {Leisenring}, Jarron and {Llop-Sayson}, Jorge and {Ressler}, Michael and {Wagner}, Kevin and {Wyatt}, Mark},
	doi = {10.3847/2041-8213/adf53e},
	eid = {L23},
	eprint = {2508.03812},
	journal = {\apjl},
	month = aug,
	number = {2},
	pages = {L23},
	primaryclass = {astro-ph.EP},
	title = {{Worlds Next Door: A Candidate Giant Planet Imaged in the Habitable Zone of {\ensuremath{\alpha}} Centauri A. II. Binary Star Modeling, Planet and Exozodi Search, and Sensitivity Analysis}},
	volume = {989},
	year = 2025
    }

@article{Sanghi2026a,
	adsurl = {https://ui.adsabs.harvard.edu/abs/2026AJ....171..225S},
	archiveprefix = {arXiv},
	author = {{Sanghi}, Aniket and {Mang}, James and {Llop-Sayson}, Jorge and {Mamajek}, Eric E. and {Thompson}, William and {Sur}, Ankan and {Beichman}, Charles and {Bryden}, Geoffrey and {G{\'a}sp{\'a}r}, Andr{\'a}s and {Leisenring}, Jarron and {Mawet}, Dimitri and {Morley}, Caroline V. and {Ruffio}, Jean-Baptiste and {Wolff}, Schuyler G. and {Ygouf}, Marie},
	doi = {10.3847/1538-3881/ae4909},
	eid = {225},
	eprint = {2602.23423},
	journal = {\aj},
	month = apr,
	number = {4},
	pages = {225},
	primaryclass = {astro-ph.EP},
	title = {{Worlds Next Door. III. Indirect Evidence for Enhanced Atmospheric Metallicity and/or the Presence of Water Clouds in the Nearest Jupiter-analog $\epsilon$ Eri b}},
	volume = {171},
	year = 2026
    }

@article{Sanghi2026b,
	adsurl = {https://ui.adsabs.harvard.edu/abs/2026AJ....172..115S},
	archiveprefix = {arXiv},
	author = {{Sanghi}, Aniket and {Thompson}, William and {Mang}, James and {Xuan}, Jerry W. and {Mawet}, Dimitri and {Ruffio}, Jean-Baptiste and {Zhang}, Yapeng and {Wang}, Jason J. and {Morley}, Caroline V. and {Nielsen}, Eric and {Roberson}, William and {Matthews}, Elisabeth and {Carter}, Aarynn L. and {Crossfield}, Ian J.~M. and {M{\^a}lin}, Mathilde and {Benneke}, Bj{\"o}rn and {Bidot}, Alexis and {G{\'a}sp{\'a}r}, Andr{\'a}s and {He}, Carrie and {Horstman}, Katelyn and {Madurowicz}, Alexander and {Marois}, Christian and {Oppenheimer}, Rebecca and {Perrin}, Marshall},
	doi = {10.3847/1538-3881/ae74c8},
	eid = {115},
	eprint = {2603.08787},
	journal = {\aj},
	month = aug,
	number = {2},
	pages = {115},
	primaryclass = {astro-ph.EP},
	title = {{Worlds Next Door. IV. Mapping the Late Stages of Giant Planet Evolution with a Precise Dynamical Mass and Luminosity for $\epsilon$ Ind Ab}},
	volume = {172},
	year = 2026
    }

@article{Gibbs2026,
	adsurl = {https://ui.adsabs.harvard.edu/abs/2026ApJ..1006L..11G},
	archiveprefix = {arXiv},
	author = {{Gibbs}, Aidan and {Ruffio}, Jean-Baptiste and {Bidot}, Alexis and {Barman}, Travis S. and {Do {\'O}}, Clarissa R. and {Konopacky}, Quinn M. and {Perrin}, Marshall D. and {Baburaj}, Aneesh and {Dacus}, Beck and {Macintosh}, Bruce and {Madurowicz}, Alex and {Xuan}, Jerry W.},
	doi = {10.3847/2041-8213/ae801b},
	eid = {L11},
	eprint = {2606.23789},
	journal = {\apjl},
	month = jul,
	number = {1},
	pages = {L11},
	primaryclass = {astro-ph.EP},
	title = {{Discovery of an Exterior Third Planet Orbiting {\ensuremath{\beta}} Pictoris}},
	volume = {1006},
	year = 2026}

@article{Sutlieff2026,
	adsurl = {https://ui.adsabs.harvard.edu/abs/2026ApJ..1006L..10S},
	archiveprefix = {arXiv},
	author = {{Sutlieff}, Ben J. and {Bonse}, Markus J. and {Christiaens}, Valentin and {Fontanive}, Cl{\'e}mence and {Matthews}, Elisabeth C. and {Parker}, Luke T. and {Pearce}, Tim D. and {Birkby}, Jayne L. and {Biller}, Beth A. and {Dupuy}, Trent J. and {Garvin}, Emily O. and {Iskandarli}, Leyla and {Kammerer}, Jens and {Zhou}, Yifan and {De Rosa}, Robert J. and {Carter}, Aarynn L. and {Hinkley}, Sasha and {Kenworthy}, Matthew A. and {Balmer}, William O. and {Hammond}, Iain and {Mang}, James and {Morley}, Caroline V. and {Neeser}, Mark J. and {Absil}, Olivier and {Boccaletti}, Anthony and {Bonavita}, Mariangela and {Bowler}, Brendan P. and {Chen}, Xueqing and {Dannert}, Felix A. and {Girard}, Julien H. and {Kasper}, Markus and {Lagrange}, Anne-Marie and {Liu}, Pengyu and {Orban de Xivry}, Gilles and {Poon}, Michael and {Quanz}, Sascha P. and {Serra}, Beno{\^\i}t and {Vos}, Johanna M. and {Wagner}, Kevin and {Wang}, Jason and {Sch{\"o}lkopf}, Bernhard and {Agapito}, Guido and {Agudo Berbel}, Alex and {Apai}, D{\'a}niel and {Baruffolo}, Andrea and {Black}, Martin and {Bonaglia}, Marco and {Briguglio}, Runa and {Cao}, Yixian and {Carbonaro}, Luca and {Chapman}, Lee and {Cresci}, Giovanni and {Dallilar}, Yigit and {Davies}, Richard and {Deysenroth}, Matthias and {Di Antonio}, Ivan and {Di Cianno}, Amico and {Di Rico}, Gianluca and {Doelman}, David and {Dolci}, Mauro and {Eisenhauer}, Frank and {Esposito}, Simone and {Ferruzzi}, Debora and {Feuchtgruber}, Helmut and {F{\"o}rster-Schreiber}, Natascha and {Franson}, Kyle and {Genzel}, Reinhard and {Gillessen}, Stefan and {Gonzales}, Eileen C. and {Hartl}, Michael and {Hayoz}, Jean and {Huber}, Heinrich and {Keller}, Christoph and {Kravchenko}, Kateryna and {Leisenring}, Jarron and {Lightfoot}, John and {Lunney}, David and {Lutz}, Dieter and {Macintosh}, Mike and {Mannucci}, Filippo and {Metchev}, Stanimir and {Ott}, Thomas and {Pearson}, David and {Puglisi}, Alfio and {Rabien}, Sebastian and {Rau}, Christian and {Riccardi}, Armando and {Salasnich}, Bernardo and {Shimizu}, Taro and {Snik}, Frans and {Sturm}, Eckhard and {Su{\'a}rez}, Genaro and {Tacconi}, Linda and {Tan}, Xianyu and {Taylor}, William and {Waring}, Christopher and {Xompero}, Marco},
	doi = {10.3847/2041-8213/ae80a0},
	eid = {L10},
	eprint = {2606.23801},
	journal = {\apjl},
	month = jul,
	number = {1},
	pages = {L10},
	primaryclass = {astro-ph.EP},
	title = {{Direct Imaging Discovery of Giant Exoplanet {\ensuremath{\beta}} Pictoris d: A Decade-long Game of Hide-and-seek}},
	volume = {1006},
	year = 2026}

@article{Bendahan-West2026,
	adsurl = {https://ui.adsabs.harvard.edu/abs/2026MNRAS.546f2255B},
	author = {{Bendahan-West}, R. and {Marino}, S. and {Carter}, A.~L. and {Squicciarini}, V. and {James}, A.~D. and {Sefilian}, A.~A. and {Pearce}, T.~D. and {Friebe}, M.~F. and {Lazzoni}, C. and {Lakeland}, B. and {Ray}, S. and {Wyatt}, M.~C. and {Matr{\`a}}, L. and {Milli}, J. and {Faramaz}, V.~C. and {Henning}, Th and {Hinkley}, S. and {Kennedy}, G.~M. and {Mesa}, D. and {Zurlo}, A.},
	doi = {10.1093/mnras/staf2255},
	eid = {staf2255},
	journal = {\mnras},
	month = mar,
	number = {3},
	pages = {staf2255},
	title = {{JWST/MIRI coronagraphic search for planets in systems with gapped exoKuiper belts and proper-motion anomalies}},
	volume = {546},
	year = 2026}

@phdthesis{Bonse2025thesis,
	author = {{Bonse}, M.~J.},
	month = {December},
	school = {ETH Zurich},
	title = {Decoding Speckles, Discovering Treasures Statistical Methods and Machine Learning for Exoplanet High-contrast Imaging},
	year = {2025}}

@article{Bonse2025,
	adsurl = {https://ui.adsabs.harvard.edu/abs/2025AJ....169..194B},
	archiveprefix = {arXiv},
	author = {{Bonse}, Markus J. and {Gebhard}, Timothy D. and {Dannert}, Felix A. and {Absil}, Olivier and {Cantalloube}, Faustine and {Christiaens}, Valentin and {Cugno}, Gabriele and {Garvin}, Emily O. and {Hayoz}, Jean and {Kasper}, Markus and {Matthews}, Elisabeth and {Sch{\"o}lkopf}, Bernhard and {Quanz}, Sascha P.},
	doi = {10.3847/1538-3881/adab79},
	eid = {194},
	eprint = {2406.01809},
	journal = {\aj},
	month = apr,
	number = {4},
	pages = {194},
	primaryclass = {astro-ph.EP},
	title = {{Use the 4S (Signal-Safe Speckle Subtraction): Explainable Machine Learning Reveals the Giant Exoplanet AF Lep b in High-contrast Imaging Data from 2011}},
	volume = {169},
	year = 2025}

@article{Squicciarini2025a,
	adsurl = {https://ui.adsabs.harvard.edu/abs/2025A&A...693A..54S},
	archiveprefix = {arXiv},
	author = {{Squicciarini}, V. and {Mazoyer}, J. and {Lagrange}, A. -M. and {Chomez}, A. and {Delorme}, P. and {Flasseur}, O. and {Kiefer}, F. and {Bergeon}, S. and {Albert}, D. and {Meunier}, N.},
	doi = {10.1051/0004-6361/202452310},
	eid = {A54},
	eprint = {2411.06157},
	journal = {\aap},
	month = jan,
	pages = {A54},
	primaryclass = {astro-ph.EP},
	title = {{The COBREX archival survey: Improved constraints on the occurrence rate of wide-orbit substellar companions: I. A uniform re-analysis of 400 stars from the GPIES survey}},
	volume = {693},
	year = 2025}

@misc{Dressler1996,
	adsurl = {https://ui.adsabs.harvard.edu/abs/1996hstb.rept.....D},
	author = {{Dressler}, Alan and {''HST and Beyond'' Committee}},
	howpublished = {Association of Universities for Research in Astronomy, 1996, xiv, p. 89, Washington, D.C.},
	month = jan,
	pages = {89},
	title = {{HST and beyond.}},
	year = 1996}

@article{Crotts2025,
	adsurl = {https://ui.adsabs.harvard.edu/abs/2025ApJ...987L..41C},
	archiveprefix = {arXiv},
	author = {{Crotts}, Katie A. and {Carter}, Aarynn L. and {Lawson}, Kellen and {Mang}, James and {Biller}, Beth and {Booth}, Mark and {Ferrer-Chavez}, Rodrigo and {Girard}, Julien H. and {Lagrange}, Anne-Marie and {Liu}, Michael C. and {Marino}, Sebastian and {Millar-Blanchaer}, Maxwell A. and {Skemer}, Andy and {Strampelli}, Giovanni M. and {Wang}, Jason and {Absil}, Olivier and {Balmer}, William O. and {Bendahan-West}, Rapha{\"e}l and {Bogat}, Ellis and {Bowens-Rubin}, Rachel and {Chauvin}, Ga{\"e}l and {Fontanive}, Cl{\'e}mence and {Franson}, Kyle and {Kammerer}, Jens and {Leisenring}, Jarron and {Morley}, Caroline V. and {Rebollido}, Isabel and {Skaf}, Nour and {Sutlieff}, Ben J. and {Bruinsma}, Evelyn L. and {Hinkley}, Sasha and {Hoch}, Kielan and {James}, Andrew D. and {Kane}, Rohan and {Mawet}, Dimitri and {Meyer}, Michael R. and {Palatnick}, Skyler and {Perrin}, Marshall D. and {Ray}, Shrishmoy and {Rickman}, Emily and {Sanghi}, Aniket and {Stephenson}, Klaus Subbotina},
	doi = {10.3847/2041-8213/ade798},
	eid = {L41},
	eprint = {2506.19932},
	journal = {\apjl},
	month = jul,
	number = {2},
	pages = {L41},
	primaryclass = {astro-ph.EP},
	title = {{Follow-up Exploration of the TWA 7 Planet{\textendash}Disk System with JWST NIRCam}},
	volume = {987},
	year = 2025
    }

@article{Kratter2016,
	adsurl = {https://ui.adsabs.harvard.edu/abs/2016ARA&A..54..271K},
	archiveprefix = {arXiv},
	author = {{Kratter}, Kaitlin and {Lodato}, Giuseppe},
	doi = {10.1146/annurev-astro-081915-023307},
	eprint = {1603.01280},
	journal = {\araa},
	month = sep,
	pages = {271-311},
	primaryclass = {astro-ph.SR},
	title = {{Gravitational Instabilities in Circumstellar Disks}},
	volume = {54},
	year = 2016}

@article{Pollack1996,
	adsurl = {http://adsabs.harvard.edu/abs/1996Icar..124...62P},
	author = {{Pollack}, J.~B. and {Hubickyj}, O. and {Bodenheimer}, P. and {Lissauer}, J.~J. and {Podolak}, M. and {Greenzweig}, Y.},
	journal = {Icarus},
	month = nov,
	pages = {62-85},
	title = {{Formation of the Giant Planets by Concurrent Accretion of Solids and Gas}},
	volume = 124,
	year = 1996}

@article{Dattilo2023,
	adsurl = {https://ui.adsabs.harvard.edu/abs/2023AJ....166..122D},
	archiveprefix = {arXiv},
	author = {{Dattilo}, Anne and {Batalha}, Natalie M. and {Bryson}, Steve},
	doi = {10.3847/1538-3881/acebc8},
	eid = {122},
	eprint = {2308.00103},
	journal = {\aj},
	month = sep,
	number = {3},
	pages = {122},
	primaryclass = {astro-ph.EP},
	title = {{A Unified Treatment of Kepler Occurrence to Trace Planet Evolution. I. Methodology}},
	volume = {166},
	year = 2023
    }

@article{McElwain2007,
	adsurl = {http://adsabs.harvard.edu/abs/2007ApJ...656..505M},
	author = {{McElwain}, M.~W. and {Metchev}, S.~A. and {Larkin}, J.~E. and {Barczys}, M. and {Iserlohe}, C. and {Krabbe}, A. and {Quirrenbach}, A. and {Weiss}, J. and {Wright}, S.~A.},
	doi = {10.1086/510063},
	eprint = {arXiv:astro-ph/0610265},
	journal = {\apj},
	month = feb,
	pages = {505-514},
	title = {{First High-Contrast Science with an Integral Field Spectrograph: The Substellar Companion to GQ Lupi}},
	volume = 656,
	year = 2007
    }

@article{Neuhauser2005,
	adsurl = {http://adsabs.harvard.edu/abs/2005A%26A...435L..13N},
	author = {{Neuh{\"a}user}, R. and {Guenther}, E.~W. and {Wuchterl}, G. and {Mugrauer}, M. and {Bedalov}, A. and {Hauschildt}, P.~H.},
	eprint = {arXiv:astro-ph/0503691},
	journal = {\aap},
	month = may,
	pages = {L13-L16},
	title = {{Evidence for a co-moving sub-stellar companion of GQ Lup}},
	volume = 435,
	year = 2005
    }

@article{Bowler2017,
	adsurl = {https://ui.adsabs.harvard.edu/abs/2017AJ....153...18B},
	archiveprefix = {arXiv},
	author = {{Bowler}, Brendan P. and {Liu}, Michael C. and {Mawet}, Dimitri and {Ngo}, Henry and {Malo}, Lison and {Mace}, Gregory N. and {McLane}, Jacob N. and {Lu}, Jessica R. and {Tristan}, Isaiah I. and {Hinkley}, Sasha and {Hillenbrand}, Lynne A. and {Shkolnik}, Evgenya L. and {Benneke}, Bj{\"o}rn and {Best}, William M.~J.},
	doi = {10.3847/1538-3881/153/1/18},
	eid = {18},
	eprint = {1611.00364},
	journal = {\aj},
	month = jan,
	number = {1},
	pages = {18},
	primaryclass = {astro-ph.EP},
	title = {{Planets around Low-mass Stars (PALMS). VI. Discovery of a Remarkably Red Planetary-mass Companion to the AB Dor Moving Group Candidate 2MASS J22362452+4751425*}},
	volume = {153},
	year = 2017
    }

@article{Crepp2015,
	adsurl = {https://ui.adsabs.harvard.edu/abs/2015ApJ...798L..43C},
	archiveprefix = {arXiv},
	author = {{Crepp}, Justin R. and {Rice}, Emily L. and {Veicht}, Aaron and {Aguilar}, Jonathan and {Pueyo}, Laurent and {Giorla}, Paige and {Nilsson}, Ricky and {Luszcz-Cook}, Statia H. and {Oppenheimer}, Rebecca and {Hinkley}, Sasha and {Brenner}, Douglas and {Vasisht}, Gautam and {Cady}, Eric and {Beichman}, Charles A. and {Hillenbrand}, Lynne A. and {Lockhart}, Thomas and {Matthews}, Christopher T. and {Roberts}, Jr., Lewis C. and {Sivaramakrishnan}, Anand and {Soummer}, Remi and {Zhai}, Chengxing},
	doi = {10.1088/2041-8205/798/2/L43},
	eid = {L43},
	eprint = {1412.6101},
	journal = {\apjl},
	month = jan,
	number = {2},
	pages = {L43},
	primaryclass = {astro-ph.SR},
	title = {{Direct Spectrum of the Benchmark T Dwarf HD 19467 B}},
	volume = {798},
	year = 2015
    }

@article{Crepp2014,
	adsurl = {https://ui.adsabs.harvard.edu/abs/2014ApJ...781...29C},
	archiveprefix = {arXiv},
	author = {{Crepp}, Justin R. and {Johnson}, John Asher and {Howard}, Andrew W. and {Marcy}, Geoffrey W. and {Brewer}, John and {Fischer}, Debra A. and {Wright}, Jason T. and {Isaacson}, Howard},
	doi = {10.1088/0004-637X/781/1/29},
	eid = {29},
	eprint = {1311.0280},
	journal = {\apj},
	month = jan,
	number = {1},
	pages = {29},
	primaryclass = {astro-ph.EP},
	title = {{The TRENDS High-contrast Imaging Survey. V. Discovery of an Old and Cold Benchmark T-dwarf Orbiting the Nearby G-star HD 19467}},
	volume = {781},
	year = 2014
    }

@article{Bonnefoy2014,
	adsurl = {https://ui.adsabs.harvard.edu/abs/2014A&A...562A.127B},
	archiveprefix = {arXiv},
	author = {{Bonnefoy}, M. and {Chauvin}, G. and {Lagrange}, A.-M. and {Rojo}, P. and {Allard}, F. and {Pinte}, C. and {Dumas}, C. and {Homeier}, D.},
	doi = {10.1051/0004-6361/201118270},
	eid = {A127},
	eprint = {1306.3709},
	journal = {\aap},
	month = feb,
	pages = {A127},
	primaryclass = {astro-ph.SR},
	title = {{A library of near-infrared integral field spectra of young M-L dwarfs}},
	volume = {562},
	year = 2014
    }

@article{Bowler2018,
	adsurl = {https://ui.adsabs.harvard.edu/abs/2018AJ....155..159B},
	archiveprefix = {arXiv},
	author = {{Bowler}, Brendan P. and {Dupuy}, Trent J. and {Endl}, Michael and {Cochran}, William D. and {MacQueen}, Phillip J. and {Fulton}, Benjamin J. and {Petigura}, Erik A. and {Howard}, Andrew W. and {Hirsch}, Lea and {Kratter}, Kaitlin M. and {Crepp}, Justin R. and {Biller}, Beth A. and {Johnson}, Marshall C. and {Wittenmyer}, Robert A.},
	doi = {10.3847/1538-3881/aab2a6},
	eid = {159},
	eprint = {1802.10126},
	journal = {\aj},
	month = apr,
	number = {4},
	pages = {159},
	primaryclass = {astro-ph.EP},
	title = {{Orbit and Dynamical Mass of the Late-T Dwarf GL 758 B}},
	volume = {155},
	year = 2018
    }

@article{Thalmann2009,
	adsurl = {http://adsabs.harvard.edu/abs/2009ApJ...707L.123T},
	archiveprefix = {arXiv},
	author = {{Thalmann}, C. and {Carson}, J. and {Janson}, M. and {Goto}, M. and {McElwain}, M. and {Egner}, S. and {Feldt}, M. and {Hashimoto}, J. and {Hayano}, Y. and {Henning}, T. and {Hodapp}, K.~W. and {Kandori}, R. and {Klahr}, H. and {Kudo}, T. and {Kusakabe}, N. and {Mordasini}, C. and {Morino}, {J.-I.} and {Suto}, H. and {Suzuki}, R. and {Tamura}, M.},
	doi = {10.1088/0004-637X/707/2/L123},
	eprint = {0911.1127},
	journal = {\apjl},
	month = dec,
	pages = {L123-L127},
	title = {{Discovery of the Coldest Imaged Companion of a Sun-like Star}},
	volume = 707,
	year = 2009
    }

@article{Naud2014,
	adsurl = {https://ui.adsabs.harvard.edu/abs/2014ApJ...787....5N},
	archiveprefix = {arXiv},
	author = {{Naud}, Marie-Eve and {Artigau}, {\'E}tienne and {Malo}, Lison and {Albert}, Lo{\"\i}c and {Doyon}, Ren{\'e} and {Lafreni{\`e}re}, David and {Gagn{\'e}}, Jonathan and {Saumon}, Didier and {Morley}, Caroline V. and {Allard}, France and {Homeier}, Derek and {Beichman}, Charles A. and {Gelino}, Christopher R. and {Boucher}, Anne},
	doi = {10.1088/0004-637X/787/1/5},
	eid = {5},
	eprint = {1405.2932},
	journal = {\apj},
	month = may,
	number = {1},
	pages = {5},
	primaryclass = {astro-ph.EP},
	title = {{Discovery of a Wide Planetary-mass Companion to the Young M3 Star GU Psc}},
	volume = {787},
	year = 2014
    }

@article{Burningham2011,
	adsurl = {https://ui.adsabs.harvard.edu/abs/2011MNRAS.414.3590B},
	archiveprefix = {arXiv},
	author = {{Burningham}, Ben and {Leggett}, S.~K. and {Homeier}, D. and {Saumon}, D. and {Lucas}, P.~W. and {Pinfield}, D.~J. and {Tinney}, C.~G. and {Allard}, F. and {Marley}, M.~S. and {Jones}, H.~R.~A. and {Murray}, D.~N. and {Ishii}, M. and {Day-Jones}, A. and {Gomes}, J. and {Zhang}, Z.~H.},
	doi = {10.1111/j.1365-2966.2011.18664.x},
	eprint = {1103.1617},
	journal = {\mnras},
	month = jul,
	number = {4},
	pages = {3590-3598},
	primaryclass = {astro-ph.SR},
	title = {{The properties of the T8.5p dwarf Ross 458C}},
	volume = {414},
	year = 2011
    }

@article{Chauvin2005,
	adsurl = {http://adsabs.harvard.edu/abs/2005A%26A...438L..29C},
	author = {{Chauvin}, G. and {Lagrange}, A.-M. and {Zuckerman}, B. and {Dumas}, C. and {Mouillet}, D. and {Song}, I. and {Beuzit}, J.-L. and {Lowrance}, P. and {Bessell}, M.~S.},
	eprint = {arXiv:astro-ph/0504658},
	journal = {\aap},
	month = aug,
	pages = {L29-L32},
	title = {{A companion to AB Pic at the planet/brown dwarf boundary}},
	volume = 438,
	year = 2005
    }

@article{Kuzuhara2011,
	adsurl = {https://ui.adsabs.harvard.edu/abs/2011AJ....141..119K},
	author = {{Kuzuhara}, M. and {Tamura}, M. and {Ishii}, M. and {Kudo}, T. and {Nishiyama}, S. and {Kandori}, R.},
	doi = {10.1088/0004-6256/141/4/119},
	eid = {119},
	journal = {\aj},
	month = apr,
	number = {4},
	pages = {119},
	title = {{The Widest-separation Substellar Companion Candidate to a Binary T Tauri Star}},
	volume = {141},
	year = 2011
    }

@article{Ireland2011,
	adsurl = {http://adsabs.harvard.edu/abs/2011ApJ...726..113I},
	archiveprefix = {arXiv},
	author = {{Ireland}, M.~J. and {Kraus}, A. and {Martinache}, F. and {Law}, N. and {Hillenbrand}, L.~A.},
	doi = {10.1088/0004-637X/726/2/113},
	eprint = {1011.2201},
	journal = {\apj},
	month = jan,
	pages = {113-+},
	primaryclass = {astro-ph.SR},
	title = {{Two Wide Planetary-mass Companions to Solar-type Stars in Upper Scorpius}},
	volume = 726,
	year = 2011
    }

@article{Kraus2008,
	adsurl = {http://adsabs.harvard.edu/abs/2008ApJ...679..762K},
	archiveprefix = {arXiv},
	author = {{Kraus}, A.~L. and {Ireland}, M.~J. and {Martinache}, F. and {Lloyd}, J.~P.},
	doi = {10.1086/587435},
	eprint = {0801.2387},
	journal = {\apj},
	month = may,
	pages = {762-782},
	title = {{Mapping the Shores of the Brown Dwarf Desert. I. Upper Scorpius}},
	volume = 679,
	year = 2008
    }

@article{Schmidt2008,
	adsurl = {https://ui.adsabs.harvard.edu/abs/2008A&A...491..311S},
	archiveprefix = {arXiv},
	author = {{Schmidt}, T.~O.~B. and {Neuh{\"a}user}, R. and {Seifahrt}, A. and {Vogt}, N. and {Bedalov}, A. and {Helling}, Ch. and {Witte}, S. and {Hauschildt}, P.~H.},
	doi = {10.1051/0004-6361:20078840},
	eprint = {0809.2812},
	journal = {\aap},
	month = nov,
	number = {1},
	pages = {311-320},
	primaryclass = {astro-ph},
	title = {{Direct evidence of a sub-stellar companion around CT Chamaeleontis}},
	volume = {491},
	year = 2008
    }

@article{Lafreniere2008,
	adsurl = {http://adsabs.harvard.edu/abs/2008ApJ...689L.153L},
	archiveprefix = {arXiv},
	author = {{Lafreni{\`e}re}, D. and {Jayawardhana}, R. and {van Kerkwijk}, M.~H.},
	doi = {10.1086/595870},
	eprint = {0809.1424},
	journal = {\apjl},
	month = dec,
	pages = {L153-L156},
	title = {{Direct Imaging and Spectroscopy of a Planetary-Mass Candidate Companion to a Young Solar Analog}},
	volume = 689,
	year = 2008
    }

@article{Luhman2009,
	adsurl = {https://ui.adsabs.harvard.edu/abs/2009ApJ...691.1265L},
	archiveprefix = {arXiv},
	author = {{Luhman}, K.~L. and {Mamajek}, E.~E. and {Allen}, P.~R. and {Muench}, A.~A. and {Finkbeiner}, D.~P.},
	doi = {10.1088/0004-637X/691/2/1265},
	eprint = {0902.0425},
	journal = {\apj},
	month = feb,
	number = {2},
	pages = {1265-1275},
	primaryclass = {astro-ph.GA},
	title = {{Discovery of a Wide Binary Brown Dwarf Born in Isolation}},
	volume = {691},
	year = 2009
    }

@article{Martinez2019,
	adsurl = {https://ui.adsabs.harvard.edu/abs/2019AJ....158..134M},
	archiveprefix = {arXiv},
	author = {{Martinez}, Raquel A. and {Kraus}, Adam L.},
	doi = {10.3847/1538-3881/ab32e6},
	eid = {134},
	eprint = {1907.06767},
	journal = {\aj},
	month = oct,
	number = {4},
	pages = {134},
	primaryclass = {astro-ph.SR},
	title = {{Searching for Wide Companions and Identifying Circum(sub)stellar Disks through PSF Fitting of Spitzer/IRAC Archival Images}},
	volume = {158},
	year = 2019
    }

@article{Milli2017,
	adsurl = {https://ui.adsabs.harvard.edu/abs/2017A&A...597L...2M},
	archiveprefix = {arXiv},
	author = {{Milli}, J. and {Hibon}, P. and {Christiaens}, V. and {Choquet}, {\'E}. and {Bonnefoy}, M. and {Kennedy}, G.~M. and {Wyatt}, M.~C. and {Absil}, O. and {G{\'o}mez Gonz{\'a}lez}, C.~A. and {del Burgo}, C.},
	doi = {10.1051/0004-6361/201629908},
	eid = {L2},
	eprint = {1612.00333},
	journal = {Astronomy \& Astrophysics},
	month = {Jan},
	pages = {L2},
	primaryclass = {astro-ph.EP},
	title = {{Discovery of a low-mass companion inside the debris ring surrounding the F5V star HD 206893}},
	volume = {597},
	year = {2017}
    }

@article{Godoy2025,
	adsurl = {https://ui.adsabs.harvard.edu/abs/2025A&A...702A...4G},
	archiveprefix = {arXiv},
	author = {{Godoy}, N. and {Choquet}, E. and {Serabyn}, E. and {M{\^a}lin}, M. and {Tremblin}, P. and {Danielski}, C. and {Lagage}, P.~O. and {Boccaletti}, A. and {Charnay}, B. and {Ressler}, M.~E.},
	doi = {10.1051/0004-6361/202554652},
	eid = {A4},
	eprint = {2509.03624},
	journal = {\aap},
	month = oct,
	pages = {A4},
	primaryclass = {astro-ph.EP},
	title = {{A JWST/MIRI view of {\ensuremath{\kappa}} Andromedae b: Refining its mass, age, and physical parameters}},
	volume = {702},
	year = 2025
    }

@article{Patapis2025,
	adsurl = {https://ui.adsabs.harvard.edu/abs/2025A&A...704A...5P},
	archiveprefix = {arXiv},
	author = {{Patapis}, P. and {Morales-Calder{\'o}n}, M. and {Arabhavi}, A.~M. and {K{\"u}hnle}, H. and {Gasman}, D. and {Cugno}, G. and {Molli{\`e}re}, P. and {Matthews}, E. and {M{\^a}lin}, M. and {Whiteford}, N. and {Lagage}, P.-O. and {Waters}, R. and {Guedel}, M. and {Henning}, Th. and {Vandenbussche}, B. and {Absil}, O. and {Argyriou}, I. and {Barrado}, D. and {Baudoz}, P. and {Boccaletti}, A. and {Bouwman}, J. and {Cossou}, C. and {Coulais}, A. and {Decin}, L. and {Gastaud}, R. and {Glasse}, A. and {Glauser}, A.~M. and {Grant}, S. and {Min}, M. and {Kamp}, I. and {Olofsson}, G. and {Pye}, J. and {Rouan}, D. and {Royer}, P. and {Scheithauer}, S. and {Sun}, X. and {Tremblin}, P. and {Colina}, L. and {Ray}, T.~P. and {{\"O}stlin}, G. and {van Dishoeck}, E.~F. and {Wright}, G.},
	doi = {10.1051/0004-6361/202556296},
	eid = {A5},
	eprint = {2507.08961},
	journal = {\aap},
	month = nov,
	pages = {A5},
	primaryclass = {astro-ph.EP},
	title = {{JWST/MIRI observations of the young TWA 27 system: Hydrocarbon disk chemistry, silicate clouds, and evidence of a circumplanetary disk}},
	volume = {704},
	year = 2025
    }

@article{Carson2013,
	adsurl = {http://adsabs.harvard.edu/abs/2013ApJ...763L..32C},
	archiveprefix = {arXiv},
	author = {{Carson}, J. and {Thalmann}, C. and {Janson}, M. and {Kozakis}, T. and {Bonnefoy}, M. and {Biller}, B. and {Schlieder}, J. and {Currie}, T. and {McElwain}, M. and {Goto}, M. and {Henning}, T. and {Brandner}, W. and {Feldt}, M. and {Kandori}, R. and {Kuzuhara}, M. and {Stevens}, L. and {Wong}, P. and {Gainey}, K. and {Fukagawa}, M. and {Kuwada}, Y. and {Brandt}, T. and {Kwon}, J. and {Abe}, L. and {Egner}, S. and {Grady}, C. and {Guyon}, O. and {Hashimoto}, J. and {Hayano}, Y. and {Hayashi}, M. and {Hayashi}, S. and {Hodapp}, K. and {Ishii}, M. and {Iye}, M. and {Knapp}, G. and {Kudo}, T. and {Kusakabe}, N. and {Matsuo}, T. and {Miyama}, S. and {Morino}, J. and {Moro-Martin}, A. and {Nishimura}, T. and {Pyo}, T. and {Serabyn}, E. and {Suto}, H. and {Suzuki}, R. and {Takami}, M. and {Takato}, N. and {Terada}, H. and {Tomono}, D. and {Turner}, E. and {Watanabe}, M. and {Wisniewski}, J. and {Yamada}, T. and {Takami}, H. and {Usuda}, T. and {Tamura}, M.},
	doi = {10.1088/2041-8205/763/2/L32},
	eid = {L32},
	eprint = {1211.3744},
	journal = {\apjl},
	month = feb,
	pages = {L32},
	primaryclass = {astro-ph.SR},
	title = {{Direct Imaging Discovery of a ''Super-Jupiter'' around the Late B-type Star {$\kappa$} And}},
	volume = 763,
	year = 2013
    }

@article{Luhman2023,
	adsurl = {https://ui.adsabs.harvard.edu/abs/2023ApJ...949L..36L},
	archiveprefix = {arXiv},
	author = {{Luhman}, K.~L. and {Tremblin}, P. and {Birkmann}, S.~M. and {Manjavacas}, E. and {Valenti}, J. and {Alves de Oliveira}, C. and {Beck}, T.~L. and {Giardino}, G. and {L{\"u}tzgendorf}, N. and {Rauscher}, B.~J. and {Sirianni}, M.},
	doi = {10.3847/2041-8213/acd635},
	eid = {L36},
	eprint = {2305.18603},
	journal = {\apjl},
	month = jun,
	number = {2},
	pages = {L36},
	primaryclass = {astro-ph.EP},
	title = {{JWST/NIRSpec Observations of the Planetary Mass Companion TWA 27B}},
	volume = {949},
	year = 2023
    }

@article{Chauvin2004,
	adsurl = {http://adsabs.harvard.edu/abs/2004A%26A...425L..29C},
	author = {{Chauvin}, G. and {Lagrange}, {A.-M.} and {Dumas}, C. and {Zuckerman}, B. and {Mouillet}, D. and {Song}, I. and {Beuzit}, {J.-L.} and {Lowrance}, P.},
	doi = {10.1051/0004-6361:200400056},
	eprint = {arXiv:astro-ph/0409323},
	journal = {\aap},
	month = oct,
	pages = {L29-L32},
	title = {{A giant planet candidate near a young brown dwarf. Direct VLT/NACO observations using IR wavefront sensing}},
	volume = 425,
	year = 2004}

@article{Zhang2021Coconuts,
	adsurl = {https://ui.adsabs.harvard.edu/abs/2021ApJ...916L..11Z},
	archiveprefix = {arXiv},
	author = {{Zhang}, Zhoujian and {Liu}, Michael C. and {Claytor}, Zachary R. and {Best}, William M.~J. and {Dupuy}, Trent J. and {Siverd}, Robert J.},
	doi = {10.3847/2041-8213/ac1123},
	eid = {L11},
	eprint = {2107.02805},
	journal = {\apjl},
	month = aug,
	number = {2},
	pages = {L11},
	primaryclass = {astro-ph.EP},
	title = {{The Second Discovery from the COCONUTS Program: A Cold Wide-orbit Exoplanet around a Young Field M Dwarf at 10.9 pc}},
	volume = {916},
	year = 2021
    }

@article{Delorme2013,
	adsurl = {https://ui.adsabs.harvard.edu/abs/2013A&A...553L...5D},
	archiveprefix = {arXiv},
	author = {{Delorme}, P. and {Gagn{\'e}}, J. and {Girard}, J.~H. and {Lagrange}, A.~M. and {Chauvin}, G. and {Naud}, M. -E. and {Lafreni{\`e}re}, D. and {Doyon}, R. and {Riedel}, A. and {Bonnefoy}, M. and {Malo}, L.},
	doi = {10.1051/0004-6361/201321169},
	eid = {L5},
	eprint = {1303.4525},
	journal = {\aap},
	month = may,
	pages = {L5},
	primaryclass = {astro-ph.SR},
	title = {{Direct-imaging discovery of a 12-14 Jupiter-mass object orbiting a young binary system of very low-mass stars}},
	volume = {553},
	year = 2013
    }

@article{Bailey2014,
	adsurl = {http://adsabs.harvard.edu/abs/2014ApJ...780L...4B},
	archiveprefix = {arXiv},
	author = {{Bailey}, V. and {Meshkat}, T. and {Reiter}, M. and {Morzinski}, K. and {Males}, J. and {Su}, K.~Y.~L. and {Hinz}, P.~M. and {Kenworthy}, M. and {Stark}, D. and {Mamajek}, E. and {Briguglio}, R. and {Close}, L.~M. and {Follette}, K.~B. and {Puglisi}, A. and {Rodigas}, T. and {Weinberger}, A.~J. and {Xompero}, M.},
	doi = {10.1088/2041-8205/780/1/L4},
	eid = {L4},
	eprint = {1312.1265},
	journal = {\apjl},
	month = jan,
	pages = {L4},
	primaryclass = {astro-ph.EP},
	title = {{HD 106906 b: A Planetary-mass Companion Outside a Massive Debris Disk}},
	volume = 780,
	year = 2014
    }

@article{Kuzuhara2013,
	adsurl = {http://adsabs.harvard.edu/abs/2013ApJ...774...11K},
	archiveprefix = {arXiv},
	author = {{Kuzuhara}, M. and {Tamura}, M. and {Kudo}, T. and {Janson}, M. and {Kandori}, R. and {Brandt}, T.~D. and {Thalmann}, C. and {Spiegel}, D. and {Biller}, B. and {Carson}, J. and {Hori}, Y. and {Suzuki}, R. and {Burrows}, A. and {Henning}, T. and {Turner}, E.~L. and {McElwain}, M.~W. and {Moro-Mart{\'{\i}}n}, A. and {Suenaga}, T. and {Takahashi}, Y.~H. and {Kwon}, J. and {Lucas}, P. and {Abe}, L. and {Brandner}, W. and {Egner}, S. and {Feldt}, M. and {Fujiwara}, H. and {Goto}, M. and {Grady}, C.~A. and {Guyon}, O. and {Hashimoto}, J. and {Hayano}, Y. and {Hayashi}, M. and {Hayashi}, S.~S. and {Hodapp}, K.~W. and {Ishii}, M. and {Iye}, M. and {Knapp}, G.~R. and {Matsuo}, T. and {Mayama}, S. and {Miyama}, S. and {Morino}, J.-I. and {Nishikawa}, J. and {Nishimura}, T. and {Kotani}, T. and {Kusakabe}, N. and {Pyo}, T.-S. and {Serabyn}, E. and {Suto}, H. and {Takami}, M. and {Takato}, N. and {Terada}, H. and {Tomono}, D. and {Watanabe}, M. and {Wisniewski}, J.~P. and {Yamada}, T. and {Takami}, H. and {Usuda}, T.},
	doi = {10.1088/0004-637X/774/1/11},
	eid = {11},
	eprint = {1307.2886},
	journal = {\apj},
	month = sep,
	pages = {11},
	primaryclass = {astro-ph.EP},
	title = {{Direct Imaging of a Cold Jovian Exoplanet in Orbit around the Sun-like Star GJ 504}},
	volume = 774,
	year = 2013
    }

@article{rameau95086,
	adsurl = {http://adsabs.harvard.edu/abs/2013ApJ...772L..15R},
	archiveprefix = {arXiv},
	author = {{Rameau}, J. and {Chauvin}, G. and {Lagrange}, A.-M. and {Boccaletti}, A. and {Quanz}, S.~P. and {Bonnefoy}, M. and {Girard}, J.~H. and {Delorme}, P. and {Desidera}, S. and {Klahr}, H. and {Mordasini}, C. and {Dumas}, C. and {Bonavita}, M.},
	doi = {10.1088/2041-8205/772/2/L15},
	eid = {L15},
	eprint = {1305.7428},
	journal = {\apjl},
	month = aug,
	pages = {L15},
	primaryclass = {astro-ph.EP},
	title = {{Discovery of a Probable 4-5 Jupiter-mass Exoplanet to HD 95086 by Direct Imaging}},
	volume = 772,
	year = 2013
    }

@article{Bardalez-Gagliuffi2021,
	adsurl = {https://ui.adsabs.harvard.edu/abs/2021ApJ...922L..43B},
	archiveprefix = {arXiv},
	author = {{Bardalez Gagliuffi}, Daniella C. and {Faherty}, Jacqueline K. and {Li}, Yiting and {Brandt}, Timothy D. and {Williams}, Lauryn and {Brandt}, G. Mirek and {Gelino}, Christopher R.},
	doi = {10.3847/2041-8213/ac382c},
	eid = {L43},
	eprint = {2111.06004},
	journal = {\apjl},
	month = dec,
	number = {2},
	pages = {L43},
	primaryclass = {astro-ph.EP},
	title = {{14 Her: A Likely Case of Planet-Planet Scattering}},
	volume = {922},
	year = 2021
    }

@article{Bohn2020,
	adsurl = {https://ui.adsabs.harvard.edu/abs/2020ApJ...898L..16B},
	archiveprefix = {arXiv},
	author = {{Bohn}, Alexander J. and {Kenworthy}, Matthew A. and {Ginski}, Christian and {Rieder}, Steven and {Mamajek}, Eric E. and {Meshkat}, Tiffany and {Pecaut}, Mark J. and {Reggiani}, Maddalena and {de Boer}, Jozua and {Keller}, Christoph U. and {Snik}, Frans and {Southworth}, John},
	doi = {10.3847/2041-8213/aba27e},
	eid = {L16},
	eprint = {2007.10991},
	journal = {\apjl},
	month = jul,
	number = {1},
	pages = {L16},
	primaryclass = {astro-ph.EP},
	title = {{Two Directly Imaged, Wide-orbit Giant Planets around the Young, Solar Analog TYC 8998-760-1}},
	volume = {898},
	year = 2020
    }

@article{Godoy2024,
	adsurl = {https://ui.adsabs.harvard.edu/abs/2024A&A...689A.185G},
	archiveprefix = {arXiv},
	author = {{Godoy}, N. and {Choquet}, E. and {Serabyn}, E. and {Danielski}, C. and {Stolker}, T. and {Charnay}, B. and {Hinkley}, S. and {Lagage}, P.~O. and {Ressler}, M.~E. and {Tremblin}, P. and {Vigan}, A.},
	doi = {10.1051/0004-6361/202449951},
	eid = {A185},
	eprint = {2409.04524},
	journal = {\aap},
	month = sep,
	pages = {A185},
	primaryclass = {astro-ph.EP},
	title = {{A new atmospheric characterization of the sub-stellar companion HR 2562 B with JWST/MIRI observations}},
	volume = {689},
	year = 2024
    }

@article{Konopacky2016,
	adsurl = {https://ui.adsabs.harvard.edu/abs/2016ApJ...829L...4K},
	archiveprefix = {arXiv},
	author = {{Konopacky}, Quinn M. and {Rameau}, Julien and {Duch{\^e}ne}, Gaspard and {Filippazzo}, Joseph C. and {Giorla Godfrey}, Paige A. and {Marois}, Christian and {Nielsen}, Eric L. and {Pueyo}, Laurent and {Rafikov}, Roman R. and {Rice}, Emily L.},
	doi = {10.3847/2041-8205/829/1/L4},
	eid = {L4},
	eprint = {1608.06660},
	journal = {\apj},
	month = {Sep},
	number = {1},
	pages = {L4},
	primaryclass = {astro-ph.EP},
	title = {{Discovery of a Substellar Companion to the Nearby Debris Disk Host HR 2562}},
	volume = {829},
	year = {2016}
    }

@article{Hughes2018,
	adsurl = {https://ui.adsabs.harvard.edu/abs/2018ARA&A..56..541H},
	archiveprefix = {arXiv},
	author = {{Hughes}, A. Meredith and {Duch{\^e}ne}, Gaspard and {Matthews}, Brenda C.},
	doi = {10.1146/annurev-astro-081817-052035},
	eprint = {1802.04313},
	journal = {\araa},
	month = sep,
	pages = {541-591},
	primaryclass = {astro-ph.EP},
	title = {{Debris Disks: Structure, Composition, and Variability}},
	volume = {56},
	year = 2018
    }

@article{Wyatt2008,
	adsurl = {http://adsabs.harvard.edu/abs/2008ARA%26A..46..339W},
	author = {{Wyatt}, M.~C.},
	journal = {\araa},
	month = sep,
	pages = {339-383},
	title = {{Evolution of Debris Disks}},
	volume = 46,
	year = 2008
    }

@article{Malin2025,
	adsurl = {https://ui.adsabs.harvard.edu/abs/2025A&A...693A.315M},
	archiveprefix = {arXiv},
	author = {{M{\^a}lin}, Mathilde and {Boccaletti}, Anthony and {Perrot}, Cl{\'e}ment and {Baudoz}, Pierre and {Rouan}, Daniel and {Lagage}, Pierre-Olivier and {Waters}, Rens and {G{\"u}del}, Manuel and {Henning}, Thomas and {Vandenbussche}, Bart and {Absil}, Olivier and {Barrado}, David and {Charnay}, Benjamin and {Choquet}, Elodie and {Cossou}, Christophe and {Danielski}, Camilla and {Decin}, Leen and {Glauser}, Adrian M. and {Pye}, John and {Olofsson}, Goran and {Glasse}, Alistair and {Patapis}, Polychronis and {Royer}, Pierre and {Scheithauer}, Silvia and {Serabyn}, Eugene and {Tremblin}, Pascal and {Whiteford}, Niall and {van Dishoeck}, Ewine F. and {Ostlin}, G{\"o}ran and {Ray}, Tom P. and {Wright}, Gillian},
	doi = {10.1051/0004-6361/202452695},
	eid = {A315},
	eprint = {2501.00104},
	journal = {\aap},
	month = jan,
	pages = {A315},
	primaryclass = {astro-ph.EP},
	title = {{First unambiguous detection of ammonia in the atmosphere of a planetary mass companion with JWST/MIRI coronagraphs}},
	volume = {693},
	year = 2025
    }

@article{Malin2026,
	adsurl = {https://ui.adsabs.harvard.edu/abs/2026arXiv260313543M},
	archiveprefix = {arXiv},
	author = {{M{\^a}lin}, Mathilde and {Boccaletti}, Anthony and {Charnay}, Benjamin and {Pueyo}, Laurent and {Bidot}, Alexis and {Patapis}, Polychronis and {Hinkley}, Sasha and {Petrus}, Simon and {Whiteford}, Niall and {Perrin}, Marshall and {Biller}, Beth A. and {Cugno}, Gabriele and {Currie}, Thayne and {Danielski}, Camilla and {Henning}, Thomas and {Hoch}, Kielan K.~W. and {Janson}, Markus and {Kammerer}, Jens and {Matthews}, Elisabeth C and {Nasedkin}, Evert and {Palma-Bifani}, Paulina and {Rebollido}, Isabel and {Samland}, Matthias and {Skemer}, Andrew and {Stone}, Jordan M. and {Su{\'a}rez}, Genaro and {Sutlieff}, Ben J. and {Tamura}, Motohide and {Theissen}, Christopher A. and {Vos}, Johanna M. and {Zhang}, Zhoujian and {Zurlo}, Alice},
	doi = {10.48550/arXiv.2603.13543},
	eid = {arXiv:2603.13543},
	eprint = {2603.13543},
	journal = {arXiv e-prints},
	month = mar,
	pages = {arXiv:2603.13543},
	primaryclass = {astro-ph.EP},
	title = {{The JWST Early Release Science Program for Direct Observations of Exoplanetary Systems. VIII. Molecular Mapping Performance with JWST/MIRI MRS: VHS 1256 b as a case study}},
	year = 2026
    }

@article{Mang2026,
	adsurl = {https://ui.adsabs.harvard.edu/abs/2026ApJ..1000...98M},
	archiveprefix = {arXiv},
	author = {{Mang}, James and {Batalha}, Natasha E. and {Morley}, Caroline V. and {Wogan}, Nicholas F. and {Mukherjee}, Sagnick and {Visscher}, Channon and {Marley}, Mark S. and {Fortney}, Jonathan J. and {Chubb}, Katy L. and {Gao}, Peter and {Malsky}, Isaac},
	doi = {10.3847/1538-4357/ae47ff},
	eid = {98},
	eprint = {2602.22468},
	journal = {\apj},
	month = mar,
	number = {1},
	pages = {98},
	primaryclass = {astro-ph.EP},
	title = {{PICASO 4.0: Clouds and Photochemistry in Climate Models of Brown Dwarfs and Exoplanets}},
	volume = {1000},
	year = 2026
    }

@article{Matthews2025,
	adsurl = {https://ui.adsabs.harvard.edu/abs/2025ApJ...981L..31M},
	archiveprefix = {arXiv},
	author = {{Matthews}, Elisabeth C. and {Molli{\`e}re}, Paul and {K{\"u}hnle}, Helena and {Patapis}, Polychronis and {Whiteford}, Niall and {Samland}, Matthias and {Lagage}, Pierre-Olivier and {Waters}, Rens and {Tsai}, Shang-Min and {Zahnle}, Kevin and {Guedel}, Manuel and {Henning}, Thomas and {Vandenbussche}, Bart and {Absil}, Olivier and {Argyriou}, Ioannis and {Barrado}, David and {Coulais}, Alain and {Glauser}, Adrian M. and {Olofsson}, Goran and {Pye}, John P. and {Rouan}, Daniel and {Royer}, Pierre and {van Dishoeck}, Ewine F. and {Ray}, T.~P. and {{\"O}stlin}, G{\"o}ran},
	doi = {10.3847/2041-8213/adb4ec},
	eid = {L31},
	eprint = {2502.13610},
	journal = {\apjl},
	month = mar,
	number = {2},
	pages = {L31},
	primaryclass = {astro-ph.EP},
	title = {{HCN and C$_{2}$H$_{2}$ in the Atmosphere of a T8.5+T9 Brown Dwarf Binary}},
	volume = {981},
	year = 2025
    }

@incollection{Espinoza2026,
	adsurl = {https://ui.adsabs.harvard.edu/abs/2026haex.book..216E},
	author = {{Espinoza}, N{\'e}stor and {Perrin}, Marshall D.},
	booktitle = {Handbook of Exoplanets},
	doi = {10.1007/978-3-319-30648-3_216-1},
	eid = {216},
	pages = {216},
	title = {{Highlights from Exoplanet Observations by the James Webb Space Telescope}},
	year = 2026}

@article{Sorahana2012,
	adsurl = {https://ui.adsabs.harvard.edu/abs/2012ApJ...760..151S},
	archiveprefix = {arXiv},
	author = {{Sorahana}, S. and {Yamamura}, I.},
	doi = {10.1088/0004-637X/760/2/151},
	eid = {151},
	eprint = {1210.3828},
	journal = {\apj},
	month = {Dec},
	number = {2},
	pages = {151},
	primaryclass = {astro-ph.SR},
	title = {{AKARI Observations of Brown Dwarfs. III. CO, CO$_{2}$, and CH$_{4}$ Fundamental Bands and Physical Parameters}},
	volume = {760},
	year = {2012}}

@article{Cronin-Coltsmann2021,
	adsurl = {https://ui.adsabs.harvard.edu/abs/2021MNRAS.504.4497C},
	archiveprefix = {arXiv},
	author = {{Cronin-Coltsmann}, Patrick F. and {Kennedy}, Grant M. and {Kalas}, Paul and {Milli}, Julien and {Clarke}, Cathie J. and {Duch{\^e}ne}, Gaspard and {Greaves}, Jane and {Lawler}, Samantha M. and {Lestrade}, Jean-Fran{\c{c}}ois and {Matthews}, Brenda C. and {Shannon}, Andrew and {Wyatt}, Mark C.},
	doi = {10.1093/mnras/stab1237},
	eprint = {2104.13396},
	journal = {\mnras},
	month = jul,
	number = {3},
	pages = {4497-4510},
	primaryclass = {astro-ph.EP},
	title = {{ALMA imaging of the M-dwarf Fomalhaut C's debris disc}},
	volume = {504},
	year = 2021
    }

@article{Lawson2024,
	adsurl = {https://ui.adsabs.harvard.edu/abs/2024ApJ...967L...8L},
	archiveprefix = {arXiv},
	author = {{Lawson}, Kellen and {Schlieder}, Joshua E. and {Leisenring}, Jarron M. and {Bogat}, Ell and {Beichman}, Charles A. and {Bryden}, Geoffrey and {G{\'a}sp{\'a}r}, Andr{\'a}s and {Groff}, Tyler D. and {McElwain}, Michael W. and {Meyer}, Michael R. and {Barclay}, Thomas and {Calissendorff}, Per and {De Furio}, Matthew and {Li}, Yiting and {Rieke}, Marcia J. and {Ygouf}, Marie and {Greene}, Thomas P. and {Girard}, Julien H. and {Gennaro}, Mario and {Kammerer}, Jens and {Rest}, Armin and {Roellig}, Thomas L. and {Sunnquist}, Ben},
	doi = {10.3847/2041-8213/ad4496},
	eid = {L8},
	eprint = {2405.00573},
	journal = {\apjl},
	month = may,
	number = {1},
	pages = {L8},
	primaryclass = {astro-ph.EP},
	title = {{JWST/NIRCam Detection of the Fomalhaut C Debris Disk in Scattered Light}},
	volume = {967},
	year = 2024
    }

@article{Lawson2023,
	adsurl = {https://ui.adsabs.harvard.edu/abs/2023AJ....166..150L},
	archiveprefix = {arXiv},
	author = {{Lawson}, Kellen and {Schlieder}, Joshua E. and {Leisenring}, Jarron M. and {Bogat}, Ell and {Beichman}, Charles A. and {Bryden}, Geoffrey and {G{\'a}sp{\'a}r}, Andr{\'a}s and {Groff}, Tyler D. and {McElwain}, Michael W. and {Meyer}, Michael R. and {Barclay}, Thomas and {Calissendorff}, Per and {De Furio}, Matthew and {Ygouf}, Marie and {Boccaletti}, Anthony and {Greene}, Thomas P. and {Krist}, John and {Plavchan}, Peter and {Rieke}, Marcia J. and {Roellig}, Thomas L. and {Stansberry}, John and {Wisniewski}, John P. and {Young}, Erick T.},
	doi = {10.3847/1538-3881/aced08},
	eid = {150},
	eprint = {2308.02486},
	journal = {\aj},
	month = oct,
	number = {4},
	pages = {150},
	primaryclass = {astro-ph.EP},
	title = {{JWST/NIRCam Coronagraphy of the Young Planet-hosting Debris Disk AU Microscopii}},
	volume = {166},
	year = 2023
    }

@article{Su2024,
	adsurl = {https://ui.adsabs.harvard.edu/abs/2024ApJ...977..277S},
	archiveprefix = {arXiv},
	author = {{Su}, Kate Y.~L. and {G{\'a}sp{\'a}r}, Andr{\'a}s and {Rieke}, George H. and {Malhotra}, Renu and {Matr{\'a}}, Luca and {Wolff}, Schuyler Grace and {Leisenring}, Jarron M. and {Beichman}, Charles and {Ygouf}, Marie},
	doi = {10.3847/1538-4357/ad8cde},
	eid = {277},
	eprint = {2410.23636},
	journal = {\apj},
	month = dec,
	number = {2},
	pages = {277},
	primaryclass = {astro-ph.EP},
	title = {{Imaging of the Vega Debris System Using JWST/MIRI}},
	volume = {977},
	year = 2024
    }

@article{Gaspar2023,
	adsurl = {https://ui.adsabs.harvard.edu/abs/2023NatAs...7..790G},
	archiveprefix = {arXiv},
	author = {{G{\'a}sp{\'a}r}, Andr{\'a}s and {Wolff}, Schuyler Grace and {Rieke}, George H. and {Leisenring}, Jarron M. and {Morrison}, Jane and {Su}, Kate Y.~L. and {Ward-Duong}, Kimberly and {Aguilar}, Jonathan and {Ygouf}, Marie and {Beichman}, Charles and {Llop-Sayson}, Jorge and {Bryden}, Geoffrey},
	doi = {10.1038/s41550-023-01962-6},
	eprint = {2305.03789},
	journal = {Nature Astronomy},
	month = jul,
	pages = {790-798},
	primaryclass = {astro-ph.EP},
	title = {{Spatially resolved imaging of the inner Fomalhaut disk using JWST/MIRI}},
	volume = {7},
	year = 2023
    }

@article{Rebollido2024,
	adsurl = {https://ui.adsabs.harvard.edu/abs/2024AJ....167...69R},
	archiveprefix = {arXiv},
	author = {{Rebollido}, Isabel and {Stark}, Christopher C. and {Kammerer}, Jens and {Perrin}, Marshall D. and {Lawson}, Kellen and {Pueyo}, Laurent and {Chen}, Christine and {Hines}, Dean and {Girard}, Julien H. and {Worthen}, Kadin and {Ingerbretsen}, Carl and {Betti}, Sarah and {Clampin}, Mark and {Golimowski}, David and {Hoch}, Kielan and {Lewis}, Nikole K. and {Lu}, Cicero X. and {van der Marel}, Roeland P. and {Rickman}, Emily and {Seager}, Sara and {Soummer}, R{\'e}mi and {Valenti}, Jeff A. and {Ward-Duong}, Kimberly and {Mountain}, C. Matt},
	doi = {10.3847/1538-3881/ad1759},
	eid = {69},
	eprint = {2401.05271},
	journal = {\aj},
	month = feb,
	number = {2},
	pages = {69},
	primaryclass = {astro-ph.EP},
	title = {{JWST-TST High Contrast: Asymmetries, Dust Populations, and Hints of a Collision in the {\ensuremath{\beta}} Pictoris Disk with NIRCam and MIRI}},
	volume = {167},
	year = 2024
    }

@article{Poleski2021,
	adsurl = {https://ui.adsabs.harvard.edu/abs/2021AcA....71....1P},
	archiveprefix = {arXiv},
	author = {{Poleski}, R. and {Skowron}, J. and {Mr{\'o}z}, P. and {Udalski}, A. and {Szyma{\'n}ski}, M.~K. and {Pietrukowicz}, P. and {Ulaczyk}, K. and {Rybicki}, K. and {Iwanek}, P. and {Wrona}, M. and {Gromadzki}, M.},
	doi = {10.32023/0001-5237/71.1.1},
	eprint = {2104.02079},
	journal = {Acta Astronomica},
	month = mar,
	number = {1},
	pages = {1-23},
	primaryclass = {astro-ph.EP},
	title = {{Wide-Orbit Exoplanets are Common. Analysis of Nearly 20 Years of OGLE Microlensing Survey Data}},
	volume = {71},
	year = 2021
    }

@article{Vigan2021,
	adsurl = {https://ui.adsabs.harvard.edu/abs/2021A&A...651A..72V},
	archiveprefix = {arXiv},
	author = {{Vigan}, A. and {Fontanive}, C. and {Meyer}, M. and {Biller}, B. and {Bonavita}, M. and {Feldt}, M. and {Desidera}, S. and {Marleau}, G. -D. and {Emsenhuber}, A. and {Galicher}, R. and {Rice}, K. and {Forgan}, D. and {Mordasini}, C. and {Gratton}, R. and {Le Coroller}, H. and {Maire}, A. -L. and {Cantalloube}, F. and {Chauvin}, G. and {Cheetham}, A. and {Hagelberg}, J. and {Lagrange}, A. -M. and {Langlois}, M. and {Bonnefoy}, M. and {Beuzit}, J. -L. and {Boccaletti}, A. and {D'Orazi}, V. and {Delorme}, P. and {Dominik}, C. and {Henning}, Th. and {Janson}, M. and {Lagadec}, E. and {Lazzoni}, C. and {Ligi}, R. and {Menard}, F. and {Mesa}, D. and {Messina}, S. and {Moutou}, C. and {M{\"u}ller}, A. and {Perrot}, C. and {Samland}, M. and {Schmid}, H.~M. and {Schmidt}, T. and {Sissa}, E. and {Turatto}, M. and {Udry}, S. and {Zurlo}, A. and {Abe}, L. and {Antichi}, J. and {Asensio-Torres}, R. and {Baruffolo}, A. and {Baudoz}, P. and {Baudrand}, J. and {Bazzon}, A. and {Blanchard}, P. and {Bohn}, A.~J. and {Brown Sevilla}, S. and {Carbillet}, M. and {Carle}, M. and {Cascone}, E. and {Charton}, J. and {Claudi}, R. and {Costille}, A. and {De Caprio}, V. and {Delboulb{\'e}}, A. and {Dohlen}, K. and {Engler}, N. and {Fantinel}, D. and {Feautrier}, P. and {Fusco}, T. and {Gigan}, P. and {Girard}, J.~H. and {Giro}, E. and {Gisler}, D. and {Gluck}, L. and {Gry}, C. and {Hubin}, N. and {Hugot}, E. and {Jaquet}, M. and {Kasper}, M. and {Le Mignant}, D. and {Llored}, M. and {Madec}, F. and {Magnard}, Y. and {Martinez}, P. and {Maurel}, D. and {M{\"o}ller-Nilsson}, O. and {Mouillet}, D. and {Moulin}, T. and {Orign{\'e}}, A. and {Pavlov}, A. and {Perret}, D. and {Petit}, C. and {Pragt}, J. and {Puget}, P. and {Rabou}, P. and {Ramos}, J. and {Rickman}, E.~L. and {Rigal}, F. and {Rochat}, S. and {Roelfsema}, R. and {Rousset}, G. and {Roux}, A. and {Salasnich}, B. and {Sauvage}, J. -F. and {Sevin}, A. and {Soenke}, C. and {Stadler}, E. and {Suarez}, M. and {Wahhaj}, Z. and {Weber}, L. and {Wildi}, F.},
	doi = {10.1051/0004-6361/202038107},
	eid = {A72},
	eprint = {2007.06573},
	journal = {\aap},
	month = jul,
	pages = {A72},
	primaryclass = {astro-ph.EP},
	title = {{The SPHERE infrared survey for exoplanets (SHINE). III. The demographics of young giant exoplanets below 300 au with SPHERE}},
	volume = {651},
	year = 2021
    }

@article{Wagner2019,
	adsurl = {https://ui.adsabs.harvard.edu/abs/2019ApJ...877...46W},
	archiveprefix = {arXiv},
	author = {{Wagner}, Kevin and {Apai}, D{\'a}niel and {Kratter}, Kaitlin M.},
	doi = {10.3847/1538-4357/ab1904},
	eid = {46},
	eprint = {1904.06438},
	journal = {\apj},
	month = may,
	number = {1},
	pages = {46},
	primaryclass = {astro-ph.EP},
	title = {{On the Mass Function, Multiplicity, and Origins of Wide-orbit Giant Planets}},
	volume = {877},
	year = 2019
    }

@article{Bardalez-Gagliuffi2025,
	adsurl = {https://ui.adsabs.harvard.edu/abs/2025ApJ...988L..18B},
	archiveprefix = {arXiv},
	author = {{Bardalez Gagliuffi}, Daniella C. and {Balmer}, William O. and {Pueyo}, Laurent and {Brandt}, Timothy D. and {Giovinazzi}, Mark R. and {Millholland}, Sarah and {Black}, Brennen and {Lu}, Tiger and {Rice}, Malena and {Mang}, James and {Morley}, Caroline and {Lacy}, Brianna and {Girard}, Julien H. and {Matthews}, Elisabeth C. and {Carter}, Aarynn L. and {Bowler}, Brendan P. and {Faherty}, Jacqueline K. and {Fontanive}, Clemence and {Rickman}, Emily},
	doi = {10.3847/2041-8213/ade30f},
	eid = {L18},
	eprint = {2506.09201},
	journal = {\apjl},
	month = jul,
	number = {1},
	pages = {L18},
	primaryclass = {astro-ph.EP},
	title = {{JWST Coronagraphic Images of 14 Her c: A Cold Giant Planet in a Dynamically Hot Multiplanet System}},
	volume = {988},
	year = 2025}

@article{Matthews2026,
	adsurl = {https://ui.adsabs.harvard.edu/abs/2026arXiv260308780M},
	archiveprefix = {arXiv},
	author = {{Matthews}, Elisabeth C. and {Mang}, James and {Carter}, Aarynn L. and {M{\^a}lin}, Mathlide and {Morley}, Caroline V. and {Rajpoot}, Bhavesh and {Boogaard}, Leindert A. and {Burt}, Jennifer A. and {Crossfield}, Ian J.~M. and {Feng}, Fabo and {Lagrange}, Anne-Marie and {Phillips}, Mark W},
	doi = {10.48550/arXiv.2603.08780},
	eid = {arXiv:2603.08780},
	eprint = {2603.08780},
	journal = {arXiv e-prints},
	month = mar,
	pages = {arXiv:2603.08780},
	primaryclass = {astro-ph.EP},
	title = {{A second visit to Eps Ind Ab with JWST: new photometry confirms ammonia and suggests thick clouds in the exoplanet atmosphere of the closest super-Jupiter}},
	year = 2026
    }

@article{Franson2024,
	adsurl = {https://ui.adsabs.harvard.edu/abs/2024ApJ...974L..11F},
	archiveprefix = {arXiv},
	author = {{Franson}, Kyle and {Balmer}, William O. and {Bowler}, Brendan P. and {Pueyo}, Laurent and {Zhou}, Yifan and {Rickman}, Emily and {Zhang}, Zhoujian and {Mukherjee}, Sagnick and {Pearce}, Tim D. and {Bardalez Gagliuffi}, Daniella C. and {Biddle}, Lauren I. and {Brandt}, Timothy D. and {Bowens-Rubin}, Rachel and {Crepp}, Justin R. and {Davidson}, James W. and {Faherty}, Jacqueline and {Ginski}, Christian and {Horch}, Elliott P. and {Morgan}, Marvin and {Morley}, Caroline V. and {Perrin}, Marshall D. and {Sanghi}, Aniket and {Salama}, Ma{\"\i}ssa and {Theissen}, Christopher A. and {Tran}, Quang H. and {Wolf}, Trevor N.},
	doi = {10.3847/2041-8213/ad736a},
	eid = {L11},
	eprint = {2406.09528},
	journal = {\apjl},
	month = oct,
	number = {1},
	pages = {L11},
	primaryclass = {astro-ph.EP},
	title = {{JWST/NIRCam 4{\textendash}5 {\ensuremath{\mu}}m Imaging of the Giant Planet AF Lep b}},
	volume = {974},
	year = 2024
    }

@article{Franson2023,
	adsurl = {https://ui.adsabs.harvard.edu/abs/2023ApJ...950L..19F},
	archiveprefix = {arXiv},
	author = {{Franson}, Kyle and {Bowler}, Brendan P. and {Zhou}, Yifan and {Pearce}, Tim D. and {Bardalez Gagliuffi}, Daniella C. and {Biddle}, Lauren I. and {Brandt}, Timothy D. and {Crepp}, Justin R. and {Dupuy}, Trent J. and {Faherty}, Jacqueline and {Jensen-Clem}, Rebecca and {Morgan}, Marvin and {Sanghi}, Aniket and {Theissen}, Christopher A. and {Tran}, Quang H. and {Wolf}, Trevor N.},
	doi = {10.3847/2041-8213/acd6f6},
	eid = {L19},
	eprint = {2302.05420},
	journal = {\apjl},
	month = jun,
	number = {2},
	pages = {L19},
	primaryclass = {astro-ph.EP},
	title = {{Astrometric Accelerations as Dynamical Beacons: A Giant Planet Imaged inside the Debris Disk of the Young Star AF Lep}},
	volume = {950},
	year = 2023
    }

@article{Hinkley2023,
	adsurl = {https://ui.adsabs.harvard.edu/abs/2023A&A...671L...5H},
	archiveprefix = {arXiv},
	author = {{Hinkley}, S. and {Lacour}, S. and {Marleau}, G. -D. and {Lagrange}, A. -M. and {Wang}, J.~J. and {Kammerer}, J. and {Cumming}, A. and {Nowak}, M. and {Rodet}, L. and {Stolker}, T. and {Balmer}, W. -O. and {Ray}, S. and {Bonnefoy}, M. and {Molli{\`e}re}, P. and {Lazzoni}, C. and {Kennedy}, G. and {Mordasini}, C. and {Abuter}, R. and {Aigrain}, S. and {Amorim}, A. and {Asensio-Torres}, R. and {Babusiaux}, C. and {Benisty}, M. and {Berger}, J. -P. and {Beust}, H. and {Blunt}, S. and {Boccaletti}, A. and {Bohn}, A. and {Bonnet}, H. and {Bourdarot}, G. and {Brandner}, W. and {Cantalloube}, F. and {Caselli}, P. and {Charnay}, B. and {Chauvin}, G. and {Chomez}, A. and {Choquet}, E. and {Christiaens}, V. and {Cl{\'e}net}, Y. and {Coud{\'e} du Foresto}, V. and {Cridland}, A. and {Delorme}, P. and {Dembet}, R. and {Drescher}, A. and {Duvert}, G. and {Eckart}, A. and {Eisenhauer}, F. and {Feuchtgruber}, H. and {Galland}, F. and {Garcia}, P. and {Garcia Lopez}, R. and {Gardner}, T. and {Gendron}, E. and {Genzel}, R. and {Gillessen}, S. and {Girard}, J.~H. and {Grandjean}, A. and {Haubois}, X. and {Hei{\ss}el}, G. and {Henning}, Th. and {Hippler}, S. and {Horrobin}, M. and {Houll{\'e}}, M. and {Hubert}, Z. and {Jocou}, L. and {Keppler}, M. and {Kervella}, P. and {Kreidberg}, L. and {Lapeyr{\`e}re}, V. and {Le Bouquin}, J. -B. and {L{\'e}na}, P. and {Lutz}, D. and {Maire}, A. -L. and {Mang}, F. and {M{\'e}rand}, A. and {Meunier}, N. and {Monnier}, J.~D. and {Mouillet}, D. and {Nasedkin}, E. and {Ott}, T. and {Otten}, G.~P.~P.~L. and {Paladini}, C. and {Paumard}, T. and {Perraut}, K. and {Perrin}, G. and {Philipot}, F. and {Pfuhl}, O. and {Pourr{\'e}}, N. and {Pueyo}, L. and {Rameau}, J. and {Rickman}, E. and {Rubini}, P. and {Rustamkulov}, Z. and {Samland}, M. and {Shangguan}, J. and {Shimizu}, T. and {Sing}, D. and {Straubmeier}, C. and {Sturm}, E. and {Tacconi}, L.~J. and {van Dishoeck}, E.~F. and {Vigan}, A. and {Vincent}, F. and {Ward-Duong}, K. and {Widmann}, F. and {Wieprecht}, E. and {Wiezorrek}, E. and {Woillez}, J. and {Yazici}, S. and {Young}, A. and {Zicher}, N.},
	doi = {10.1051/0004-6361/202244727},
	eid = {L5},
	eprint = {2208.04867},
	journal = {\aap},
	month = mar,
	pages = {L5},
	primaryclass = {astro-ph.EP},
	title = {{Direct discovery of the inner exoplanet in the HD 206893 system. Evidence for deuterium burning in a planetary-mass companion}},
	volume = {671},
	year = 2023
    }

@article{Desdoigts2025,
	adsurl = {https://ui.adsabs.harvard.edu/abs/2025arXiv251009806D},
	archiveprefix = {arXiv},
	author = {{Desdoigts}, Louis and {Pope}, Benjamin and {Charles}, Max and {Tuthill}, Peter and {Blakely}, Dori and {Johnstone}, Doug and {Ray}, Shrishmoy and {Sivaramakrishnan}, Anand and {Kammerer}, Jens and {Thatte}, Deepashri and {Cooper}, Rachel},
	doi = {10.48550/arXiv.2510.09806},
	eid = {arXiv:2510.09806},
	eprint = {2510.09806},
	journal = {arXiv e-prints},
	month = oct,
	pages = {arXiv:2510.09806},
	primaryclass = {astro-ph.IM},
	title = {{AMIGO: a Data-Driven Calibration of the JWST Interferometer}},
	year = 2025
    }

@article{Charles2025,
	adsurl = {https://ui.adsabs.harvard.edu/abs/2025arXiv251010924C},
	archiveprefix = {arXiv},
	author = {{Charles}, Max and {Desdoigts}, Louis and {Pope}, Benjamin and {Tuthill}, Peter and {Blakely}, Dori and {Johnstone}, Doug and {Ray}, Shrishmoy and {Ford}, K.~E. Saavik and {McKernan}, Barry and {Sivaramakrishnan}, Anand},
	doi = {10.48550/arXiv.2510.10924},
	eid = {arXiv:2510.10924},
	eprint = {2510.10924},
	journal = {arXiv e-prints},
	month = oct,
	pages = {arXiv:2510.10924},
	primaryclass = {astro-ph.IM},
	title = {{Image reconstruction with the JWST Interferometer}},
	year = 2025
    }

@article{Sallum2024,
	adsurl = {https://ui.adsabs.harvard.edu/abs/2024ApJ...963L...2S},
	archiveprefix = {arXiv},
	author = {{Sallum}, Steph and {Ray}, Shrishmoy and {Kammerer}, Jens and {Sivaramakrishnan}, Anand and {Cooper}, Rachel and {Greebaum}, Alexandra Z. and {Thatte}, Deepashri and {De Furio}, Matthew and {Factor}, Samuel M. and {Meyer}, Michael R. and {Stone}, Jordan M. and {Carter}, Aarynn and {Biller}, Beth and {Hinkley}, Sasha and {Skemer}, Andrew and {Su{\'a}rez}, Genaro and {Leisenring}, Jarron M. and {Perrin}, Marshall D. and {Kraus}, Adam L. and {Absil}, Olivier and {Balmer}, William O. and {Betti}, Sarah K. and {Boccaletti}, Anthony and {Bonavita}, Mariangela and {Bonnefoy}, Mickael and {Booth}, Mark and {Bowler}, Brendan P. and {Briesemeister}, Zackery W. and {Bryan}, Marta L. and {Calissendorff}, Per and {Cantalloube}, Faustine and {Chauvin}, Gael and {Chen}, Christine H. and {Choquet}, Elodie and {Christiaens}, Valentin and {Cugno}, Gabriele and {Currie}, Thayne and {Danielski}, Camilla and {Dupuy}, Trent J. and {Faherty}, Jacqueline K. and {Fitzgerald}, Michael P. and {Fortney}, Jonathan J. and {Franson}, Kyle and {Girard}, Julien H. and {Grady}, Carol A. and {Gonzales}, Eileen C. and {Henning}, Thomas and {Hines}, Dean C. and {Hoch}, Kielan K.~W. and {Hood}, Callie E. and {Howe}, Alex R. and {Janson}, Markus and {Kalas}, Paul and {Kennedy}, Grant M. and {Kenworthy}, Matthew A. and {Kervella}, Pierre and {Kitzmann}, Daniel and {Kuzuhara}, Masayuki and {Lagrange}, Anne-Marie and {Lagage}, Pierre-Olivier and {Lawson}, Kellen and {Lazzoni}, Cecilia and {Lew}, Ben W.~P. and {Liu}, Michael C. and {Liu}, Pengyu and {Llop-Sayson}, Jorge and {Lloyd}, James P. and {Lueber}, Anna and {Macintosh}, Bruce and {Manjavacas}, Elena and {Marino}, Sebastian and {Marley}, Mark S. and {Marois}, Christian and {Martinez}, Raquel A. and {Matthews}, Brenda C. and {Matthews}, Elisabeth C. and {Mawet}, Dimitri and {Mazoyer}, Johan and {McElwain}, Michael W. and {Metchev}, Stanimir and {Miles}, Brittany E. and {Millar-Blanchaer}, Maxwell A. and {Molliere}, Paul and {Moran}, Sarah E. and {Morley}, Caroline V. and {Mukherjee}, Sagnick and {Palma-Bifani}, Paulina and {Pantin}, Eric and {Patapis}, Polychronis and {Petrus}, Simon and {Pueyo}, Laurent and {Quanz}, Sascha P. and {Quirrenbach}, Andreas and {Rebollido}, Isabel and {Redai}, Jea Adams and {Ren}, Bin B. and {Rickman}, Emily and {Samland}, Matthias and {Sargent}, B.~A. and {Schlieder}, Joshua E. and {Schneider}, Glenn and {Stapelfeldt}, Karl R. and {Sutlieff}, Ben J. and {Tamura}, Motohide and {Tan}, Xianyu and {Theissen}, Christopher A. and {Uyama}, Taichi and {Vigan}, Arthur and {Vasist}, Malavika and {Vos}, Johanna M. and {Wagner}, Kevin and {Wang}, Jason J. and {Ward-Duong}, Kimberly and {Whiteford}, Niall and {Wolff}, Schuyler G. and {Worthen}, Kadin and {Wyatt}, Mark C. and {Ygouf}, Marie and {Zhang}, Xi and {Zhang}, Keming and {Zhang}, Zhoujian and {Zhou}, Yifan and {Zurlo}, Alice},
	doi = {10.3847/2041-8213/ad21fb},
	eid = {L2},
	eprint = {2310.11499},
	journal = {\apjl},
	month = mar,
	number = {1},
	pages = {L2},
	primaryclass = {astro-ph.EP},
	title = {{The JWST Early Release Science Program for Direct Observations of Exoplanetary Systems. IV. NIRISS Aperture Masking Interferometry Performance and Lessons Learned}},
	volume = {963},
	year = 2024
    }

@article{Ray2025,
	adsurl = {https://ui.adsabs.harvard.edu/abs/2025ApJ...983L..25R},
	archiveprefix = {arXiv},
	author = {{Ray}, Shrishmoy and {Sallum}, Steph and {Hinkley}, Sasha and {Sivaramkrishnan}, Anand and {Cooper}, Rachel and {Kammerer}, Jens and {Greebaum}, Alexandra Z. and {Thatte}, Deeparshi and {Stolker}, Tomas and {Lazzoni}, Cecilia and {Tokovinin}, Andrei and {de Furio}, Matthew and {Factor}, Samuel and {Meyer}, Michael and {Stone}, Jordan M. and {Carter}, Aarynn and {Biller}, Beth and {Skemer}, Andrew and {Su{\'a}rez}, Genaro and {Leisenring}, Jarron M. and {Perrin}, Marshall D. and {Kraus}, Adam L. and {Absil}, Olivier and {Balmer}, William O. and {Boccaletti}, Anthony and {Bonavita}, Mariangela and {Bonnefoy}, Mickael and {Booth}, Mark and {Bowler}, Brendan P. and {Briesemeister}, Zackery W. and {Bryan}, Marta L. and {Calissendorff}, Per and {Cantalloube}, Faustine and {Chauvin}, Gael and {Chen}, Christine H. and {Choquet}, Elodie and {Christiaens}, Valentin and {Cugno}, Gabriele and {Currie}, Thayne and {Danielski}, Camilla and {Dupuy}, Trent J. and {Faherty}, Jacqueline K. and {Fitzgerald}, Michael P. and {Fortney}, Jonathan J. and {Franson}, Kyle and {Girard}, Julien H. and {Grady}, Carol A. and {Gonzales}, Eileen C. and {Henning}, Thomas and {Hines}, Dean C. and {Hoch}, Kielan K.~W. and {Hood}, Callie E. and {Howe}, Alex R. and {Janson}, Markus and {Kalas}, Paul and {Kennedy}, Grant M. and {Kenworthy}, Matthew A. and {Kervella}, Pierre and {Kuzuhara}, Masayuki and {Lagrange}, Anne-Marie and {Lagage}, Pierre-Olivier and {Lawson}, Kellen and {Lew}, Ben W.~P. and {Liu}, Michael C. and {Liu}, Pengyu and {Llop-Sayson}, Jorge and {Lloyd}, James P. and {Macintosh}, Bruce and {Marino}, Sebastian and {Marley}, Mark S. and {Marois}, Christian and {Martinez}, Raquel A. and {Matthews}, Brenda C. and {Matthews}, Elisabeth C. and {Mawet}, Dimitri and {Mazoyer}, Johan and {McElwain}, Michael W. and {Metchev}, Stanimir and {Meyer}, Michael R. and {Miles}, Brittany E. and {Millar-Blanchaer}, Maxwell A. and {Molliere}, Paul and {Moran}, Sarah E. and {Morley}, Caroline V. and {Mukherjee}, Sagnick and {Palma-Bifani}, Paulina and {Pantin}, Eric and {Patapis}, Polychronis and {Petrus}, Simon and {Pueyo}, Laurent and {Quanz}, Sascha P. and {Quirrenbach}, Andreas and {Rebollido}, Isabel and {Adams Redai}, Jea and {Ren}, Bin B. and {Rickman}, Emily and {Samland}, Matthias and {Schlieder}, Joshua E. and {Schneider}, Glenn and {Stapelfeldt}, Karl R. and {Tamura}, Motohide and {Tan}, Xianyu and {Uyama}, Taichi and {Vigan}, Arthur and {Vasist}, Malavika and {Vos}, Johanna M. and {Wagner}, Kevin and {Wang}, Jason J. and {Ward-Duong}, Kimberly and {Whiteford}, Niall and {Wolff}, Schuyler G. and {Worthen}, Kadin and {Wyatt}, Mark C. and {Ygouf}, Marie and {Zhang}, Xi and {Zhang}, Keming and {Zhang}, Zhoujian and {Zhou}, Yifan and {Zurlo}, Alice and {Sargent}, B.~A. and {Theissen}, Christopher A. and {Manjavacas}, Elena and {Lueber}, Anna and {Kitzmann}, Daniel and {Sutlieff}, Ben J. and {Betti}, Sarah K.},
	doi = {10.3847/2041-8213/adaeb7},
	eid = {L25},
	eprint = {2310.11508},
	journal = {\apjl},
	month = apr,
	number = {1},
	pages = {L25},
	primaryclass = {astro-ph.EP},
	title = {{The JWST Early Release Science Program for Direct Observations of Exoplanetary Systems. III. Aperture Masking Interferometric Observations of the Star HIP 65426 at 3.8 {\ensuremath{\mu}}m}},
	volume = {983},
	year = 2025
    }

@article{Blakely2025,
	adsurl = {https://ui.adsabs.harvard.edu/abs/2025AJ....169..137B},
	archiveprefix = {arXiv},
	author = {{Blakely}, Dori and {Johnstone}, Doug and {Cugno}, Gabriele and {Sivaramakrishnan}, Anand and {Tuthill}, Peter and {Dong}, Ruobing and {Pope}, Benjamin J.~S. and {Albert}, Lo{\"\i}c and {Charles}, Max and {Cooper}, Rachel A. and {Furio}, Matthew De and {Desdoigts}, Louis and {Doyon}, Ren{\'e} and {Francis}, Logan and {Greenbaum}, Alexandra Z. and {Lafreni{\'e}re}, David and {Lloyd}, James P. and {Meyer}, Michael R. and {Pueyo}, Laurent and {Ray}, Shrishmoy and {S{\'a}nchez-Berm{\'u}dez}, Joel and {Soulain}, Anthony and {Thatte}, Deepashri and {Thompson}, William and {Vandal}, Thomas},
	doi = {10.3847/1538-3881/ad9b94},
	eid = {137},
	eprint = {2404.13032},
	journal = {\aj},
	month = mar,
	number = {3},
	pages = {137},
	primaryclass = {astro-ph.EP},
	title = {{The James Webb Interferometer: Space-based Interferometric Detections of PDS 70 b and c at 4.8 {\ensuremath{\mu}}m}},
	volume = {169},
	year = 2025
    }

@article{Haffert2019,
	adsurl = {https://ui.adsabs.harvard.edu/abs/2019NatAs...3..749H},
	archiveprefix = {arXiv},
	author = {{Haffert}, S.~Y. and {Bohn}, A.~J. and {de Boer}, J. and {Snellen}, I.~A.~G. and {Brinchmann}, J. and {Girard}, J.~H. and {Keller}, C.~U. and {Bacon}, R.},
	doi = {10.1038/s41550-019-0780-5},
	eprint = {1906.01486},
	journal = {Nature Astronomy},
	month = jun,
	pages = {749-754},
	primaryclass = {astro-ph.EP},
	title = {{Two accreting protoplanets around the young star PDS 70}},
	volume = {3},
	year = 2019
    }

@article{Keppler2018,
	adsurl = {https://ui.adsabs.harvard.edu/abs/2018A&A...617A..44K},
	archiveprefix = {arXiv},
	author = {{Keppler}, M. and {Benisty}, M. and {M{\"u}ller}, A. and {Henning}, Th. and {van Boekel}, R. and {Cantalloube}, F. and {Ginski}, C. and {van Holstein}, R.~G. and {Maire}, A. -L. and {Pohl}, A. and {Samland }, M. and {Avenhaus}, H. and {Baudino}, J. -L. and {Boccaletti}, A. and {de Boer}, J. and {Bonnefoy}, M. and {Chauvin}, G. and {Desidera}, S. and {Langlois}, M. and {Lazzoni}, C. and {Marleau}, G. -D. and {Mordasini}, C. and {Pawellek}, N. and {Stolker}, T. and {Vigan}, A. and {Zurlo}, A. and {Birnstiel}, T. and {Brandner}, W. and {Feldt}, M. and {Flock}, M. and {Girard}, J. and {Gratton}, R. and {Hagelberg}, J. and {Isella}, A. and {Janson}, M. and {Juhasz}, A. and {Kemmer}, J. and {Kral}, Q. and {Lagrange}, A. -M. and {Launhardt}, R. and {Matter}, A. and {M{\'e}nard}, F. and {Milli}, J. and {Molli{\`e}re}, P. and {Olofsson}, J. and {P{\'e}rez}, L. and {Pinilla}, P. and {Pinte}, C. and {Quanz}, S.~P. and {Schmidt}, T. and {Udry}, S. and {Wahhaj}, Z. and {Williams}, J.~P. and {Buenzli}, E. and {Cudel}, M. and {Dominik}, C. and {Galicher}, R. and {Kasper}, M. and {Lannier}, J. and {Mesa}, D. and {Mouillet}, D. and {Peretti}, S. and {Perrot}, C. and {Salter}, G. and {Sissa}, E. and {Wildi}, F. and {Abe}, L. and {Antichi}, J. and {Augereau}, J. -C. and {Baruffolo}, A. and {Baudoz}, P. and {Bazzon}, A. and {Beuzit}, J. -L. and {Blanchard}, P. and {Brems}, S.~S. and {Buey}, T. and {De Caprio}, V. and {Carbillet}, M. and {Carle}, M. and {Cascone}, E. and {Cheetham}, A. and {Claudi}, R. and {Costille}, A. and {Delboulb{\'e}}, A. and {Dohlen}, K. and {Fantinel}, D. and {Feautrier}, P. and {Fusco}, T. and {Giro}, E. and {Gluck}, L. and {Gry}, C. and {Hubin}, N. and {Hugot}, E. and {Jaquet}, M. and {Le Mignant}, D. and {Llored}, M. and {Madec}, F. and {Magnard}, Y. and {Martinez}, P. and {Maurel}, D. and {Meyer}, M. and {M{\"o}ller-Nilsson}, O. and {Moulin}, T. and {Mugnier}, L. and {Orign{\'e}}, A. and {Pavlov}, A. and {Perret}, D. and {Petit}, C. and {Pragt}, J. and {Puget}, P. and {Rabou}, P. and {Ramos}, J. and {Rigal}, F. and {Rochat}, S. and {Roelfsema}, R. and {Rousset}, G. and {Roux}, A. and {Salasnich}, B. and {Sauvage}, J. -F. and {Sevin}, A. and {Soenke}, C. and {Stadler}, E. and {Suarez}, M. and {Turatto}, M. and {Weber}, L.},
	doi = {10.1051/0004-6361/201832957},
	eid = {A44},
	eprint = {1806.11568},
	journal = {\aap},
	month = sep,
	pages = {A44},
	primaryclass = {astro-ph.EP},
	title = {{Discovery of a planetary-mass companion within the gap of the transition disk around PDS 70}},
	volume = {617},
	year = 2018
    }

@article{Christiaens2024,
	adsurl = {https://ui.adsabs.harvard.edu/abs/2024A&A...685L...1C},
	archiveprefix = {arXiv},
	author = {{Christiaens}, V. and {Samland}, M. and {Henning}, Th. and {Portilla-Revelo}, B. and {Perotti}, G. and {Matthews}, E. and {Absil}, O. and {Decin}, L. and {Kamp}, I. and {Boccaletti}, A. and {Tabone}, B. and {Marleau}, G. -D. and {van Dishoeck}, E.~F. and {G{\"u}del}, M. and {Lagage}, P. -O. and {Barrado}, D. and {Caratti o Garatti}, A. and {Glauser}, A.~M. and {Olofsson}, G. and {Ray}, T.~P. and {Scheithauer}, S. and {Vandenbussche}, B. and {Waters}, L.~B.~F.~M. and {Arabhavi}, A.~M. and {Grant}, S.~L. and {Jang}, H. and {Kanwar}, J. and {Schreiber}, J. and {Schwarz}, K. and {Temmink}, M. and {{\"O}stlin}, G.},
	doi = {10.1051/0004-6361/202349089},
	eid = {L1},
	eprint = {2403.04855},
	journal = {\aap},
	month = may,
	pages = {L1},
	primaryclass = {astro-ph.EP},
	title = {{MINDS: JWST/NIRCam imaging of the protoplanetary disk PDS 70. A spiral accretion stream and a potential third protoplanet}},
	volume = {685},
	year = 2024
    }

@article{Madurowicz2025,
	adsurl = {https://ui.adsabs.harvard.edu/abs/2025AJ....170..326M},
	archiveprefix = {arXiv},
	author = {{Madurowicz}, Alexander and {Ruffio}, Jean-Baptiste and {Macintosh}, Bruce and {Perrin}, Marshall and {Konopacky}, Quinn M. and {Baburaj}, Aneesh and {Hoch}, Kielan},
	doi = {10.3847/1538-3881/ae1028},
	eid = {326},
	eprint = {2510.08327},
	journal = {\aj},
	month = dec,
	number = {6},
	pages = {326},
	primaryclass = {astro-ph.EP},
	title = {{Direct Spectroscopy of 51 Eridani b with JWST NIRSpec}},
	volume = {170},
	year = 2025
    }

@article{Ruffio2024,
	adsurl = {https://ui.adsabs.harvard.edu/abs/2024AJ....168...73R},
	archiveprefix = {arXiv},
	author = {{Ruffio}, Jean-Baptiste and {Perrin}, Marshall D. and {Hoch}, Kielan K.~W. and {Kammerer}, Jens and {Konopacky}, Quinn M. and {Pueyo}, Laurent and {Madurowicz}, Alex and {Rickman}, Emily and {Theissen}, Christopher A. and {Agrawal}, Shubh and {Greenbaum}, Alexandra Z. and {Miles}, Brittany E. and {Barman}, Travis S. and {Balmer}, William O. and {Llop-Sayson}, Jorge and {Girard}, Julien H. and {Rebollido}, Isabel and {Soummer}, R{\'e}mi and {Allen}, Natalie H. and {Anderson}, Jay and {Beichman}, Charles A. and {Bellini}, Andrea and {Bryden}, Geoffrey and {Espinoza}, N{\'e}stor and {Glidden}, Ana and {Huang}, Jingcheng and {Lewis}, Nikole K. and {Libralato}, Mattia and {Louie}, Dana R. and {Sohn}, Sangmo Tony and {Seager}, Sara and {van der Marel}, Roeland P. and {Wakeford}, Hannah R. and {Watkins}, Laura L. and {Ygouf}, Marie and {Mountain}, C. Matt},
	doi = {10.3847/1538-3881/ad5281},
	eid = {73},
	eprint = {2310.09902},
	journal = {\aj},
	month = aug,
	number = {2},
	pages = {73},
	primaryclass = {astro-ph.EP},
	title = {{JWST-TST High Contrast: Achieving Direct Spectroscopy of Faint Substellar Companions Next to Bright Stars with the NIRSpec Integral Field Unit}},
	volume = {168},
	year = 2024
    }

@article{Cugno2024,
	adsurl = {https://ui.adsabs.harvard.edu/abs/2024ApJ...966L..21C},
	archiveprefix = {arXiv},
	author = {{Cugno}, Gabriele and {Patapis}, Polychronis and {Banzatti}, Andrea and {Meyer}, Michael and {Dannert}, Felix A. and {Stolker}, Tomas and {MacDonald}, Ryan J. and {Pontoppidan}, Klaus M.},
	doi = {10.3847/2041-8213/ad3cbc},
	eid = {L21},
	eprint = {2404.07086},
	journal = {\apjl},
	month = may,
	number = {1},
	pages = {L21},
	primaryclass = {astro-ph.EP},
	title = {{Mid-infrared Spectrum of the Disk around the Forming Companion GQ Lup B Revealed by JWST/MIRI}},
	volume = {966},
	year = 2024
    }

@article{Hoch2025,
	adsurl = {https://ui.adsabs.harvard.edu/abs/2025Natur.643..938H},
	archiveprefix = {arXiv},
	author = {{Hoch}, K.~K.~W. and {Rowland}, M. and {Petrus}, S. and {Nasedkin}, E. and {Ingebretsen}, C. and {Kammerer}, J. and {Perrin}, M. and {D'Orazi}, V. and {Balmer}, W.~O. and {Barman}, T. and {Bonnefoy}, M. and {Chauvin}, G. and {Chen}, C. and {De Rosa}, R.~J. and {Girard}, J. and {Gonzales}, E. and {Kenworthy}, M. and {Konopacky}, Q.~M. and {Macintosh}, B. and {Moran}, S.~E. and {Morley}, C.~V. and {Palma-Bifani}, P. and {Pueyo}, L. and {Ren}, B. and {Rickman}, E. and {Ruffio}, J. -B. and {Theissen}, C.~A. and {Ward-Duong}, K. and {Zhang}, Y.},
	doi = {10.1038/s41586-025-09174-w},
	eprint = {2507.18861},
	journal = {\nat},
	month = jul,
	number = {8073},
	pages = {938-942},
	primaryclass = {astro-ph.EP},
	title = {{Silicate clouds and a circumplanetary disk in the YSES-1 exoplanet system}},
	volume = {643},
	year = 2025
    }

@article{Malin2025b,
	adsurl = {https://ui.adsabs.harvard.edu/abs/2025A&A...704A.181M},
	archiveprefix = {arXiv},
	author = {{M{\^a}lin}, Mathilde and {Ward-Duong}, Kimberly and {Grant}, Sierra L. and {Arulanantham}, Nicole and {Tabone}, Beno{\^\i}t and {Pueyo}, Laurent and {Perrin}, Marshall and {Balmer}, William O. and {Betti}, Sarah and {Chen}, Christine H. and {Debes}, John H. and {Girard}, Julien H. and {Hoch}, Kielan K.~W. and {Kammerer}, Jens and {Lu}, Cicero and {Rebollido}, Isabel and {Rickman}, Emily and {Robinson}, Connor and {Worthen}, Kadin and {van der Marel}, Roeland P. and {Lewis}, Nikole K. and {Seager}, Sara and {Valenti}, Jeff A. and {Soummer}, Remi},
	doi = {10.1051/0004-6361/202556792},
	eid = {A181},
	eprint = {2510.07253},
	journal = {\aap},
	month = dec,
	pages = {A181},
	primaryclass = {astro-ph.EP},
	title = {{JWST-TST High Contrast: Medium-resolution spectroscopy reveals a carbon-rich circumplanetary disk around the young accreting exoplanet Delorme 1 AB b}},
	volume = {704},
	year = 2025
    }

@article{Kammerer2024,
	adsurl = {https://ui.adsabs.harvard.edu/abs/2024AJ....168...51K},
	archiveprefix = {arXiv},
	author = {{Kammerer}, Jens and {Lawson}, Kellen and {Perrin}, Marshall D. and {Rebollido}, Isabel and {Stark}, Christopher C. and {Stolker}, Tomas and {Girard}, Julien H. and {Pueyo}, Laurent and {Balmer}, William O. and {Worthen}, Kadin and {Chen}, Christine and {van der Marel}, Roeland P. and {Lewis}, Nikole K. and {Ward-Duong}, Kimberly and {Valenti}, Jeff A. and {Clampin}, Mark and {Mountain}, C. Matt},
	doi = {10.3847/1538-3881/ad4ffe},
	eid = {51},
	eprint = {2405.18422},
	journal = {\aj},
	month = aug,
	number = {2},
	pages = {51},
	primaryclass = {astro-ph.EP},
	title = {{JWST-TST High Contrast: JWST/NIRCam Observations of the Young Giant Planet {\ensuremath{\beta}} Pic b}},
	volume = {168},
	year = 2024
    }

@article{Malin2024,
	adsurl = {https://ui.adsabs.harvard.edu/abs/2024A&A...690A.316M},
	archiveprefix = {arXiv},
	author = {{M{\^a}lin}, Mathilde and {Boccaletti}, Anthony and {Perrot}, Cl{\'e}ment and {Baudoz}, Pierre and {Rouan}, Daniel and {Lagage}, Pierre-Olivier and {Waters}, Rens and {G{\"u}del}, Manuel and {Henning}, Thomas and {Vandenbussche}, Bart and {Absil}, Olivier and {Barrado}, David and {Cossou}, Christophe and {Decin}, Leen and {Glauser}, Adrian M. and {Pye}, John and {Olofsson}, Goran and {Glasse}, Alistair and {Lahuis}, Fred and {Patapis}, Polychronis and {Royer}, Pierre and {Scheithauer}, Silvia and {Whiteford}, Niall and {Serabyn}, Eugene and {Choquet}, Elodie and {Colina}, Luis and {Ostlin}, G{\"o}ran and {Ray}, Tom P. and {Wright}, Gillian},
	doi = {10.1051/0004-6361/202450470},
	eid = {A316},
	eprint = {2408.16843},
	journal = {\aap},
	month = oct,
	pages = {A316},
	primaryclass = {astro-ph.EP},
	title = {{Unveiling the HD 95086 system at mid-infrared wavelengths with JWST/MIRI}},
	volume = {690},
	year = 2024
    }

@article{Balmer2025,
	adsurl = {https://ui.adsabs.harvard.edu/abs/2025AJ....169..209B},
	archiveprefix = {arXiv},
	author = {{Balmer}, William O. and {Kammerer}, Jens and {Pueyo}, Laurent and {Perrin}, Marshall D. and {Girard}, Julien H. and {Leisenring}, Jarron M. and {Lawson}, Kellen and {Dennen}, Henry and {van der Marel}, Roeland P. and {Beichman}, Charles A. and {Bryden}, Geoffrey and {Llop-Sayson}, Jorge and {Valenti}, Jeff A. and {Lothringer}, Joshua D. and {Lewis}, Nikole K. and {M{\^a}lin}, Mathilde and {Rebollido}, Isabel and {Rickman}, Emily and {Hoch}, Kielan K.~W. and {Soummer}, R{\'e}mi and {Clampin}, Mark and {Mountain}, C. Matt},
	doi = {10.3847/1538-3881/adb1c6},
	eid = {209},
	eprint = {2503.13608},
	journal = {\aj},
	month = apr,
	number = {4},
	pages = {209},
	primaryclass = {astro-ph.EP},
	title = {{JWST-TST High Contrast: Living on the Wedge, or, NIRCam Bar Coronagraphy Reveals CO$_{2}$ in the HR 8799 and 51 Eri Exoplanets' Atmospheres}},
	volume = {169},
	year = 2025
    }

@article{Boccaletti2024,
	adsurl = {https://ui.adsabs.harvard.edu/abs/2024A&A...686A..33B},
	archiveprefix = {arXiv},
	author = {{Boccaletti}, Anthony and {M{\^a}lin}, Mathilde and {Baudoz}, Pierre and {Tremblin}, Pascal and {Perrot}, Cl{\'e}ment and {Rouan}, Daniel and {Lagage}, Pierre-Olivier and {Whiteford}, Niall and {Molli{\`e}re}, Paul and {Waters}, Rens and {Henning}, Thomas and {Decin}, Leen and {G{\"u}del}, Manuel and {Vandenbussche}, Bart and {Absil}, Olivier and {Argyriou}, Ioannis and {Bouwman}, Jeroen and {Cossou}, Christophe and {Coulais}, Alain and {Gastaud}, Ren{\'e} and {Glasse}, Alistair and {Glauser}, Adrian M. and {Kamp}, Inga and {Kendrew}, Sarah and {Krause}, Oliver and {Lahuis}, Fred and {Mueller}, Michael and {Olofsson}, Goran and {Patapis}, Polychronis and {Pye}, John and {Royer}, Pierre and {Serabyn}, Eugene and {Scheithauer}, Silvia and {Colina}, Luis and {van Dishoeck}, Ewine F. and {Ostlin}, G{\"o}ran and {Ray}, Tom P. and {Wright}, Gillian},
	doi = {10.1051/0004-6361/202347912},
	eid = {A33},
	eprint = {2310.13414},
	journal = {\aap},
	month = jun,
	pages = {A33},
	primaryclass = {astro-ph.EP},
	title = {{Imaging detection of the inner dust belt and the four exoplanets in the HR 8799 system with JWST's MIRI coronagraph}},
	volume = {686},
	year = 2024
    }

@article{Skemer2014,
	adsurl = {http://adsabs.harvard.edu/abs/2014ApJ...792...17S},
	archiveprefix = {arXiv},
	author = {{Skemer}, A.~J. and {Marley}, M.~S. and {Hinz}, P.~M. and {Morzinski}, K.~M. and {Skrutskie}, M.~F. and {Leisenring}, J.~M. and {Close}, L.~M. and {Saumon}, D. and {Bailey}, V.~P. and {Briguglio}, R. and {Defrere}, D. and {Esposito}, S. and {Follette}, K.~B. and {Hill}, J.~M. and {Males}, J.~R. and {Puglisi}, A. and {Rodigas}, T.~J. and {Xompero}, M.},
	doi = {10.1088/0004-637X/792/1/17},
	eid = {17},
	eprint = {1311.2085},
	journal = {\apj},
	month = sep,
	pages = {17},
	primaryclass = {astro-ph.EP},
	title = {{Directly Imaged L-T Transition Exoplanets in the Mid-infrared}},
	volume = 792,
	year = 2014
    }

@article{Lagrange2010,
	adsurl = {http://adsabs.harvard.edu/abs/2010Sci...329...57L},
	archiveprefix = {arXiv},
	author = {{Lagrange}, {A.-M.} and {Bonnefoy}, M. and {Chauvin}, G. and {Apai}, D. and {Ehrenreich}, D. and {Boccaletti}, A. and {Gratadour}, D. and {Rouan}, D. and {Mouillet}, D. and {Lacour}, S. and {Kasper}, M.},
	doi = {10.1126/science.1187187},
	eprint = {1006.3314},
	journal = {Science},
	month = jul,
	pages = {57-},
	primaryclass = {astro-ph.EP},
	title = {{A Giant Planet Imaged in the Disk of the Young Star {$\beta$} Pictoris}},
	volume = 329,
	year = 2010
    }

@article{Marois2010,
	adsurl = {http://adsabs.harvard.edu/abs/2010Natur.468.1080M},
	archiveprefix = {arXiv},
	author = {{Marois}, C. and {Zuckerman}, B. and {Konopacky}, Q.~M. and {Macintosh}, B. and {Barman}, T.},
	doi = {10.1038/nature09684},
	eprint = {1011.4918},
	journal = {\nat},
	month = dec,
	pages = {1080-1083},
	primaryclass = {astro-ph.EP},
	title = {{Images of a fourth planet orbiting HR 8799}},
	volume = 468,
	year = 2010
    }

@article{Marois2008,
	adsurl = {http://adsabs.harvard.edu/abs/2008Sci...322.1348M},
	archiveprefix = {arXiv},
	author = {{Marois}, C. and {Macintosh}, B. and {Barman}, T. and {Zuckerman}, B. and {Song}, I. and {Patience}, J. and {Lafreni{\`e}re}, D. and {Doyon}, R.},
	eprint = {0811.2606},
	journal = {Science},
	month = nov,
	pages = {1348-},
	title = {{Direct Imaging of Multiple Planets Orbiting the Star HR 8799}},
	volume = 322,
	year = 2008
    }

@article{Zhou2022,
	adsurl = {https://ui.adsabs.harvard.edu/abs/2022AJ....164..239Z},
	archiveprefix = {arXiv},
	author = {{Zhou}, Yifan and {Bowler}, Brendan P. and {Apai}, D{\'a}niel and {Kataria}, Tiffany and {Morley}, Caroline V. and {Bryan}, Marta L. and {Skemer}, Andrew J. and {Benneke}, Bj{\"o}rn},
	doi = {10.3847/1538-3881/ac9905},
	eid = {239},
	eprint = {2210.02464},
	journal = {\aj},
	month = dec,
	number = {6},
	pages = {239},
	primaryclass = {astro-ph.EP},
	title = {{Roaring Storms in the Planetary-mass Companion VHS 1256-1257 b: Hubble Space Telescope Multiepoch Monitoring Reveals Vigorous Evolution in an Ultracool Atmosphere}},
	volume = {164},
	year = 2022
    }

@article{Zhou2020,
	adsurl = {https://ui.adsabs.harvard.edu/abs/2020AJ....160...77Z},
	archiveprefix = {arXiv},
	author = {{Zhou}, Yifan and {Bowler}, Brendan P. and {Morley}, Caroline V. and {Apai}, D{\'a}niel and {Kataria}, Tiffany and {Bryan}, Marta L. and {Benneke}, Bj{\"o}rn},
	doi = {10.3847/1538-3881/ab9e04},
	eid = {77},
	eprint = {2004.05168},
	journal = {\aj},
	month = aug,
	number = {2},
	pages = {77},
	primaryclass = {astro-ph.EP},
	title = {{Spectral Variability of VHS J1256-1257b from 1 to 5 {\ensuremath{\mu}}m}},
	volume = {160},
	year = 2020
    }

@article{Bowler2020,
	adsurl = {https://ui.adsabs.harvard.edu/abs/2020ApJ...893L..30B},
	archiveprefix = {arXiv},
	author = {{Bowler}, Brendan P. and {Zhou}, Yifan and {Morley}, Caroline V. and {Kataria}, Tiffany and {Bryan}, Marta L. and {Benneke}, Bj{\"o}rn and {Batygin}, Konstantin},
	doi = {10.3847/2041-8213/ab8197},
	eid = {L30},
	eprint = {2004.05170},
	journal = {\apjl},
	month = apr,
	number = {2},
	pages = {L30},
	primaryclass = {astro-ph.EP},
	title = {{Strong Near-infrared Spectral Variability of the Young Cloudy L Dwarf Companion VHS J1256-1257 b}},
	volume = {893},
	year = 2020
    }

@article{Cushing2006,
	adsurl = {https://ui.adsabs.harvard.edu/abs/2006ApJ...648..614C},
	archiveprefix = {arXiv},
	author = {{Cushing}, Michael C. and {Roellig}, Thomas L. and {Marley}, Mark S. and {Saumon}, D. and {Leggett}, S.~K. and {Kirkpatrick}, J. Davy and {Wilson}, John C. and {Sloan}, G.~C. and {Mainzer}, Amy K. and {Van Cleve}, Jeff E.},
	doi = {10.1086/505637},
	eprint = {astro-ph/0605639},
	journal = {\apj},
	month = {Sep},
	number = {1},
	pages = {614-628},
	primaryclass = {astro-ph},
	title = {{A Spitzer Infrared Spectrograph Spectral Sequence of M, L, and T Dwarfs}},
	volume = {648},
	year = {2006}
    }

@article{Gauza2015,
	adsurl = {http://adsabs.harvard.edu/abs/2015ApJ...804...96G},
	archiveprefix = {arXiv},
	author = {{Gauza}, B. and {B{\'e}jar}, V.~J.~S. and {P{\'e}rez-Garrido}, A. and {Rosa Zapatero Osorio}, M. and {Lodieu}, N. and {Rebolo}, R. and {Pall{\'e}}, E. and {Nowak}, G.},
	doi = {10.1088/0004-637X/804/2/96},
	eid = {96},
	eprint = {1505.00806},
	journal = {\apj},
	month = may,
	pages = {96},
	primaryclass = {astro-ph.EP},
	title = {{Discovery of a Young Planetary Mass Companion to the Nearby M Dwarf VHS J125601.92-125723.9}},
	volume = 804,
	year = 2015
    }

@article{Perrot2016,
	adsurl = {https://ui.adsabs.harvard.edu/abs/2016A&A...590L...7P},
	archiveprefix = {arXiv},
	author = {{Perrot}, C. and {Boccaletti}, A. and {Pantin}, E. and {Augereau}, J. -C. and {Lagrange}, A. -M. and {Galicher}, R. and {Maire}, A. -L. and {Mazoyer}, J. and {Milli}, J. and {Rousset}, G. and {Gratton}, R. and {Bonnefoy}, M. and {Brandner}, W. and {Buenzli}, E. and {Langlois}, M. and {Lannier}, J. and {Mesa}, D. and {Peretti}, S. and {Salter}, G. and {Sissa}, E. and {Chauvin}, G. and {Desidera}, S. and {Feldt}, M. and {Vigan}, A. and {Di Folco}, E. and {Dutrey}, A. and {P{\'e}ricaud}, J. and {Baudoz}, P. and {Benisty}, M. and {De Boer}, J. and {Garufi}, A. and {Girard}, J.~H. and {Menard}, F. and {Olofsson}, J. and {Quanz}, S.~P. and {Mouillet}, D. and {Christiaens}, V. and {Casassus}, S. and {Beuzit}, J. -L. and {Blanchard}, P. and {Carle}, M. and {Fusco}, T. and {Giro}, E. and {Hubin}, N. and {Maurel}, D. and {Moeller-Nilsson}, O. and {Sevin}, A. and {Weber}, L.},
	doi = {10.1051/0004-6361/201628396},
	eid = {L7},
	eprint = {1605.00468},
	journal = {\aap},
	month = may,
	pages = {L7},
	primaryclass = {astro-ph.EP},
	title = {{Discovery of concentric broken rings at sub-arcsec separations in the HD 141569A gas-rich, debris disk with VLT/SPHERE}},
	volume = {590},
	year = 2016
    }

@article{Mouillet2001,
	adsurl = {https://ui.adsabs.harvard.edu/abs/2001A&A...372L..61M},
	author = {{Mouillet}, D. and {Lagrange}, A.~M. and {Augereau}, J.~C. and {M{\'e}nard}, F.},
	doi = {10.1051/0004-6361:20010660},
	journal = {\aap},
	month = jun,
	pages = {L61-L64},
	title = {{Asymmetries in the HD 141569 circumstellar disk}},
	volume = {372},
	year = 2001
    }

@article{Clampin2003,
	adsurl = {http://adsabs.harvard.edu/abs/2003AJ....126..385C},
	author = {{Clampin}, M. and {Krist}, J.~E. and {Ardila}, D.~R. and {Golimowski}, D.~A. and {Hartig}, G.~F. and {Ford}, H.~C. and {Illingworth}, G.~D. and {Bartko}, F. and {Ben{\'{\i}}tez}, N. and {Blakeslee}, J.~P. and {Bouwens}, R.~J. and {Broadhurst}, T.~J. and {Brown}, R.~A. and {Burrows}, C.~J. and {Cheng}, E.~S. and {Cross}, N.~J.~G. and {Feldman}, P.~D. and {Franx}, M. and {Gronwall}, C. and {Infante}, L. and {Kimble}, R.~A. and {Lesser}, M.~P. and {Martel}, A.~R. and {Menanteau}, F. and {Meurer}, G.~R. and {Miley}, G.~K. and {Postman}, M. and {Rosati}, P. and {Sirianni}, M. and {Sparks}, W.~B. and {Tran}, H.~D. and {Tsvetanov}, Z.~I. and {White}, R.~L. and {Zheng}, W.},
	eprint = {arXiv:astro-ph/0303605},
	journal = {\aj},
	month = jul,
	pages = {385-392},
	title = {{Hubble Space Telescope ACS Coronagraphic Imaging of the Circumstellar Disk around HD 141569A}},
	volume = 126,
	year = 2003
    }

@article{Millar-Blanchaer2025,
	adsurl = {https://ui.adsabs.harvard.edu/abs/2025ApJ...994..199M},
	author = {{Millar-Blanchaer}, Maxwell A. and {Choquet}, {\'E}lodie and {Lawson}, Kellen and {Marino}, Sebasti{\'a}n and {Kammerer}, Jens and {Carter}, Aarynn L. and {Rebollido}, Isabel and {Leisenring}, Jarron M. and {Kim}, Minjae and {Kalas}, Paul and {Stapelfeldt}, Karl R. and {Hinkley}, Sasha and {Booth}, Mark and {Grady}, Carol A. and {Matthews}, Elisabeth C. and {Biller}, Beth A. and {Skemer}, Andrew and {Girard}, Julien H. and {Wolff}, Schuyler G. and {Ward-Duong}, Kimberly and {Meyer}, Michael R. and {Boccaletti}, Anthony and {Pantin}, Eric and {Matthews}, Brenda C. and {Metchev}, Stanimir and {Perrin}, Marshall D. and {Chen}, Christine H. and {Crotts}, Katie and {Absil}, Olivier and {Balmer}, William O. and {Calissendorff}, Per and {Cugno}, Gabriele and {Currie}, Thayne and {Danielski}, Camilla and {Hoch}, Kielan K.~W. and {Janson}, Markus and {Manjavacas}, Elena and {Lagage}, Pierre-Olivier and {Ben J. Sutlieff} and {Ray}, Shrishmoy and {Ren}, Bin B. and {Rickman}, Emily and {Su{\'a}rez}, Genaro and {Theissen}, Christopher A. and {Uyama}, Taichi and {Quirrenbach}, Andreas and {Wang}, Jason J. and {Whiteford}, Niall and {Wyatt}, Mark C. and {Zurlo}, Alice and {JWST ERS Collaboration}},
	doi = {10.3847/1538-4357/ae0615},
	eid = {199},
	journal = {\apj},
	month = dec,
	number = {2},
	pages = {199},
	title = {{The JWST Early Release Science Program for Direct Observations of Exoplanetary Systems. VI. Evidence for Radially Evolving Icy Grains in the HD 141569A Disk via NIRCam Coronagraphic Imaging}},
	volume = {994},
	year = 2025
    }

@article{Hinkley2022,
	adsurl = {https://ui.adsabs.harvard.edu/abs/2022PASP..134i5003H},
	archiveprefix = {arXiv},
	author = {{Hinkley}, Sasha and {Carter}, Aarynn L. and {Ray}, Shrishmoy and {Skemer}, Andrew and {Biller}, Beth and {Choquet}, Elodie and {Millar-Blanchaer}, Maxwell A. and {Sallum}, Stephanie and {Miles}, Brittany and {Whiteford}, Niall and {Patapis}, Polychronis and {Perrin}, Marshall and {Pueyo}, Laurent and {Schneider}, Glenn and {Stapelfeldt}, Karl and {Wang}, Jason and {Ward-Duong}, Kimberly and {Bowler}, Brendan P. and {Boccaletti}, Anthony and {H. Girard}, Julien and {Hines}, Dean and {Kalas}, Paul and {Kammerer}, Jens and {Kervella}, Pierre and {Leisenring}, Jarron and {Pantin}, Eric and {Zhou}, Yifan and {Meyer}, Michael and {Liu}, Michael C. and {Bonnefoy}, Mickael and {Currie}, Thayne and {McElwain}, Michael and {Metchev}, Stanimir and {Wyatt}, Mark and {Absil}, Olivier and {Adams}, Jea and {Barman}, Travis and {Baraffe}, Isabelle and {Bonavita}, Mariangela and {Booth}, Mark and {Bryan}, Marta and {Chauvin}, Gael and {Chen}, Christine and {Danielski}, Camilla and {De Furio}, Matthew and {Factor}, Samuel M. and {Fitzgerald}, Michael P. and {Fortney}, Jonathan J. and {Grady}, Carol and {Greenbaum}, Alexandra and {Henning}, Thomas and {Hoch}, Kielan K.~W. and {Janson}, Markus and {Kennedy}, Grant and {Kenworthy}, Matthew and {Kraus}, Adam and {Kuzuhara}, Masayuki and {Lagage}, Pierre-Olivier and {Lagrange}, Anne-Marie and {Launhardt}, Ralf and {Lazzoni}, Cecilia and {Lloyd}, James and {Marino}, Sebastian and {Marley}, Mark and {Martinez}, Raquel and {Marois}, Christian and {Matthews}, Brenda and {Matthews}, Elisabeth C. and {Mawet}, Dimitri and {Mazoyer}, Johan and {Phillips}, Mark and {Petrus}, Simon and {Quanz}, Sascha P. and {Quirrenbach}, Andreas and {Rameau}, Julien and {Rebollido}, Isabel and {Rickman}, Emily and {Samland}, Matthias and {Sargent}, B. and {Schlieder}, Joshua E. and {Sivaramakrishnan}, Anand and {Stone}, Jordan M. and {Tamura}, Motohide and {Tremblin}, Pascal and {Uyama}, Taichi and {Vasist}, Malavika and {Vigan}, Arthur and {Wagner}, Kevin and {Ygouf}, Marie},
	doi = {10.1088/1538-3873/ac77bd},
	eid = {095003},
	eprint = {2205.12972},
	journal = {\pasp},
	month = sep,
	number = {1039},
	pages = {095003},
	primaryclass = {astro-ph.EP},
	title = {{The JWST Early Release Science Program for the Direct Imaging and Spectroscopy of Exoplanetary Systems}},
	volume = {134},
	year = 2022
    }

@article{Feinberg2024,
	adsurl = {https://ui.adsabs.harvard.edu/abs/2024JATIS..10a1204F},
	author = {{Feinberg}, Lee D. and {McElwain}, Michael W. and {Bowers}, Charles W. and {Johnston}, John D. and {Mosier}, Gary E. and {Kimble}, Randy A. and {Levi}, Joshua S. and {Lightsey}, Paul and {Scott Knight}, J. and {Bluth}, Marcel and {Jurling}, Alden S. and {Levine}, Marie B. and {Scott Acton}, D. and {Atkinson}, Charles and {Barto}, Allison and {Bergkoetter}, Matthew D. and {Brady}, Gregory R. and {Carey}, Larkin and {Cohen}, Lester and {Coyle}, Laura and {Dean}, Bruce H. and {Eisenhower}, Michael and {Flagey}, Nicolas and {Hartig}, George F. and {Havey}, Keith A. and {Hicks}, Brian and {Howard}, Joseph M. and {Keski-Kuha}, Ritva A. and {Lajoie}, Charles-Philippe and {Lallo}, Matthew D. and {Matthews}, Gary W. and {Mel{\'e}ndez}, Marcio and {Menzel}, Michael T. and {Park}, Sang and {Perrin}, Marshall D. and {Pueyo}, Laurent and {Quesnel}, Lisbeth and {Reynolds}, Paul and {Rigby}, Jane R. and {Saif}, Babak N. and {Stark}, Christopher C. and {Telfer}, Randal and {Texter}, Scott C. and {Van Campen}, Julie M. and {Vila}, Bego{\~n}a. and {West}, Garrett and {Wolf}, Erin and {Whitman}, Tony L. and {Zielinski}, Thomas P.},
	doi = {10.1117/1.JATIS.10.1.011204},
	eid = {011204},
	journal = {Journal of Astronomical Telescopes, Instruments, and Systems},
	month = jan,
	pages = {011204},
	title = {{James Webb Space Telescope optical stability lessons learned for future great observatories}},
	volume = {10},
	year = 2024
    }

@inproceedings{Telfer2024,
	adsurl = {https://ui.adsabs.harvard.edu/abs/2024SPIE13092E..10T},
	author = {{Telfer}, Randal C. and {Mel{\'e}ndez}, Marcio and {Flagey}, Nicolas and {Lajoie}, Charles-Philippe and {Brady}, Gregory R. and {Beck}, Tracy and {Comeau}, Thomas M. and {Kulp}, Bernard A. and {Perrin}, Marshall D. and {Lallo}, Matthew D.},
	booktitle = {Space Telescopes and Instrumentation 2024: Optical, Infrared, and Millimeter Wave},
	doi = {10.1117/12.3020273},
	editor = {{Coyle}, Laura E. and {Matsuura}, Shuji and {Perrin}, Marshall D.},
	eid = {1309210},
	month = aug,
	pages = {1309210},
	series = {Society of Photo-Optical Instrumentation Engineers (SPIE) Conference Series},
	title = {{Empirical characterization of JWST wavefront error variations}},
	volume = {13092},
	year = 2024
    }

@misc{Biller2025,
	adsurl = {https://ui.adsabs.harvard.edu/abs/2025jwst.prop.6005B},
	author = {{Biller}, Beth and {Carter}, Aarynn L. and {Kraus}, Adam L. and {Quirrenbach}, Andreas and {James}, Andrew David and {Sutlieff}, Ben and {Bowler}, Brendan and {Miles}, Brittany E. and {Marois}, Christian and {Matthews}, Elisabeth C. and {Suarez}, Genaro and {Rebollido}, Isabel and {Vos}, Johanna and {Leisenring}, Jarron and {Kammerer}, Jens and {Stone}, Jordan Michael and {Girard}, Julien and {Wagner}, Kevin and {Franson}, Kyle and {Booth}, Mark and {Kenworthy}, Matthew and {Samland}, Matthias and {Meyer}, Michael R. and {Liu}, Pengyu and {Bowens-Rubin}, Rachel and {Martinez}, Raquel and {De Rosa}, Robert J. and {Factor}, Samuel M. and {Hinkley}, Sasha and {Marino}, Sebastian and {Petrus}, Simon and {Currie}, Thayne M. and {Dupuy}, Trent J. and {Christiaens}, Valentin and {Balmer}, William and {Zhou}, Yifan and {Li}, Yiting and {Zhang}, Zhoujian and {Lazzoni}, Cecilia},
	howpublished = {JWST Proposal. Cycle 3, ID. \#6005},
	month = may,
	pages = {6005},
	title = {{Imaging Young Sub-Jupiter Planets down to Solar-System Scales}},
	year = 2025}

@misc{Carter2024,
	adsurl = {https://ui.adsabs.harvard.edu/abs/2024jwst.prop.5835C},
	author = {{Carter}, Aarynn and {Absil}, Olivier and {Balmer}, William and {Biller}, Beth and {Bogat}, Ell and {Bonavita}, Mariangela and {Booth}, Mark and {Bowens-Rubin}, Rachel and {Bowler}, Brendan and {Calissendorff}, Per and {Chauvin}, Gael and {Fontanive}, Clemence and {Franson}, Kyle and {Gagne}, Jonathan and {Girard}, Julien and {Hinkley}, Sasha and {Hoch}, Kielan K.~W. and {Kammerer}, Jens and {Kennedy}, Grant and {Leisenring}, Jarron Michael and {Li}, Yiting and {Limbach}, Mary Anne and {Liu}, Michael C. and {Macintosh}, Bruce A. and {Matthews}, Elisabeth C. and {Meyer}, Michael R. and {Millar-Blanchaer}, Maxwell Andrew and {Morley}, Caroline and {Perrin}, Marshall and {Pueyo}, Laurent and {Ray}, Shrishmoy and {Rebollido}, Isabel and {Rickman}, Emily and {Skemer}, Andrew and {Wang}, Jason J. and {Ward-Duong}, Kimberly and {Whiteford}, Niall},
	howpublished = {JWST Proposal. Cycle 3, ID. \#5835},
	month = feb,
	pages = {5835},
	title = {{Into The Spotlight: Unveiling Wide-Separation Sub-Jupiters for Future JWST Characterization}},
	year = 2024}

@misc{Kammerer2025,
	adsurl = {https://ui.adsabs.harvard.edu/abs/2025jwst.prop.7651K},
	author = {{Kammerer}, Jens and {Pueyo}, Laurent and {Balmer}, William and {Bogat}, Ell and {Bonavita}, Mariangela and {Burr}, Zach and {Carter}, Aarynn L. and {Chen}, Christine and {Cugno}, Gabriele and {Fontanive}, Clemence and {Ginski}, Christian and {Girard}, Julien and {Hinkley}, Sasha and {Hoch}, Kielan K.~W. and {Kenworthy}, Matthew and {Manjavacas}, Elena and {Perrin}, Marshall and {Rebollido}, Isabel and {Rickman}, Emily and {van Capelleveen}, Richelle Felicia},
	howpublished = {JWST Proposal. Cycle 4, ID. \#7651},
	month = mar,
	pages = {7651},
	title = {{JWST-YSES: a Young Suns Exoplanet Survey to study the demographics of sub-Jovian planets around Sun-like stars and unveil the formation and evolution history of widely separated companions}},
	year = 2025}

@article{Beuzit2019,
	adsurl = {https://ui.adsabs.harvard.edu/abs/2019A&A...631A.155B},
	archiveprefix = {arXiv},
	author = {{Beuzit}, J. -L. and {Vigan}, A. and {Mouillet}, D. and {Dohlen}, K. and {Gratton}, R. and {Boccaletti}, A. and {Sauvage}, J. -F. and {Schmid}, H.~M. and {Langlois}, M. and {Petit}, C. and {Baruffolo}, A. and {Feldt}, M. and {Milli}, J. and {Wahhaj}, Z. and {Abe}, L. and {Anselmi}, U. and {Antichi}, J. and {Barette}, R. and {Baudrand}, J. and {Baudoz}, P. and {Bazzon}, A. and {Bernardi}, P. and {Blanchard}, P. and {Brast}, R. and {Bruno}, P. and {Buey}, T. and {Carbillet}, M. and {Carle}, M. and {Cascone}, E. and {Chapron}, F. and {Charton}, J. and {Chauvin}, G. and {Claudi}, R. and {Costille}, A. and {De Caprio}, V. and {de Boer}, J. and {Delboulb{\'e}}, A. and {Desidera}, S. and {Dominik}, C. and {Downing}, M. and {Dupuis}, O. and {Fabron}, C. and {Fantinel}, D. and {Farisato}, G. and {Feautrier}, P. and {Fedrigo}, E. and {Fusco}, T. and {Gigan}, P. and {Ginski}, C. and {Girard}, J. and {Giro}, E. and {Gisler}, D. and {Gluck}, L. and {Gry}, C. and {Henning}, T. and {Hubin}, N. and {Hugot}, E. and {Incorvaia}, S. and {Jaquet}, M. and {Kasper}, M. and {Lagadec}, E. and {Lagrange}, A. -M. and {Le Coroller}, H. and {Le Mignant}, D. and {Le Ruyet}, B. and {Lessio}, G. and {Lizon}, J. -L. and {Llored}, M. and {Lundin}, L. and {Madec}, F. and {Magnard}, Y. and {Marteaud}, M. and {Martinez}, P. and {Maurel}, D. and {M{\'e}nard}, F. and {Mesa}, D. and {M{\"o}ller-Nilsson}, O. and {Moulin}, T. and {Moutou}, C. and {Orign{\'e}}, A. and {Parisot}, J. and {Pavlov}, A. and {Perret}, D. and {Pragt}, J. and {Puget}, P. and {Rabou}, P. and {Ramos}, J. and {Reess}, J. -M. and {Rigal}, F. and {Rochat}, S. and {Roelfsema}, R. and {Rousset}, G. and {Roux}, A. and {Saisse}, M. and {Salasnich}, B. and {Santambrogio}, E. and {Scuderi}, S. and {Segransan}, D. and {Sevin}, A. and {Siebenmorgen}, R. and {Soenke}, C. and {Stadler}, E. and {Suarez}, M. and {Tiph{\`e}ne}, D. and {Turatto}, M. and {Udry}, S. and {Vakili}, F. and {Waters}, L.~B.~F.~M. and {Weber}, L. and {Wildi}, F. and {Zins}, G. and {Zurlo}, A.},
	doi = {10.1051/0004-6361/201935251},
	eid = {A155},
	eprint = {1902.04080},
	journal = {\aap},
	month = nov,
	pages = {A155},
	primaryclass = {astro-ph.IM},
	title = {{SPHERE: the exoplanet imager for the Very Large Telescope}},
	volume = {631},
	year = 2019
    }

@article{Konopacky2013,
	adsurl = {http://adsabs.harvard.edu/abs/2013Sci...339.1398K},
	archiveprefix = {arXiv},
	author = {{Konopacky}, Q.~M. and {Barman}, T.~S. and {Macintosh}, B.~A. and {Marois}, C.},
	doi = {10.1126/science.1232003},
	eprint = {1303.3280},
	journal = {Science},
	month = mar,
	pages = {1398-1401},
	primaryclass = {astro-ph.EP},
	title = {{Detection of Carbon Monoxide and Water Absorption Lines in an Exoplanet Atmosphere}},
	volume = 339,
	year = 2013
    }

@article{Chabrier2007,
	adsurl = {http://adsabs.harvard.edu/abs/2007prpl.conf..623C},
	author = {{Chabrier}, G. and {Baraffe}, I. and {Selsis}, F. and {Barman}, T.~S. and {Hennebelle}, P. and {Alibert}, Y.},
	eprint = {arXiv:astro-ph/0602291},
	journal = {Protostars and Planets V},
	pages = {623-638},
	title = {{Gaseous Planets, Protostars, and Young Brown Dwarfs: Birth and Fate}},
	year = 2007
    }

@article{Barman2011,
	adsurl = {http://adsabs.harvard.edu/abs/2011ApJ...733...65B},
	archiveprefix = {arXiv},
	author = {{Barman}, T.~S. and {Macintosh}, B. and {Konopacky}, Q.~M. and {Marois}, C.},
	doi = {10.1088/0004-637X/733/1/65},
	eid = {65},
	eprint = {1103.3895},
	journal = {\apj},
	month = may,
	pages = {65},
	primaryclass = {astro-ph.EP},
	title = {{Clouds and Chemistry in the Atmosphere of Extrasolar Planet HR8799b}},
	volume = 733,
	year = 2011
    }

@article{Hubeny2007,
	adsurl = {https://ui.adsabs.harvard.edu/abs/2007ApJ...669.1248H},
	archiveprefix = {arXiv},
	author = {{Hubeny}, Ivan and {Burrows}, Adam},
	doi = {10.1086/522107},
	eprint = {0705.3922},
	journal = {\apj},
	month = nov,
	number = {2},
	pages = {1248-1261},
	primaryclass = {astro-ph},
	title = {{A Systematic Study of Departures from Chemical Equilibrium in the Atmospheres of Substellar Mass Objects}},
	volume = {669},
	year = 2007
    }

@article{Miles2020,
	adsurl = {https://ui.adsabs.harvard.edu/abs/2020AJ....160...63M},
	archiveprefix = {arXiv},
	author = {{Miles}, Brittany E. and {Skemer}, Andrew J.~I. and {Morley}, Caroline V. and {Marley}, Mark S. and {Fortney}, Jonathan J. and {Allers}, Katelyn N. and {Faherty}, Jacqueline K. and {Geballe}, Thomas R. and {Visscher}, Channon and {Schneider}, Adam C. and {Lupu}, Roxana and {Freedman}, Richard S. and {Bjoraker}, Gordon L.},
	doi = {10.3847/1538-3881/ab9114},
	eid = {63},
	eprint = {2004.10770},
	journal = {\aj},
	month = aug,
	number = {2},
	pages = {63},
	primaryclass = {astro-ph.EP},
	title = {{Observations of Disequilibrium CO Chemistry in the Coldest Brown Dwarfs}},
	volume = {160},
	year = 2020
    }

@article{Skemer2012,
	adsurl = {http://adsabs.harvard.edu/abs/2012ApJ...753...14S},
	archiveprefix = {arXiv},
	author = {{Skemer}, A.~J. and {Hinz}, P.~M. and {Esposito}, S. and {Burrows}, A. and {Leisenring}, J. and {Skrutskie}, M. and {Desidera}, S. and {Mesa}, D. and {Arcidiacono}, C. and {Mannucci}, F. and {Rodigas}, T.~J. and {Close}, L. and {McCarthy}, D. and {Kulesa}, C. and {Agapito}, G. and {Apai}, D. and {Argomedo}, J. and {Bailey}, V. and {Boutsia}, K. and {Briguglio}, R. and {Brusa}, G. and {Busoni}, L. and {Claudi}, R. and {Eisner}, J. and {Fini}, L. and {Follette}, K.~B. and {Garnavich}, P. and {Gratton}, R. and {Guerra}, J.~C. and {Hill}, J.~M. and {Hoffmann}, W.~F. and {Jones}, T. and {Krejny}, M. and {Males}, J. and {Masciadri}, E. and {Meyer}, M.~R. and {Miller}, D.~L. and {Morzinski}, K. and {Nelson}, M. and {Pinna}, E. and {Puglisi}, A. and {Quanz}, S.~P. and {Quiros-Pacheco}, F. and {Riccardi}, A. and {Stefanini}, P. and {Vaitheeswaran}, V. and {Wilson}, J.~C. and {Xompero}, M.},
	doi = {10.1088/0004-637X/753/1/14},
	eid = {14},
	eprint = {1203.2615},
	journal = {\apj},
	month = jul,
	pages = {14},
	primaryclass = {astro-ph.EP},
	title = {{First Light LBT AO Images of HR 8799 bcde at 1.6 and 3.3{$\mu$}m: New Discrepancies between Young Planets and Old Brown Dwarfs}},
	volume = 753,
	year = 2012
    }

@article{Matthews2024,
	adsurl = {https://ui.adsabs.harvard.edu/abs/2024Natur.633..789M},
	archiveprefix = {arXiv},
	author = {{Matthews}, E.~C. and {Carter}, A.~L. and {Pathak}, P. and {Morley}, C.~V. and {Phillips}, M.~W. and {P.~M.}, S. Krishanth and {Feng}, F. and {Bonse}, M.~J. and {Boogaard}, L.~A. and {Burt}, J.~A. and {Crossfield}, I.~J.~M. and {Douglas}, E.~S. and {Henning}, Th. and {Hom}, J. and {Ko}, C. -L. and {Kasper}, M. and {Lagrange}, A. -M. and {Petit dit de la Roche}, D. and {Philipot}, F.},
	doi = {10.1038/s41586-024-07837-8},
	eprint = {2503.01599},
	journal = {\nat},
	month = sep,
	number = {8031},
	pages = {789-792},
	primaryclass = {astro-ph.EP},
	title = {{A temperate super-Jupiter imaged with JWST in the mid-infrared}},
	volume = {633},
	year = 2024
    }

@article{Golimowski2006,
	adsurl = {http://adsabs.harvard.edu/abs/2006AJ....131.3109G},
	author = {{Golimowski}, D.~A. and {Ardila}, D.~R. and {Krist}, J.~E. and {Clampin}, M. and {Ford}, H.~C. and {Illingworth}, G.~D. and {Bartko}, F. and {Ben{\'{\i}}tez}, N. and {Blakeslee}, J.~P. and {Bouwens}, R.~J. and {Bradley}, L.~D. and {Broadhurst}, T.~J. and {Brown}, R.~A. and {Burrows}, C.~J. and {Cheng}, E.~S. and {Cross}, N.~J.~G. and {Demarco}, R. and {Feldman}, P.~D. and {Franx}, M. and {Goto}, T. and {Gronwall}, C. and {Hartig}, G.~F. and {Holden}, B.~P. and {Homeier}, N.~L. and {Infante}, L. and {Jee}, M.~J. and {Kimble}, R.~A. and {Lesser}, M.~P. and {Martel}, A.~R. and {Mei}, S. and {Menanteau}, F. and {Meurer}, G.~R. and {Miley}, G.~K. and {Motta}, V. and {Postman}, M. and {Rosati}, P. and {Sirianni}, M. and {Sparks}, W.~B. and {Tran}, H.~D. and {Tsvetanov}, Z.~I. and {White}, R.~L. and {Zheng}, W. and {Zirm}, A.~W.},
	doi = {10.1086/503801},
	eprint = {arXiv:astro-ph/0602292},
	journal = {\aj},
	month = jun,
	pages = {3109-3130},
	title = {{Hubble Space Telescope ACS Multiband Coronagraphic Imaging of the Debris Disk around {$\beta$} Pictoris}},
	volume = 131,
	year = 2006
    }

@article{Krist2005,
	adsurl = {http://adsabs.harvard.edu/abs/2005AJ....129.1008K},
	author = {{Krist}, J.~E. and {Ardila}, D.~R. and {Golimowski}, D.~A. and {Clampin}, M. and {Ford}, H.~C. and {Illingworth}, G.~D. and {Hartig}, G.~F. and {Bartko}, F. and {Ben{\'{\i}}tez}, N. and {Blakeslee}, J.~P. and {Bouwens}, R.~J. and {Bradley}, L.~D. and {Broadhurst}, T.~J. and {Brown}, R.~A. and {Burrows}, C.~J. and {Cheng}, E.~S. and {Cross}, N.~J.~G. and {Demarco}, R. and {Feldman}, P.~D. and {Franx}, M. and {Goto}, T. and {Gronwall}, C. and {Holden}, B. and {Homeier}, N. and {Infante}, L. and {Kimble}, R.~A. and {Lesser}, M.~P. and {Martel}, A.~R. and {Mei}, S. and {Menanteau}, F. and {Meurer}, G.~R. and {Miley}, G.~K. and {Motta}, V. and {Postman}, M. and {Rosati}, P. and {Sirianni}, M. and {Sparks}, W.~B. and {Tran}, H.~D. and {Tsvetanov}, Z.~I. and {White}, R.~L. and {Zheng}, W.},
	journal = {\aj},
	month = feb,
	pages = {1008-1017},
	title = {{Hubble Space Telescope Advanced Camera for Surveys Coronagraphic Imaging of the AU Microscopii Debris Disk}},
	volume = 129,
	year = 2005
    }

@article{Stapelfeldt1998,
	adsurl = {http://adsabs.harvard.edu/abs/1998ApJ...502L..65S},
	author = {{Stapelfeldt}, K.~R. and {Krist}, J.~E. and {Menard}, F. and {Bouvier}, J. and {Padgett}, D.~L. and {Burrows}, C.~J.},
	journal = {\apjl},
	month = jul,
	pages = {L65+},
	title = {{An Edge-On Circumstellar Disk in the Young Binary System HK Tauri}},
	volume = 502,
	year = 1998
    }

@article{Davies2012,
	adsurl = {http://adsabs.harvard.edu/abs/2012arXiv1201.5741D},
	archiveprefix = {arXiv},
	author = {{Davies}, R. and {Kasper}, M.},
	eprint = {1201.5741},
	journal = {ArXiv e-prints},
	month = jan,
	primaryclass = {astro-ph.IM},
	title = {{Adaptive Optics for Astronomy}},
	year = 2012
    }

@article{Oppenheimer2009,
	adsurl = {http://adsabs.harvard.edu/abs/2009ARA%26A..47..253O},
	archiveprefix = {arXiv},
	author = {{Oppenheimer}, B.~R. and {Hinkley}, S.},
	doi = {10.1146/annurev-astro-082708-101717},
	eprint = {0903.4466},
	journal = {\araa},
	month = sep,
	pages = {253-289},
	title = {{High-Contrast Observations in Optical and Infrared Astronomy}},
	volume = 47,
	year = 2009
    }

@inproceedings{Perrin2018,
	adsurl = {https://ui.adsabs.harvard.edu/abs/2018SPIE10698E..09P},
	author = {{Perrin}, Marshall D. and {Pueyo}, Laurent and {Van Gorkom}, Kyle and {Brooks}, Keira and {Rajan}, Abhijith and {Girard}, Julien and {Lajoie}, Charles-Philippe},
	booktitle = {Space Telescopes and Instrumentation 2018: Optical, Infrared, and Millimeter Wave},
	doi = {10.1117/12.2313552},
	eid = {1069809},
	month = aug,
	pages = {1069809},
	series = {Society of Photo-Optical Instrumentation Engineers (SPIE) Conference Series},
	title = {{Updated optical modeling of JWST coronagraph performance contrast, stability, and strategies}},
	volume = {10698},
	year = 2018
    }

@article{Soummer2012,
	adsurl = {http://adsabs.harvard.edu/abs/2012ApJ...755L..28S},
	archiveprefix = {arXiv},
	author = {{Soummer}, R. and {Pueyo}, L. and {Larkin}, J.},
	doi = {10.1088/2041-8205/755/2/L28},
	eid = {L28},
	eprint = {1207.4197},
	journal = {\apjl},
	month = aug,
	pages = {L28},
	primaryclass = {astro-ph.IM},
	title = {{Detection and Characterization of Exoplanets and Disks Using Projections on Karhunen-Lo{\`e}ve Eigenimages}},
	volume = 755,
	year = 2012
    }

@article{Lafreniere2007,
	adsurl = {http://adsabs.harvard.edu/abs/2007ApJ...660..770L},
	author = {{Lafreni{\`e}re}, D. and {Marois}, C. and {Doyon}, R. and {Nadeau}, D. and {Artigau}, {\'E}.},
	eprint = {arXiv:astro-ph/0702697},
	journal = {\apj},
	month = may,
	pages = {770-780},
	title = {{A New Algorithm for Point-Spread Function Subtraction in High-Contrast Imaging: A Demonstration with Angular Differential Imaging}},
	volume = 660,
	year = 2007
    }

@article{Marois2006,
	adsurl = {http://adsabs.harvard.edu/cgi-bin/nph-bib_query?bibcode=2006ApJ...641..556M&db_key=AST},
	author = {{Marois}, C. and {Lafreni{\`e}re}, D. and {Doyon}, R. and {Macintosh}, B. and {Nadeau}, D.},
	eprint = {astro-ph/0512335},
	journal = {\apj},
	month = apr,
	pages = {556-564},
	title = {{Angular Differential Imaging: A Powerful High-Contrast Imaging Technique}},
	volume = 641,
	year = 2006
    }

@article{Phillips2020,
	adsurl = {https://ui.adsabs.harvard.edu/abs/2020A&A...637A..38P},
	archiveprefix = {arXiv},
	author = {{Phillips}, M.~W. and {Tremblin}, P. and {Baraffe}, I. and {Chabrier}, G. and {Allard}, N.~F. and {Spiegelman}, F. and {Goyal}, J.~M. and {Drummond}, B. and {H{\'e}brard}, E.},
	doi = {10.1051/0004-6361/201937381},
	eid = {A38},
	eprint = {2003.13717},
	journal = {\aap},
	month = may,
	pages = {A38},
	primaryclass = {astro-ph.SR},
	title = {{A new set of atmosphere and evolution models for cool T-Y brown dwarfs and giant exoplanets}},
	volume = {637},
	year = 2020
    }

@article{Baraffe2015,
	adsurl = {http://adsabs.harvard.edu/abs/2015A%26A...577A..42B},
	archiveprefix = {arXiv},
	author = {{Baraffe}, I. and {Homeier}, D. and {Allard}, F. and {Chabrier}, G.},
	doi = {10.1051/0004-6361/201425481},
	eid = {A42},
	eprint = {1503.04107},
	journal = {\aap},
	month = may,
	pages = {A42},
	primaryclass = {astro-ph.SR},
	title = {{New evolutionary models for pre-main sequence and main sequence low-mass stars down to the hydrogen-burning limit}},
	volume = 577,
	year = 2015
    }

@article{Carter2021,
	adsurl = {https://ui.adsabs.harvard.edu/abs/2021MNRAS.501.1999C},
	archiveprefix = {arXiv},
	author = {{Carter}, Aarynn L. and {Hinkley}, Sasha and {Bonavita}, Mariangela and {Phillips}, Mark W. and {Girard}, Julien H. and {Perrin}, Marshall and {Pueyo}, Laurent and {Vigan}, Arthur and {Gagn{\'e}}, Jonathan and {Skemer}, Andrew J.~I.},
	doi = {10.1093/mnras/staa3579},
	eprint = {2011.07075},
	journal = {\mnras},
	month = feb,
	number = {2},
	pages = {1999-2016},
	primaryclass = {astro-ph.EP},
	title = {{Direct imaging of sub-Jupiter mass exoplanets with James Webb Space Telescope coronagraphy}},
	volume = {501},
	year = 2021
    }

@article{Ruffio2026,
	adsurl = {https://ui.adsabs.harvard.edu/abs/2026NatAs.tmp...39R},
	archiveprefix = {arXiv},
	author = {{Ruffio}, Jean-Baptiste and {Xuan}, Jerry W. and {Chachan}, Yayaati and {Kesseli}, Aurora and {Lee}, Eve J. and {Beichman}, Charles and {Hodapp}, Klaus and {Balmer}, William O. and {Konopacky}, Quinn and {Perrin}, Marshall D. and {Mawet}, Dimitri and {Knutson}, Heather A. and {Bryden}, Geoffrey and {Greene}, Thomas P. and {Johnstone}, Doug and {Leisenring}, Jarron and {Meyer}, Michael and {Ygouf}, Marie},
	doi = {10.1038/s41550-026-02783-z},
	eprint = {2601.08227},
	journal = {Nature Astronomy},
	month = feb,
	primaryclass = {astro-ph.EP},
	title = {{Jupiter-like uniform metal enrichment in a system of multiple giant exoplanets}},
	year = 2026
    }

@article{Xuan2026,
	adsurl = {https://ui.adsabs.harvard.edu/abs/2026ApJ..1000...27X},
	archiveprefix = {arXiv},
	author = {{Xuan}, Jerry W. and {Ruffio}, Jean-Baptiste and {Chachan}, Yayaati and {Ohno}, Kazumasa and {Kesseli}, Aurora and {Murray-Clay}, Ruth and {Lee}, Eve J. and {Moses}, Julianne I. and {Balmer}, William O. and {Baburaj}, Aneesh and {Blake}, Geoffrey A. and {Johnstone}, Doug and {Zhang}, Yapeng and {Knutson}, Heather A. and {Mawet}, Dimitri and {Beichman}, Charles and {Hodapp}, Klaus and {Perrin}, Marshall D. and {Konopacky}, Quinn and {Meyer}, Michael and {Bryden}, Geoffrey and {Greene}, Thomas P. and {Leisenring}, Jarron and {Ygouf}, Marie and {Benneke}, Bj{\"o}rn and {Inglis}, Julie and {Wallack}, Nicole L.},
	doi = {10.3847/1538-4357/ae448f},
	eid = {27},
	eprint = {2602.09422},
	journal = {\apj},
	month = mar,
	number = {1},
	pages = {27},
	primaryclass = {astro-ph.EP},
	title = {{The Compositions of the HR 8799 Planets Reflect Accretion of Both Solids and Metal-enriched Gas}},
	volume = {1000},
	year = 2026
    }

\end{document}